\documentclass[twocolumn,trackchanges,onecolappendix]{aastex63}

\usepackage{natbib,enumitem,array,amsmath}

\graphicspath{{./}{figures/}}

\newcommand{\btwnpoint}{\hspace{-0.55ex}}

\newcommand{\pointsec}{%
  \mathrel{\vbox{\offinterlineskip\ialign{%
    ##\cr
    $\btwnpoint\scriptstyle\text{s}$\cr
    \noalign{\kern0.75ex}
    $\btwnpoint.$\cr
}}}}

\newcommand{\pointarcsec}{%
  \mathrel{\vbox{\offinterlineskip\ialign{%
    ##\cr
    $\btwnpoint\scriptstyle\prime\prime$\cr
    \noalign{\kern0.75ex}
    $\btwnpoint.$\cr
}}}}

\received{8 March 2022}
\revised{10 May 2022}
\accepted{11 May 2022}
\submitjournal{AJ}

\shorttitle{Colors of Substructures in Fornax Dwarf ETGs} 
\shortauthors{Michea et al.}

\begin{document}

\def\arraystretch{1.13} 
\setlength{\tabcolsep}{0.5em}

\title{Brought to Light III: Colors of Disk and Clump Substructures in Dwarf Early-Type Galaxies of the Fornax Cluster}

\correspondingauthor{Josefina Michea}
\email{michea@ari.uni-heidelberg.de}

\author[0000-0001-8205-945X]{Josefina Michea}
\affiliation{Astronomisches Rechen-Institut, Zentrum f\"ur Astronomie der Universit\"at Heidelberg, M\"onchhofstra{\ss}e 12-14, 69120 Heidelberg, Germany}

\author[0000-0001-5171-5629]{Anna Pasquali}
\affiliation{Astronomisches Rechen-Institut, Zentrum f\"ur Astronomie der Universit\"at Heidelberg, M\"onchhofstra{\ss}e 12-14, 69120 Heidelberg, Germany}

\author[0000-0001-5303-6830]{Rory Smith}
\affiliation{Departamento de F\'isica, Universidad T\'ecnica Federico Santa Mar\'ia, Avenida Vicu\~na Mackenna 3939, San Joaqu\'in, Santiago, Chile}

\author[0000-0002-7069-113X]{Paula Calder\'on-Castillo}
\affiliation{Departamento de Astronom\'ia, Universidad de Concepci\'on, Casilla 160-C, Concepci\'on, Chile}

\author[0000-0002-1891-3794]{Eva K. Grebel}
\affiliation{Astronomisches Rechen-Institut, Zentrum f\"ur Astronomie der Universit\"at Heidelberg, M\"onchhofstra{\ss}e 12-14, 69120 Heidelberg, Germany}

\author[0000-0001-7621-947X]{Reynier F. Peletier}
\affiliation{Kapteyn Astronomical Institute, University of Groningen, Landleven 12, NL-9747 AD Groningen, The Netherlands}

\begin{abstract} 
It has been well established that dwarf early-type galaxies (ETGs) can often exhibit a complex morphology, whereby faint spiral arms, bars, edge-on disks or clumps are embedded in their main, brighter diffuse body. In our first paper \citep[Brought to Light I:][]{Michea:2021}, we developed a new method for robustly identifying and extracting substructures in deep imaging data of dwarf ETGs in the Virgo galaxy cluster. Here we apply our method to a sample of $23$ dwarf ETGs in the Fornax galaxy cluster, out of which $9$ have disk-like and $14$ have clump-like substructures. According to Fornax Deep Survey (FDS) data, our sample constitutes $12\%$ of all dwarf ETGs in Fornax brighter than $\text{M}_{r}=-13$ mag, and contains all cases that unequivocally exhibit substructure features. We use $g$ and $r$-band FDS images to measure the relative contribution of the substructures to the total galaxy light and to estimate their $g-r$ colors. We find that substructures typically contribute $8.7\%$ and $5.3\%$ of the total galaxy light in the $g$ and $r$ bands, respectively, within two effective radii. Disk substructures are usually found in dwarf ETGs with redder global colors, and they can be either as red as or bluer than their galaxy's diffuse component. In contrast, clump substructures are found in comparatively bluer dwarf ETGs, and they are always bluer than their galaxy's diffuse component. These results provide further evidence that dwarf ETGs can hide diverse complex substructures, with stellar populations that can greatly differ from those of the dominant diffuse light in which they are embedded.
\end{abstract}


\section{Introduction}\label{sec:introduction}

The galaxies in the local Universe are dominated, in number, by dwarf galaxies fainter than M$_{V} = -18$ mag \citep{Ferguson:1994} and less massive than $\log (\text{M}_{\star}/\text{M}_{\odot}) = 9.0-9.5$. These galaxies can be subdivided into late-type (LTG) and early-type (ETG) systems mainly on the basis of their global colors and star formation activity. Indeed, dwarf LTGs, characterized by blue colors, are actively forming stars, are rich in gas and dust, and display either an irregular morphology with clump substructures  \citep[e.g.,][]{Carignan:1989,Skillman:1989,Conselice:2003b,Ann:2015} or disk structures such as spiral arms and bars similarly to L$_{\star}$ spiral galaxies \citep[e.g.,][]{Grebel:2001}. They typically reside in low density environments, like in the field and galaxy groups. In contrast, dwarf ETGs with their red colors are mostly quenched and devoid of gas. This category includes dwarf ellipticals \citep[dEs;][]{Ferguson:1994,Graham:2003b}, dwarf lenticulars \citep[dS0s;][]{Sandage:1984,Binggeli:1991}, and dwarf spheroidal systems \citep[dSphs;][]{Harbeck:2001,Grebel:2003}, all characterized by an overall smooth and featureless appearance. Dwarf ETGs predominantly populate galaxy clusters \citep{Binggeli:1985}, with a probability to find them in the field lower than $0.06\%$, according to \citet{Geha:2012}.  

Detailed studies of the morphology of dwarf ETGs have revealed a number of unexpected features. For example, while a single S\'ersic profile \citep{Sersic:1968} may approximate the overall light distribution, a large fraction of dwarf ETGs are better fit by multiple components \citep{Janz:2014,Su:2021}. In some dwarf ETGs, color imaging reveals the presence of blue cores, where star formation is taking place or has occurred in the last $1$ Gyr \citep{Hodge:1973,Vigroux:1984,Peletier:1993,Lisker:2006b,Pak:2014,Urich:2017,Hamraz:2019}. These blue-cored dwarf ETGs are usually found in the outskirts of galaxy clusters where the galaxy density is lowest. The more massive dwarf ETGs exhibit nuclei that tend to be somewhat bluer than their host galaxies \citep{Lotz:2004,Cote:2006,Paudel:2010,Paudel:2011,Neumayer:2020,Poulain:2021}.
In addition, spiral arms and bars have been detected in several dwarf ETGs in the Virgo cluster \citep{Jerjen:2001,Barazza:2002,Ferrarese:2006,Lisker:2006a}, as well as in the Fornax \citep{DeRijcke:2003,Venhola:2019}, Coma \citep{Graham:2003}, and Perseus clusters \citep{Penny:2014}. Recently, \citet{Paudel:2017} revealed the presence of shells in some cluster dwarf ETGs, possibly formed during dwarf-dwarf mergers.

In the literature, the main formation channels of dwarf ETGs are often boiled down to ``nature'' or ``nurture'', or some combination of the two, in order to try to explain their observed diversity. Nature means their properties are a result of the manner in which they formed, such as their mass growth and merger history. In the case of nurture, their properties are altered as a result of their interaction with environmental mechanisms. In the latter, dwarf ETGs could arise from the environmentally-driven transformation of dwarf irregulars and dwarf spiral galaxies, for instance after strangulation \citep{Larson:1980}, ram-pressure stripping \citep{Gunn:1972}, and galaxy-galaxy, or galaxy-cluster tidal interactions \citep{Moore:1999} have deprived them of their gas and halted their star formation \citep[see, e.g.,][and references therein]{Lisker:2009b}.

The observed substructures within dwarf ETGs may provide valuable constraints on the formation channels of these galaxies. A comparative analysis of the luminosities and colors of the small-scale features and the more diffuse component of dwarf ETGs may reveal differences in their corresponding stellar populations and star formation histories. One type of substructure that is of particular interest to this study is that which could be best described as disk-like features. These include disks that appear thin when seen near edge-on or, in more face-on cases, tightly wrapped spiral arms and in some cases bars. \citet{Lisker:2006a} conducted the first systematic study of such disk-like features hidden in the dominant diffuse light of dwarf ETGs, and attempted to measure the fraction of light contained in these features for dwarf ETGs in the Virgo cluster. They found values ranging between $5-15\%$, confirming that these features are indeed faint, and found that the fraction of dwarf ETGs with detected substructures increases from few percent at $\text{M}_{V} > -15$ mag to as high as $\sim50\%$ at $\text{M}_{V} = -17$ mag. However, as Sloan Digital Sky Survey (SDSS) Data Release $4$ \citep[DR4;][]{Adelman-McCarthy:2006} imaging was used, the data were quite shallow and poorly resolved. In order to produce deeper images, coadded images were produced by stacking multiple filters but, in the process, color information was lost.

In Brought to Light I \citep{Michea:2021}, we improved on the measurement technique employed by \citet{Lisker:2006a} by developing the ``residual method'', which we used to determine the light fraction contributed by disk substructures embedded in nine Virgo dwarf ETGs. The sample was constructed based on the fact that these galaxies already showed the presence of embedded spiral arms and bars in SDSS DR4 images, and were re-observed with higher sensitivity using the Wide Field Imager \citep[WFI;][]{Baade:1999} instrument at the MPG/ESO $2.2$m telescope coupled with a white filter. We found that disk substructures contribute only between $2.2-6.4\%$ of the total galaxy light within two effective radii, with no clear dependence on the galaxy brightness ---although the sample is too small for this trend to be statistically significant.

In Brought to Light II \citep{Smith:2021}, we compared the morphology of these substructure features with those arising from high-resolution numerical simulations of tidal harassment by a galaxy cluster potential. We found that the formation of spiral arms and bars can be effectively tidally triggered, during close pericenter passages, in dwarf ETGs that contain a fraction of their stellar disk in a cold, highly rotationally-supported component embedded in a comparatively hotter, little rotationally-supported disk that dominates the total galaxy mass. Such a result bears implications on the potential progenitors of these dwarf ETGs. At the time of their infall onto the cluster, these galaxies could have been star-forming field dwarfs dominated by a thick hot disk, or perhaps regular field dwarfs that transformed part of their thin cold disk into a thicker and hotter disk as the cluster environment quenched them. Alternatively, they could have developed a more pronounced thick disk if they formed via gas-poor galaxy mergers. Multi-wavelength photometry and spectroscopy could disclose the star formation history of these disk substructures and provide us with better constrains on their possible formation scenarios. 

In this work, we take our study one step further. We focus on the analysis of a sample of dwarf ETGs with embedded substructure features that belong to the Fornax galaxy cluster, based on deep multi-band imaging from the Fornax Deep Survey \citep[FDS;][]{Iodice:2016}. This provides us with much deeper imaging compared to the SDSS data that were used in the Virgo study of \citet{Lisker:2006a}, and taken at improved atmospheric conditions, with a typical FDS seeing of $\sim1$ arcsec. Furthermore, this Fornax sample has the advantage of tripling the number of dwarf ETGs with substructures with respect to our Virgo study in Brought to Light I \citep{Michea:2021}, and providing the optical color information for both the galaxy's substructure features and diffuse component. Therefore, we investigate how the light fractions and colors of the substructures correlate with both global galaxy parameters and substructure parameters, and use this information to shed light on the potential evolutionary paths of the dwarf ETGs in our sample.

This paper is organized as follows. First, we introduce the data in Section \ref{sec:data}, which consist in an imaging survey and a dwarf galaxy catalog of the Fornax cluster. Next, in Section \ref{sec:sample}, we process and prepare the imaging data, identify based on several criteria the dwarf ETGs that present substructure features, construct the sample, and characterize it by deriving the main properties of the galaxies. The residual method originally presented in Brought to Light I \citep{Michea:2021} is then applied to the dwarf ETG sample, with its configuration parameters and the results obtained being described in Section \ref{sec:res_method}. Then, in order to showcase a potential application of the residual method, in Section \ref{sec:color} we perform color and stellar population analyses separately on the diffuse and substructure components of the dwarf ETG sample. This is followed by a discussion in Section \ref{sec:discussion}, in which we provide a critical look into possible formation and evolution scenarios of dwarf ETGs with substructure features, based on the comparison of the Fornax sample presented in this work and the Virgo sample of Brought to Light I \citep{Michea:2021}. Finally, we conclude in Section \ref{sec:summary} with a summary of this work.

Throughout this work, all photometric measurements are given in the AB magnitude system. The assumed distance to the Fornax cluster is $20.0\pm 1.4$ Mpc, which is derived from a distance modulus of $31.51 \pm 0.15$ mag \citep{Blakeslee:2009}.

\section{Data}\label{sec:data}

Our analysis of Fornax dwarf ETGs with substructure features is 
based on the Fornax Deep Survey \citep[FDS;][]{Iodice:2016,Peletier:2020}, a deep, multi-band imaging survey of the Fornax cluster, and the Fornax Deep Survey Dwarf Catalog \citep[FDSDC;][]{Venhola:2018}, listing all dwarf galaxies identified in the FDS footprint.

\subsection{FDS Images}\label{subsec:data_images}

The Fornax Deep Survey \citep[FDS;][]{Iodice:2016,Peletier:2020} is a joint effort of the guaranteed-time observation surveys FOCUS (P.I. R. F. Peletier) and VEGAS \citep[P.I. M. Capaccioli,][]{Capaccioli:2015}, carried out with the Very Large Telescope (VLT) Survey Telescope \citep[VST;][]{Arnaboldi:1998} located at the European Southern Observatory (ESO), Paranal. Using the OmegaCAM instrument \citep{Kuijken:2002,Kuijken:2011}, the FDS targets the Fornax galaxy cluster out to its virial radius \citep[$0.7$ Mpc;][]{Drinkwater:2001}, and also the Fornax A infalling subgroup. The FDS data consist in deep, multi-band imaging in the $u$, $g$, $r$, and $i$ bands, although the Fornax A subgroup lacks $u$-band imaging. With OmegaCAM having a field-of-view of $1 \times 1$ deg$^{2}$, the FDS consists of a total of $32$ fields of this size. In practice, we only work with $25$ fields, as these are the ones covered by the FDS dwarf galaxy catalog (see following Section \ref{subsec:data_catalog}).

After assessing the available imaging data and associated metadata, we decide to discard the $u$ and $i$-band data, since, when compared to the $g$ and $r$-band data, the former \textit{(a)} are shallower, and \textit{(b)} lack an analytical parametrization of the PSF core and wings. The ideal situation would be to be able to use all available bands, as more bands means more colors, and consequently better constraints on the properties, such as age and metallicity, of the stellar populations of the galaxies. However, the aforementioned restrictions of the $u$ and $i$-band data (plus the limited spatial coverage of the Fornax cluster in the $u$ band) cements our choice to focus our analysis solely on the $g$ and $r$-band reduced, background-subtracted, and flux-calibrated FDS images. For a detailed description of the data reduction, we refer to \citet{Venhola:2018}.

The median depth and median PSF FWHM of the FDS fields in the $g$ and $r$ bands are provided in Table A.1 of \citet{Venhola:2018}. The depth is defined as the surface brightness of a source with a $\text{S}/\text{N} = 1$ at the pixel scale of the FDS data, which corresponds to $0.2$ arcsec pixel$^{-1}$. After correcting for Galactic foreground extinction (for details, see Section \ref{subsec:sample_prep}), the $g$-band depth lies in the range of $25.9-26.8$ mag arcsec$^{-2}$, with a median of $\mu_{g} = 26.6$ mag arcsec$^{-2}$, while the $r$-band depth lies in the range of $25.5-26.2$ mag arcsec$^{-2}$, with a median of $\mu_{r} = 26.0$ mag arcsec$^{-2}$. Therefore, the $g$-band imaging is on average about half a magnitude deeper. With regard to the FWHM of the PSF, the $g$ band has a median FWHM$_{g}= 1.11$ arcsec, while the $r$ band has slightly better seeing with a median FWHM$_{r}= 0.95$ arcsec.

\subsection{FDS Dwarf Galaxy Catalog}\label{subsec:data_catalog}

The FDS Dwarf Catalog \citep[FDSDC;][]{Venhola:2018} contains a total of $564$ dwarf galaxies spread throughout $26$ FDS fields. Their membership to the cluster has been established based on the distribution and correlation of their properties (such as brightness, color, size, and concentration), which allows one to distinguish them from background galaxies. The catalog considers all galaxies with an $r$-band absolute magnitude fainter than $\text{M}_{r} = -18.5$ mag to be dwarfs, and reaches a $50\%$ completeness limit at $\text{M}_{r} = -10.5$ mag and at a limiting $r$-band surface brightness of $\mu_{r} = 26.0$ mag arcsec$^{-2}$. For more details on the construction and overall characteristics of the FDSDC, we refer to \citet{Venhola:2018}.

For each dwarf galaxy, the catalog provides its ICRS coordinates, several structural and photometric parameters, and a morphological classification. The structural parameters (such as S\'ersic index, effective radius, axis ratio, and position angle) were derived from the $r$-band with GALFIT \citep{Peng:2002}, using either a single S\'ersic function \citep{Sersic:1968} or a combination of a S\'ersic function and a PSF function if a nucleus is present. The photometric parameters were then derived by using the structural information provided by the $r$-band fit, and correspond to the apparent magnitude of the galaxy in the $g$ and $r$ bands, the apparent magnitude of its nucleus (if present) in the $g$ and $r$ bands, and the apparent magnitude within the effective radius in the $u$, $g$, $r$, and $i$ bands. These photometric measurements are given in the SDSS $u$, $g$, $r$, and $i$ filters calibrated to the AB magnitude system, and are not corrected for Galactic foreground extinction. Additionally, based on the appearance and overall color of the galaxies, they are classified into early and late morphological types.

In practice, we work with the FDSDC information of $560$ dwarf galaxies located in $25$ FDS fields. Our analysis excludes the four dwarf galaxies contained in field $33$, as the inner and outer regions of the PSF were not modeled for this particular field. Nonetheless, a quick check reveals that no substructure features are embedded in these four galaxies, so no loss is incurred by excluding them.

\section{Construction of the Data Sample}\label{sec:sample}

In order to construct our data sample, we need to identify which of the dwarf ETGs in Fornax have substructure features. For this purpose, we set up an ad hoc procedure which we explain in detail in the next Subsections.

\subsection{Galaxy Cutouts}\label{subsec:sample_cutouts}

The first step is to construct cutouts of the $g$ and $r$-band FDS images of each dwarf galaxy present in the catalog using its coordinates and the World Coordinate System (WCS) provided in the FDS image headers. We choose to make the size of each cutout a variable parameter that depends on the overall extension of the galaxy. Consequently, for each of the $560$ FDSDC galaxies, we construct cutouts of their $g$ and $r$-band FDS images with dimensions $20\,\text{R}_{e}\times20\,\text{R}_{e}$, where $\text{R}_{e}$ is the effective radius\footnote{The effective radius actually corresponds to the effective semi-major axis measured by \citet{Venhola:2018}.} in the $r$ band as listed in Table \ref{tab:sample_struc}. These dimensions ensure that the galaxy is well contained within the cutout, and that the cutout samples the local background residuals.

\subsection{PSF-Matching}\label{subsec:sample_psfmatch}

To be able to perform reliable measurements in the $g$ and $r$-band galaxy cutouts simultaneously, it is necessary to take into account their different PSFs. For this purpose, we use the analytic functions describing the core and the wings of the PSF separately as provided by \citet{Venhola:2018}. The PSF core, extending over the central $10$ arcsec of the PSF, is modeled by the sum of a Gaussian function with a Moffat function. The PSF core parameters of each FDS field in the $g$ and $r$ bands are provided in Table A.2 of \citet{Venhola:2018}. In some cases, a Gaussian function is not required in order to describe properly the PSF core, so only a Moffat function is used. The PSF wings, defined as the radial region beyond $40$ arcsec, are characterized by an exponential function, whose parameters are the same for all the FDS fields, and are provided in Section 5.1.2 of \citet{Venhola:2018}.

In order to match the $g$ and $r$-band PSFs of each galaxy image cutout, we make use of the Python utilities provided by the \texttt{astropy} package and its affiliated package \texttt{photutils} \citep{Photutils:2021}. We first create a 2D model of the composite PSF for each FDS field and band. Then, for each galaxy, we translate their $g$ and $r$-band composite PSF models into kernel images, where the kernel size is a variable parameter that depends on the observable galaxy size. As the process of convolution redistributes the galaxy light across the image, it is important that the size of the convolution kernel matches the overall observable extension of the galaxy, so that no light becomes unnecessarily dispersed and lost beyond the galaxy's boundaries. For this reason, we choose to create $g$ and $r$-band PSF kernel images of dimensions $6\,\text{R}_{e}\times6\,\text{R}_{e}$ (i.e., of $3\,\text{R}_{e}$ in radial length) for each galaxy. As a consequence, all measurements and analyses in the following Sections are performed only out to a radial extension of $3\, \text{R}_{e}$.

The next step consists in identifying, for each galaxy, which band has the narrowest (i.e., best) PSF, and which has the broadest (i.e., worst) PSF. This is necessary as the quality of the FDS imaging varies from field to field, so while in some fields the $g$-band imaging has the narrowest PSF and the $r$-band has the broadest PSF, in other fields it can be the other way around. We evaluate the PSF quality by comparing the $g$ and $r$-band PSF FWHMs provided in Table A.1 of \citet{Venhola:2018} for each FDS field. Once we have identified the narrow and the broad PSF, we create a matching kernel image from the previously constructed $g$ and $r$-band PSF kernel images. Finally, to match the PSFs of the two bands, we convolve the galaxy image cutout of the band that has the narrowest PSF with the matching PSF kernel we just constructed, thus effectively degrading its PSF to match the band that has the broadest PSF. As a result, we obtain PSF-matched $g$ and $r$-band image cutouts for each of the FDSDC galaxies.

\subsection{Coaddition}\label{subsec:sample_coadds}

The PSF-matched $g$ and $r$-band galaxy cutouts we created allow us to make reliable photometric measurements in both bands simultaneously. However, we are also interested in stacking both bands in order to create a single $g+r$ galaxy image with increased S/N and depth. As we aim to analyze substructure features that are intrinsically faint, the increase in depth allows us to better detect and identify them. Moreover, the stacking ensures that both bands are taken into account during the substructure detection process. At this stage, we still have no knowledge about the brightness and color of the substructure features. This means that we do not know in which band they are brighter, or if there is a band in which they are barely visible. The impact of these unknowns is minimized when we combine the contribution of both bands into one image.

We thus construct a $g+r$ image for each galaxy by coadding their PSF-matched $g$ and $r$-band image cutouts. We adopt a simple approach when stacking: the $g$ and $r$ bands are assigned the same weights, so their contribution to the total flux of the final coadded image is the same. On average, the $g+r$ coadded images are able to reach $0.5$ magnitudes deeper when compared to the $r$-band images.

\subsection{Unsharp Masking}\label{subsec:sample_uns}

Unsharp masking constitutes an efficient way to assess if faint substructure features are embedded in the bright diffuse body of a galaxy. The creation of an unsharp mask image requires first to smooth the galaxy image, and then to divide the original galaxy image by the smoothed-out version of itself. This way, the majority of the smooth light is removed, thus revealing any non-smooth features that would normally lie hidden to the naked eye.

Using utilities from the \texttt{astropy} package in Python, we create a series of unsharp mask images from the $g+r$ coadded image of the FDSDC galaxies. As the size of the smoothing kernel affects the width of the substructures that are revealed, we adopt a range of smoothing kernel sizes. For each FDS field and band, we transform their PSF FWHM into a Gaussian standard deviation $\sigma_{\text{PSF}}$. Then, we create five different Gaussian smoothing kernels, with standard deviations corresponding to $1,3,5,7,$ and $9 \times \sigma_{\text{PSF}}$ of the PSF, respectively. With regard to the shape and orientation of the kernels, we adopt the axis ratio and position angle given by the FDSDC, in order to match the overall geometry of the galaxy when smoothing. As a result, we obtain five unsharp mask images for each FDSDC galaxy, where each unsharp mask is tuned to reveal potential substructure features of a particular width and extent.

\subsection{Sample Selection}\label{subsec:sample_sel}

For each of the $560$ FDSDC galaxies, we visually inspect their $g+r$ coadded image and their associated unsharp mask images. To become part of our sample of Fornax dwarf ETGs with substructures, a dwarf galaxy must comply with the following three criteria:

\begin{itemize}[leftmargin=1.4em]
\item[\textbf{1.}] \textbf{Red color ($\mathbf{g-r \geq 0.5}$ mag).} The galaxy must have a red global color, which we define as $g-r \geq 0.5$ mag. The total integrated $g-r$ color of the galaxy is derived from their $g$ and $r$ band apparent magnitudes reported by the FDSDC.
\item[\textbf{2.}] \textbf{Early-type morphology.} In the $g+r$ coadded image, the galaxy must exhibit an overall early-type morphology, such as that of a dwarf elliptical (dE), lenticular (dS0), or spheroidal (dSph). The morphological appearance is assessed based on a visual inspection of the image.
\item[\textbf{3.}] \textbf{Disk or clump substructure features.} In the unsharp mask images, the galaxy must show either disk or clump substructure features. Disk substructures encompass bars, spiral arms, rings, and dumbbells, as per the definition of \citet{Lisker:2006a}. Clump substructures encompass irregular light overdensities such as star forming regions, dust lanes, and off-center nuclei.
\end{itemize}

According to the morphological classification of the FDSDC, there are a total of $194$ dwarf galaxies brighter than $\text{M}_{r} = -13$ mag that simultaneously fulfill criteria \textit{(1)} and \textit{(2)}; i.e., that are red dwarf ETGs. Out of these, $23$ dwarf ETGs also show substructures as described in criterion \textit{(3)}, which amounts to $12\%$ of them. These $23$ dwarf ETGs constitute our working sample, where $9$ have disk substructures ($40\%$ of the sample) and $14$ have clump substructures ($60\%$ of the sample). For a more detailed look into the magnitude distribution of these galaxies, refer to Figure \ref{fig:plot_hist_subfrac} at the end of Section \ref{subsec:res_method_results}. Finally, for purposes of this work, we shall from now assume that the Fornax dwarf ETG population is constituted by \textit{(a)} all of the dwarf galaxies morphologically classified as early types in the FDSDC, plus \textit{(b)} all of the galaxies in our sample.

\subsection{Sample Preparation}\label{subsec:sample_prep}

In order to run the ``residual method'' \citep{Michea:2021} on our sample, we first need to mask out the foreground and background sources in the projected vicinity of our objects, so that subsequent measurements and analyses of the galaxy images will not be contaminated by their light. We do so by visually identifying interloping sources in the $g+r$ coadded images of our sample, whose positions and extents are manually registered and then translated into a bad pixel mask (BPM) image.

We use this BPM image in the IRAF STSDAS \texttt{ellipse} task \citep[based on][]{Jedrzejewski:1987} with which we independently fit the PSF-matched $g$ and $r$-band images of each sample galaxy, while allowing their center coordinates, isophote ellipticity, and isophote position angle to change freely with galactocentric radius. We then use the IRAF \texttt{bmodel} task to build the $g$ and $r$-band galaxy model images from the fitted isophotes. Finally, we replace the values of the bad pixels in the PSF-matched $g$ and $r$-band galaxy images with their corresponding values in the $g$ and $r$-band galaxy models, respectively. By masking the interloping sources, we obtain PSF-matched $g$ and $r$-band galaxy images that have a much cleaner and regular appearance. Nonetheless, we note that the regions marked on the BPM images are still omitted during any kind of measurement or analysis. 

We also need to correct for Galactic foreground extinction, as both the FDS images and the photometric measurements provided by the FDSDC do not take dust reddening into account. For this purpose, we employ the NASA/IPAC Extragalactic Database (NED) \href{https://ned.ipac.caltech.edu/extinction_calculator}{Extinction Calculator}, based on the Galactic extinction maps of \citet{Schlafly:2011} and the Galactic extinction law of \citet{Fitzpatrick:1999} with $\text{R}_V = 3.1$, and compute the $A_{g}$ and $A_{r}$ extinction coefficients for each galaxy in our sample. In the $g$ band, we find that the extinction coefficient varies between $0.015 - 0.079$ mag across our sample, with a median value of $A_{g}=0.040$ mag. In the $r$ band, it lies in the range of $0.010-0.055$ mag, with a median value of $A_{r}=0.028$ mag. In the Sections that follow, all photometric measurements are corrected for Galactic foreground extinction and are given in the AB magnitude system.

\subsection{Sample Properties}\label{subsec:sample_prop}

The photometric properties of our dwarf ETG sample are provided in Table \ref{tab:sample_coords_photo}. From the $g$ and $r$-band integrated apparent magnitudes given by the FDSDC, we derive the $g$ and $r$-band integrated absolute magnitudes by assuming a Fornax cluster distance modulus of $31.51 \pm 0.15$ mag \citep[equivalent to a distance of $20.0\pm 1.4$ Mpc;][]{Blakeslee:2009}. We assume this Fornax cluster distance from here onward. We also derive the integrated $g-r$ color of the galaxies. By construction, the galaxies in our sample were selected based on their integrated red colors, and thus span the range $0.51 \leq g-r \leq 0.76$ mag, with a median of $g-r=0.61$ mag. Being dwarf systems, they are also characteristically faint. Their absolute magnitudes in the $g$ band lie in the range between $-14.3 \geq \text{M}_{g} \geq -18.0$ mag, with a median of $\text{M}_{g} = -16.8$ mag; while in the $r$ band they cover the range between $-14.9 \geq \text{M}_{r} \geq -18.7$ mag, with a median of $\text{M}_{r} = -17.5$ mag.

\begin{deluxetable*}{ccc@{\hspace{0pt}}ccc@{\hspace{0pt}}ccc@{\hspace{0pt}}cccc}[ht!]
\rotate
\setlength{\tabcolsep}{6pt} 
\renewcommand{\arraystretch}{0.92} 
\tablewidth{0pt}
\tabletypesize{\small}
\tablecaption{Central coordinates and photometric properties of our Fornax dwarf ETG sample.\label{tab:sample_coords_photo}}
\tablehead{
\colhead{Dwarf Galaxy} & \multicolumn2c{Central Coordinates} & & \multicolumn2c{$\mu_{e}$} & & \multicolumn2c{$m$} & & \multicolumn2c{$M$} & \colhead{$g-r$} \\[-1ex]
& \multicolumn2c{(J2000.0)} & & \multicolumn2c{(mag arcsec$^{-2}$)} & & \multicolumn2c{(mag)} & & \multicolumn2c{(mag)} & \colhead{(mag)} \vspace*{1pt}\\
\cline{2-3}\cline{5-6}\cline{8-9}\cline{11-12}
 & \colhead{R.A.} & \colhead{Decl.} & & \colhead{$g$ band} & \colhead{$r$ band} & & \colhead{$g$ band} & \colhead{$r$ band} & & \colhead{$g$ band} & \colhead{$r$ band} & \\
\colhead{(1)} & \colhead{(2)} & \colhead{(3)} & & \colhead{(4)} & \colhead{(5)} & & \colhead{(6)} & \colhead{(7)} & & \colhead{(8)} & \colhead{(9)} & \colhead{(10)} \vspace*{0.5pt}
}
\decimals
\startdata
F01D145 & $03^{\text{h}}46^{\text{m}}33.38^{\text{s}}$ & $-34^{\circ}41^{\prime}10.32^{\prime\prime}$ & & $24.52$ & $23.90$ & & $17.23\pm0.24$ & $16.66\pm0.21$ & & $-14.28\pm0.28$ & $-14.85\pm0.25$ & $0.57\pm0.31$ \\ [0.5ex]
F02D000 & $03^{\text{h}}50^{\text{m}}36.72^{\text{s}}$ & $-35^{\circ}54^{\prime}33.84^{\prime\prime}$ & & $22.23$ & $21.53$ & & $14.17\pm0.11$ & $13.51\pm0.09$ & & $-17.34\pm0.19$ & $-18.00\pm0.18$ & $0.66\pm0.15$ \\ [0.5ex]
F04D000 & $03^{\text{h}}42^{\text{m}}45.55^{\text{s}}$ & $-33^{\circ}55^{\prime}12.36^{\prime\prime}$ & & $22.49$ & $21.83$ & & $14.49\pm0.11$ & $13.84\pm0.09$ & & $-17.02\pm0.19$ & $-17.67\pm0.18$ & $0.65\pm0.14$ \\ [0.5ex]
F04D001 & $03^{\text{h}}41^{\text{m}}3.60^{\text{s}}$ & $-33^{\circ}46^{\prime}44.76^{\prime\prime}$ & & $21.66$ & $20.99$ & & $13.48\pm0.10$ & $12.83\pm0.09$ & & $-18.03\pm0.18$ & $-18.68\pm0.17$ & $0.65\pm0.13$ \\ [0.5ex]
F04D002 & $03^{\text{h}}43^{\text{m}}22.66^{\text{s}}$ & $-33^{\circ}56^{\prime}19.68^{\prime\prime}$ & & $22.46$ & $21.77$ & & $15.22\pm0.15$ & $14.55\pm0.12$ & & $-16.29\pm0.21$ & $-16.96\pm0.19$ & $0.67\pm0.19$ \\ [0.5ex]
F04D053 & $03^{\text{h}}41^{\text{m}}45.41^{\text{s}}$ & $-33^{\circ}47^{\prime}29.40^{\prime\prime}$ & & $24.24$ & $23.67$ & & $15.90\pm0.16$ & $15.35\pm0.14$ & & $-15.61\pm0.22$ & $-16.16\pm0.21$ & $0.55\pm0.21$ \\ [0.5ex]
F04D061 & $03^{\text{h}}41^{\text{m}}21.19^{\text{s}}$ & $-33^{\circ}46^{\prime}9.84^{\prime\prime}$ & & $24.29$ & $23.71$ & & $15.61\pm0.17$ & $15.03\pm0.15$ & & $-15.90\pm0.23$ & $-16.48\pm0.21$ & $0.57\pm0.22$ \\ [0.5ex]
F05D000 & $03^{\text{h}}41^{\text{m}}32.54^{\text{s}}$ & $-34^{\circ}53^{\prime}19.68^{\prime\prime}$ & & $21.99$ & $21.41$ & & $13.74\pm0.11$ & $13.20\pm0.09$ & & $-17.77\pm0.18$ & $-18.31\pm0.18$ & $0.54\pm0.14$ \\ [0.5ex]
F07D000 & $03^{\text{h}}45^{\text{m}}3.58^{\text{s}}$ & $-35^{\circ}58^{\prime}21.72^{\prime\prime}$ & & $20.79$ & $20.08$ & & $13.76\pm0.07$ & $13.15\pm0.06$ & & $-17.75\pm0.17$ & $-18.36\pm0.16$ & $0.62\pm0.09$ \\ [0.5ex]
F09D255 & $03^{\text{h}}36^{\text{m}}49.73^{\text{s}}$ & $-33^{\circ}27^{\prime}39.24^{\prime\prime}$ & & $23.44$ & $22.81$ & & $16.34\pm0.16$ & $15.74\pm0.14$ & & $-15.17\pm0.22$ & $-15.77\pm0.21$ & $0.60\pm0.22$ \\ [0.5ex]
F09D492 & $03^{\text{h}}39^{\text{m}}55.03^{\text{s}}$ & $-33^{\circ}03^{\prime}11.88^{\prime\prime}$ & & $23.26$ & $22.69$ & & $16.81\pm0.14$ & $16.26\pm0.12$ & & $-14.70\pm0.21$ & $-15.25\pm0.19$ & $0.54\pm0.19$ \\ [0.5ex]
F10D189 & $03^{\text{h}}38^{\text{m}}9.17^{\text{s}}$ & $-34^{\circ}31^{\prime}7.68^{\prime\prime}$ & & $23.52$ & $22.84$ & & $15.20\pm0.15$ & $14.58\pm0.13$ & & $-16.31\pm0.21$ & $-16.93\pm0.20$ & $0.62\pm0.20$ \\ [0.5ex]
F11D279 & $03^{\text{h}}36^{\text{m}}54.31^{\text{s}}$ & $-35^{\circ}22^{\prime}28.92^{\prime\prime}$ & & $22.41$ & $21.64$ & & $14.35\pm0.09$ & $13.59\pm0.08$ & & $-17.16\pm0.18$ & $-17.92\pm0.17$ & $0.76\pm0.12$ \\ [0.5ex]
F14D144 & $03^{\text{h}}33^{\text{m}}34.06^{\text{s}}$ & $-33^{\circ}34^{\prime}17.04^{\prime\prime}$ & & $23.33$ & $22.62$ & & $14.76\pm0.13$ & $14.10\pm0.11$ & & $-16.75\pm0.20$ & $-17.41\pm0.18$ & $0.66\pm0.17$ \\ [0.5ex]
F15D384 & $03^{\text{h}}34^{\text{m}}30.82^{\text{s}}$ & $-34^{\circ}17^{\prime}50.64^{\prime\prime}$ & & $23.26$ & $22.63$ & & $15.26\pm0.15$ & $14.66\pm0.13$ & & $-16.25\pm0.21$ & $-16.85\pm0.20$ & $0.61\pm0.19$ \\ [0.5ex]
F15D417 & $03^{\text{h}}32^{\text{m}}47.69^{\text{s}}$ & $-34^{\circ}14^{\prime}19.32^{\prime\prime}$ & & $22.20$ & $21.55$ & & $14.70\pm0.11$ & $14.06\pm0.09$ & & $-16.81\pm0.18$ & $-17.45\pm0.18$ & $0.64\pm0.14$ \\ [0.5ex]
F17D227 & $03^{\text{h}}31^{\text{m}}8.28^{\text{s}}$ & $-36^{\circ}17^{\prime}24.36^{\prime\prime}$ & & $21.91$ & $21.35$ & & $14.51\pm0.09$ & $13.97\pm0.08$ & & $-17.00\pm0.17$ & $-17.54\pm0.17$ & $0.54\pm0.12$ \\ [0.5ex]
F19D001 & $03^{\text{h}}27^{\text{m}}18.02^{\text{s}}$ & $-34^{\circ}31^{\prime}35.40^{\prime\prime}$ & & $21.28$ & $20.49$ & & $13.77\pm0.10$ & $13.02\pm0.08$ & & $-17.74\pm0.18$ & $-18.49\pm0.17$ & $0.75\pm0.13$ \\ [0.5ex]
F22D244 & $03^{\text{h}}26^{\text{m}}25.03^{\text{s}}$ & $-37^{\circ}07^{\prime}40.08^{\prime\prime}$ & & $22.48$ & $21.94$ & & $15.66\pm0.12$ & $15.15\pm0.11$ & & $-15.85\pm0.19$ & $-16.36\pm0.18$ & $0.51\pm0.16$ \\ [0.5ex]
F26D000 & $03^{\text{h}}23^{\text{m}}54.41^{\text{s}}$ & $-37^{\circ}30^{\prime}36.00^{\prime\prime}$ & & $22.13$ & $21.57$ & & $13.72\pm0.11$ & $13.16\pm0.10$ & & $-17.79\pm0.19$ & $-18.35\pm0.18$ & $0.56\pm0.15$ \\ [0.5ex]
F26D003 & $03^{\text{h}}24^{\text{m}}58.37^{\text{s}}$ & $-37^{\circ}00^{\prime}34.56^{\prime\prime}$ & & $21.83$ & $21.16$ & & $14.03\pm0.12$ & $13.39\pm0.10$ & & $-17.48\pm0.19$ & $-18.12\pm0.18$ & $0.64\pm0.15$ \\ [0.5ex]
F26D141 & $03^{\text{h}}22^{\text{m}}22.70^{\text{s}}$ & $-37^{\circ}23^{\prime}51.36^{\prime\prime}$ & & $23.04$ & $22.41$ & & $15.14\pm0.15$ & $14.53\pm0.13$ & & $-16.37\pm0.21$ & $-16.98\pm0.20$ & $0.61\pm0.20$ \\ [0.5ex]
F31D196 & $03^{\text{h}}29^{\text{m}}43.25^{\text{s}}$ & $-33^{\circ}33^{\prime}25.20^{\prime\prime}$ & & $23.70$ & $23.06$ & & $14.46\pm0.16$ & $13.86\pm0.14$ & & $-17.05\pm0.22$ & $-17.65\pm0.20$ & $0.60\pm0.21$ \\ [0.2ex]
\enddata
\tablecomments{Col. (1): name of the dwarf galaxy. Cols. (2) and (3): right ascension and declination of the central coordinates in the International Celestial Reference System (ICRS), according to the FDSDC \citep{Venhola:2018}. Cols. (4) and (5): surface brightness at one effective radius in the $g$ and $r$ bands, respectively. Cols. (6) and (7): total apparent magnitude in the $g$ and $r$ bands, respectively, according to the FDSDC. Cols. (8) and (9): total absolute magnitude in the $g$ and $r$ bands, respectively, assuming a Fornax cluster distance modulus of $31.51 \pm 0.15$ mag \citep[equivalent to a distance of $20.0\pm 1.4$ Mpc;][]{Blakeslee:2009}. Col. (10): total $g-r$ color. All photometric measurements are in the AB magnitude system and have been corrected for Galactic foreground extinction.}
\end{deluxetable*}

\begin{deluxetable*}{c@{\hspace{0.6em}}DC@{\hspace{0.6em}}c@{\hspace{0.6em}}CC@{\hspace{0.6em}}c@{\hspace{0.6em}}DD}[ht!]
\setlength{\tabcolsep}{6pt} 
\renewcommand{\arraystretch}{1} 
\tablewidth{0pt}
\tabletypesize{\small}
\tablecaption{Structural properties of our Fornax dwarf ETG sample in the $r$ band.\label{tab:sample_struc}}
\tablehead{
\colhead{Dwarf Galaxy} & \multicolumn3c{Effective Radius} & & \multicolumn2c{Ellipticity} & & \multicolumn4c{Position Angle} \vspace*{1pt}\\
\cline{2-4}\cline{6-7}\cline{9-12}
 & \multicolumn2c{} & & & \colhead{At $1\,\text{R}_{e}$} & \colhead{At $2\,\text{R}_{e}$} & & \multicolumn2c{At $1\,\text{R}_{e}$} & \multicolumn2c{At $2\,\text{R}_{e}$}\\
 & \multicolumn2c{(arcsec)} & \colhead{(kpc)} & & & & & \multicolumn2c{(deg)} & \multicolumn2c{(deg)}\\
\colhead{(1)} & \multicolumn2c{(2)} & \colhead{(3)} & & \colhead{(4)} & \colhead{(5)} & & \multicolumn2c{(6)} & \multicolumn2c{(7)} \vspace*{0.5pt}
}
\decimals
\startdata
F01D145 & 14.63\pm2.76 & 1.4\pm0.3 & &  $0.705\pm0.003$ & $0.736\pm0.003$ & & 91.9\pm0.2 & 91.6\pm0.2 \\ [0.5ex]
F02D000 & 14.54\pm1.23 & 1.4\pm0.1 & & $0.413\pm0.003$ & $0.364\pm0.003$ & & 49.7\pm0.3 & 48.5\pm0.3 \\ [0.5ex]
F04D000 & 12.45\pm1.05 & 1.2\pm0.1 & & $0.114\pm0.003$ & $0.132\pm0.003$ & & 160.1\pm0.7 & 151.4\pm0.7 \\ [0.5ex]
F04D001 & 17.03\pm1.31 & 1.7\pm0.1 & & $0.571\pm0.003$ & $0.445\pm0.003$ & & 172.2\pm0.2 & 172.6\pm0.2 \\ [0.5ex]
F04D002 & 14.98\pm1.67 & 1.5\pm0.2 & & $0.706\pm0.003$ & $0.704\pm0.002$ & & 4.5\pm0.3 & 6.3\pm0.1 \\ [0.5ex]
F04D053 & 13.17\pm1.68 & 1.3\pm0.2 & & $0.135\pm0.004$ & $0.117\pm0.008$ & & 137.3\pm0.9 & 165.8\pm2.0 \\ [0.5ex]
F04D061 & 16.64\pm2.23 & 1.6\pm0.2 & & $0.190\pm0.005$ & $0.203\pm0.013$ & & 139.3\pm0.8 & 145.1\pm1.7 \\ [0.5ex]
F05D000 & 16.47\pm1.38 & 1.6\pm0.1 & & $0.492\pm0.008$ & $0.485\pm0.002$ & & 7.0\pm0.6 & 1.0\pm0.2 \\ [0.5ex]
F07D000 & 7.60\pm0.41 & 0.7\pm0.0 & & $0.414\pm0.002$ & $0.348\pm0.002$ & & 156.8\pm0.2 & 156.4\pm0.1 \\ [0.5ex]
F09D255 & 11.04\pm1.41 & 1.1\pm0.1 & & $0.612\pm0.004$ & $0.665\pm0.003$ & & 124.7\pm0.3 & 124.8\pm0.2 \\ [0.5ex]
F09D492 & 6.75\pm0.75 & 0.7\pm0.1 & & $0.317\pm0.004$ & $0.415\pm0.005$ & & 71.4\pm0.5 & 73.0\pm0.4 \\ [0.5ex]
F10D189 & 16.04\pm1.88 & 1.6\pm0.2 & & $0.445\pm0.003$ & $0.431\pm0.004$ & & 140.3\pm0.3 & 141.9\pm0.3 \\ [0.5ex]
F11D279 & 9.67\pm0.66 & 0.9\pm0.1 & & $0.058\pm0.003$ & $0.002\pm0.003$ & & 0.9\pm1.6 & 172.5\pm8.5 \\ [0.5ex]
F14D144 & 13.97\pm1.34 & 1.4\pm0.1 & & $0.131\pm0.002$ & $0.124\pm0.003$ & & 47.8\pm0.5 & 50.2\pm0.8 \\ [0.5ex]
F15D384 & 14.72\pm1.68 & 1.4\pm0.2 & & $0.476\pm0.002$ & $0.514\pm0.003$ & & 152.4\pm0.2 & 151.9\pm0.2 \\ [0.5ex]
F15D417 & 10.65\pm0.87 & 1.0\pm0.1 & & $0.466\pm0.005$ & $0.531\pm0.001$ & & 120.0\pm0.4 & 119.6\pm0.1 \\ [0.5ex]
F17D227 & 8.09\pm0.55 & 0.8\pm0.1 & & $0.213\pm0.007$ & $0.228\pm0.004$ & & 134.8\pm1.1 & 154.7\pm0.5 \\ [0.5ex]
F19D001 & 14.64\pm1.09 & 1.4\pm0.1 & & $0.592\pm0.004$ & $0.514\pm0.001$ & & 29.1\pm0.3 & 29.3\pm0.1 \\ [0.5ex]
F22D244 & 8.51\pm0.82 & 0.8\pm0.1 & & $0.381\pm0.004$ & $0.365\pm0.004$ & & 82.9\pm0.3 & 81.1\pm0.4 \\ [0.5ex]
F26D000 & 17.90\pm1.56 & 1.7\pm0.2 & & $0.197\pm0.018$ & $0.083\pm0.003$ & & 32.2\pm2.8 & 152.5\pm1.3 \\ [0.5ex]
F26D003 & 16.89\pm1.51 & 1.6\pm0.1 & & $0.625\pm0.002$ & $0.543\pm0.001$ & & 82.6\pm0.2 & 80.9\pm0.1 \\ [0.5ex]
F26D141 & 16.62\pm1.97 & 1.6\pm0.2 & & $0.579\pm0.003$ & $0.400\pm0.004$ & & 56.0\pm0.2 & 54.8\pm0.4 \\ [0.5ex]
F31D196 & 24.30\pm2.98 & 2.4\pm0.3 & & $0.433\pm0.002$ & $0.468\pm0.003$ & & 58.0\pm0.2 & 60.9\pm0.2 \\ [0.2ex]
\enddata
\tablecomments{Col. (1): name of the dwarf galaxy. Cols. (2) and (3): effective semi-major axis radius in the $r$ band, according to the FDSDC \citep{Venhola:2018}. For the conversion to kpc, the assumed Fornax cluster distance is $20.0\pm 1.4$ Mpc \citep{Blakeslee:2009}. Cols. (4) and (5): ellipticity in the $r$ band at the isophotes at one and two effective radii, respectively. Cols. (6) and (7): position angle in the $r$ band at the isophotes at one and two effective radii, respectively. The position angle is measured counterclockwise from the +\textit{y}-axis (north towards east of the images).}
\end{deluxetable*}

We also report in Table \ref{tab:sample_coords_photo} the $g$ and $r$-band surface brightness derived from the isophote whose semi-major axis corresponds to the one effective semi-major axis in the $r$ band according to the FDSDC, which are listed in Table \ref{tab:sample_struc}. Specifically, we computed the effective surface brightness $\mu_{e,g}$ and $\mu_{e,r}$ from the \texttt{ellipse} isophotes fitted to the PSF-matched $r$-band image by keeping their center coordinates, ellipticity, and position angle free to vary with galactocentric distance. We then ran \texttt{ellipse} on the PSF-matched $g$-band image by imposing the geometry of the $r$-band isophotes in a no-fit photometry-only mode. This way, the galaxy geometry stays the same in both bands, making all photometric measurements in the $g$ and $r$ bands comparable on an isophote-by-isophote basis. From Table \ref{tab:sample_coords_photo}, we can see that the effective surface brightness of our sample has a median value of $\mu_{e,g} = 22.5$ mag arcsec$^{-2}$ in the $g$ band, and $\mu_{e,r} = 21.8$ mag arcsec$^{-2}$ in the $r$ band. Therefore, both the absolute magnitude and surface brightness distributions of our sample indicate that the galaxies are brighter in the $r$ band than in the $g$ band.

The structural properties of our dwarf ETG sample in the $r$-band are provided in Table \ref{tab:sample_struc}, where the ellipticity and position angle belong to the isophote whose semi-major axis corresponds to either the one or two effective semi-major axes. For simplicity, we will simply refer to the effective semi-major axis as the effective radius ($\text{R}_{e}$) throughout this work. Our galaxy sample features an ample variety of sizes and shapes. The effective radius varies between $6.7 - 24.3$ arcsec ($0.6 - 2.4$ kpc), with a median of $\text{R}_{e}=14.6$ arcsec ($1.4$ kpc). The axis ratio at one effective radius lies in the range between $0.29-0.94$, with a median of $b/a=0.57$. The galaxy sample shows a similar distribution at two effective radii. Furthermore, while for some galaxies the ellipticity and position angle stay approximately the same at one and two effective radii, there are also cases in which the galaxy geometry changes drastically with galactocentric radius. Consequently, the structural properties of the sample are quite heterogeneous, in contrast to its more homogeneous photometric properties.

\begin{figure*}[ht]
\begin{center}
\gridline{\fig{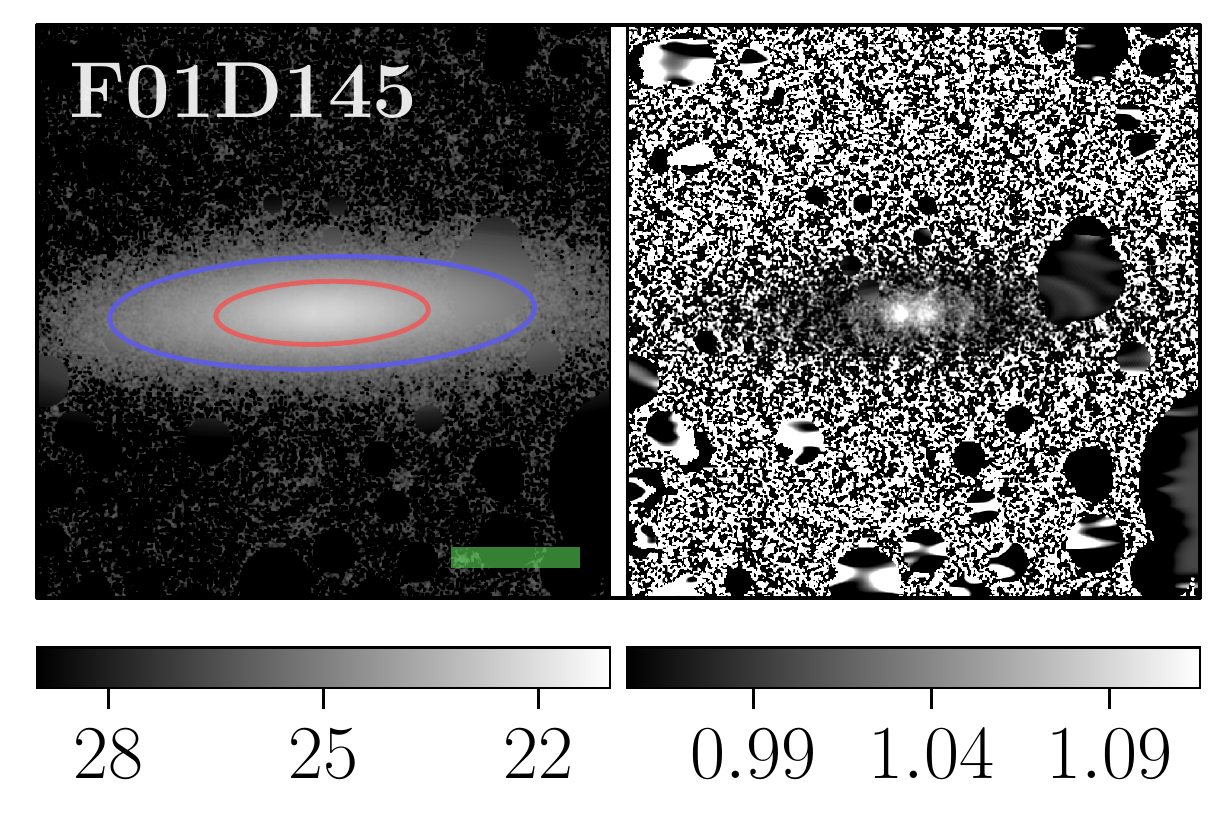}{0.29\textwidth}{}
          \hspace*{-40pt}
          \fig{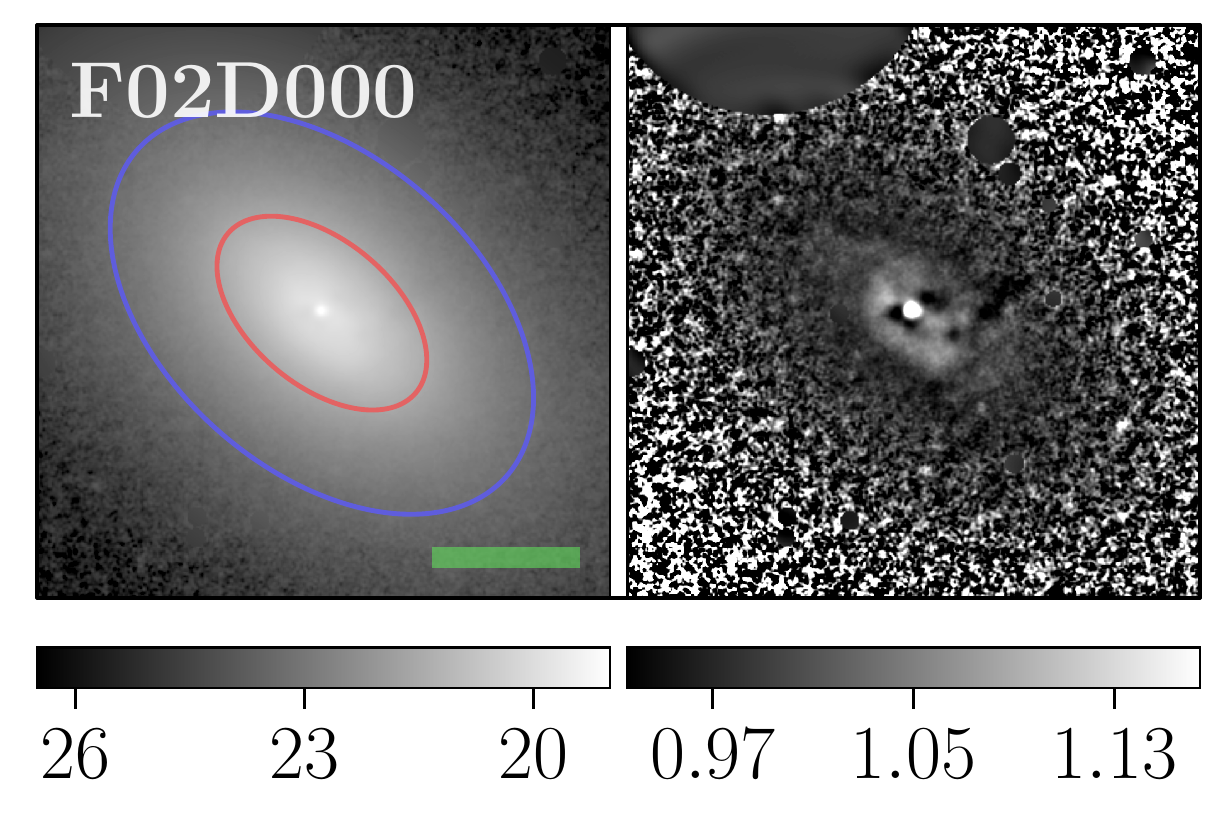}{0.29\textwidth}{}
          \hspace*{-40pt}
          \fig{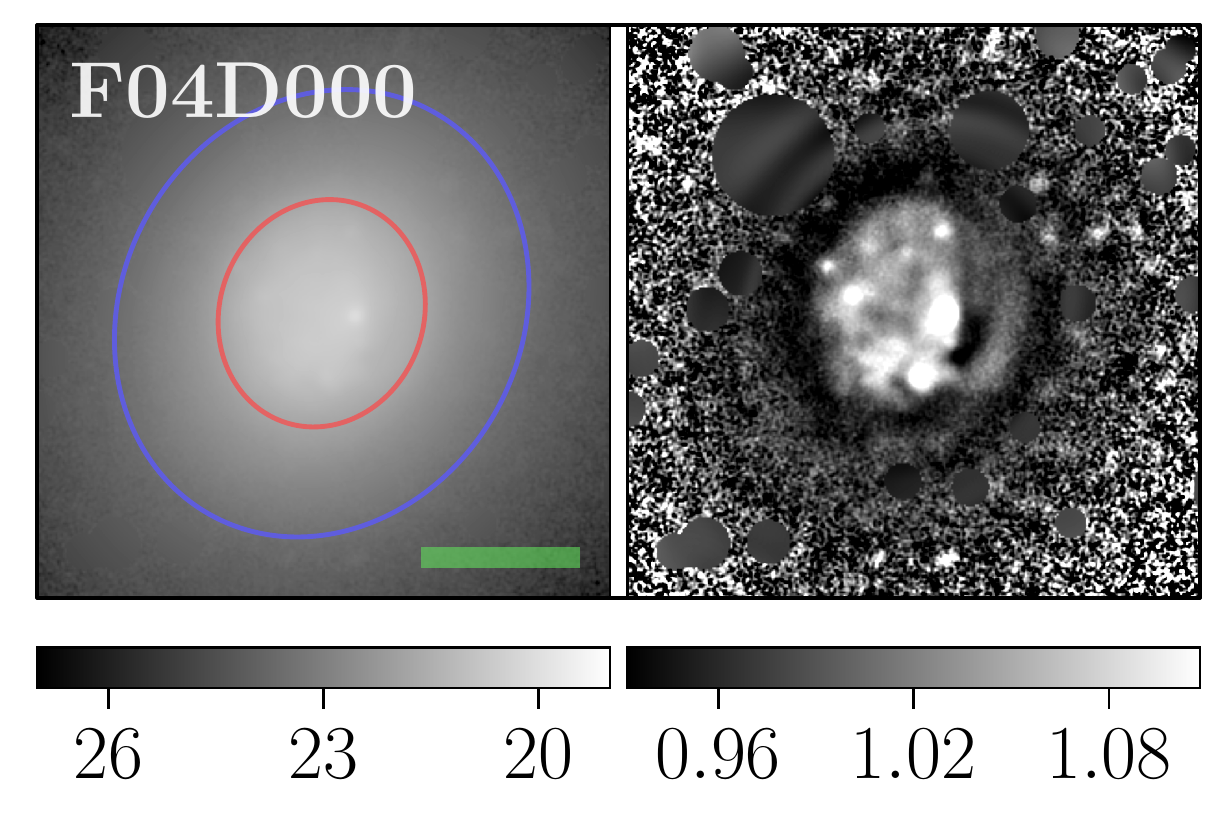}{0.29\textwidth}{}}
          \vspace*{-28.5pt}
\gridline{\fig{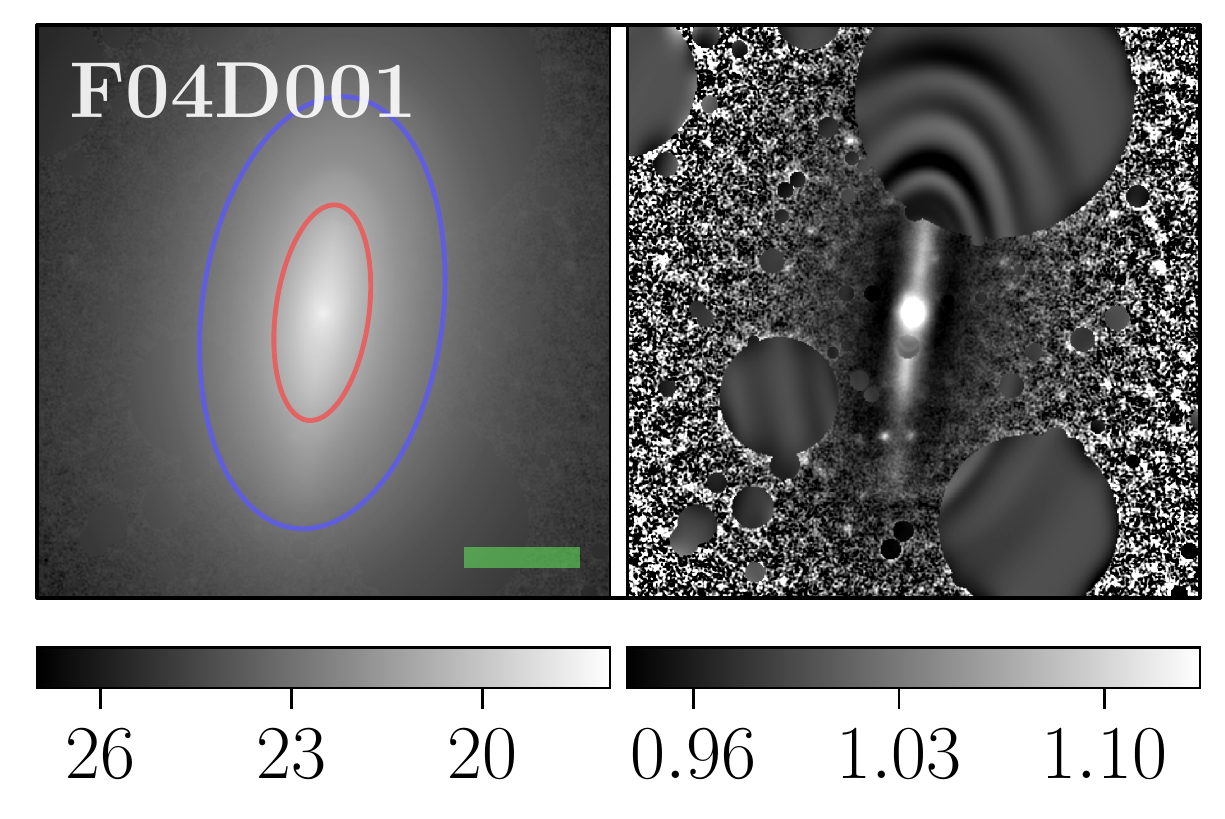}{0.29\textwidth}{}
          \hspace*{-40pt}
          \fig{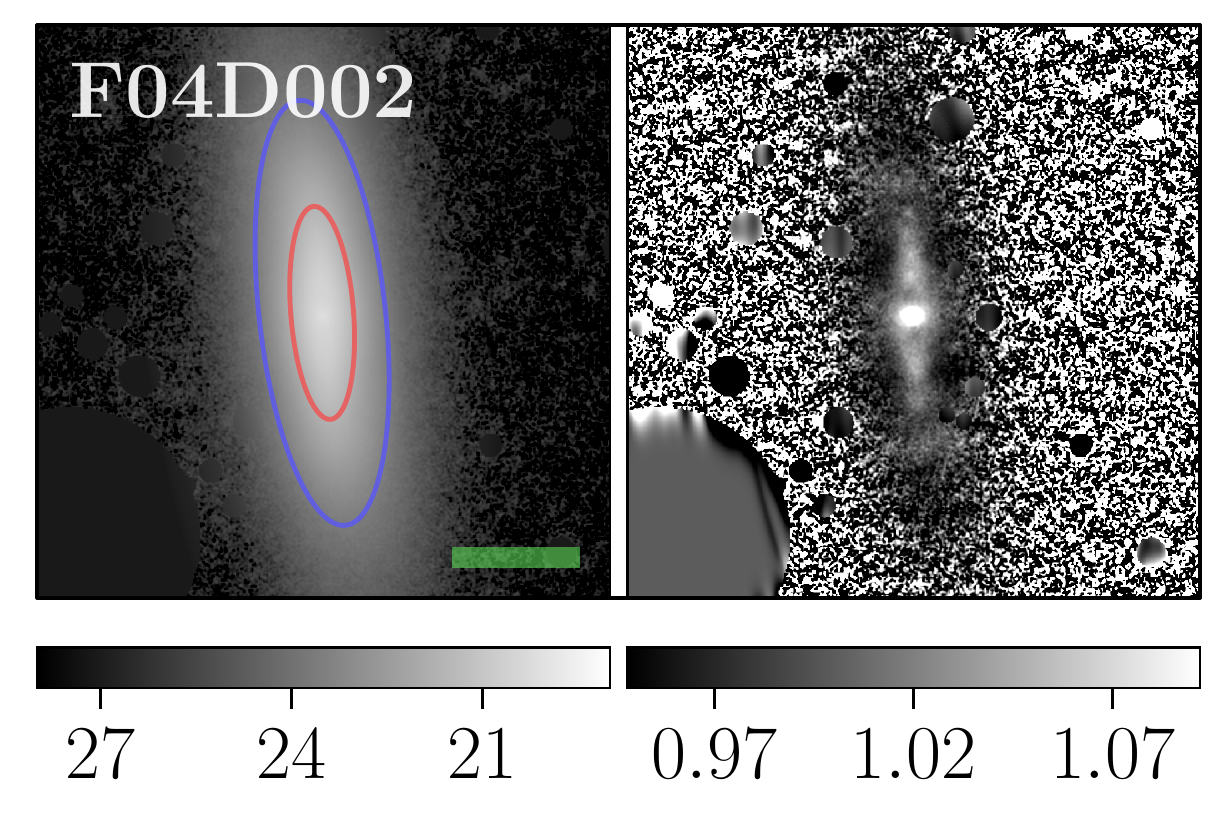}{0.29\textwidth}{}
          \hspace*{-40pt}
          \fig{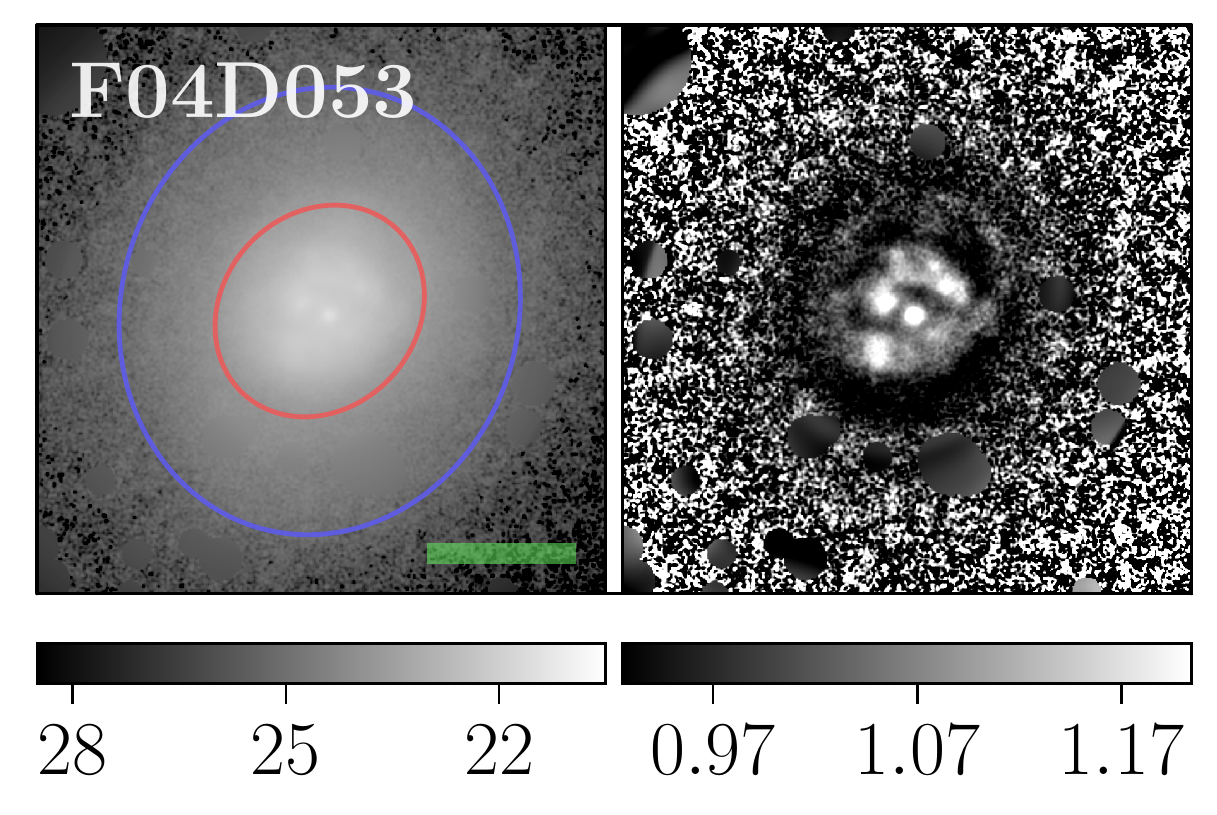}{0.29\textwidth}{}}
          \vspace*{-28.5pt}
\gridline{\fig{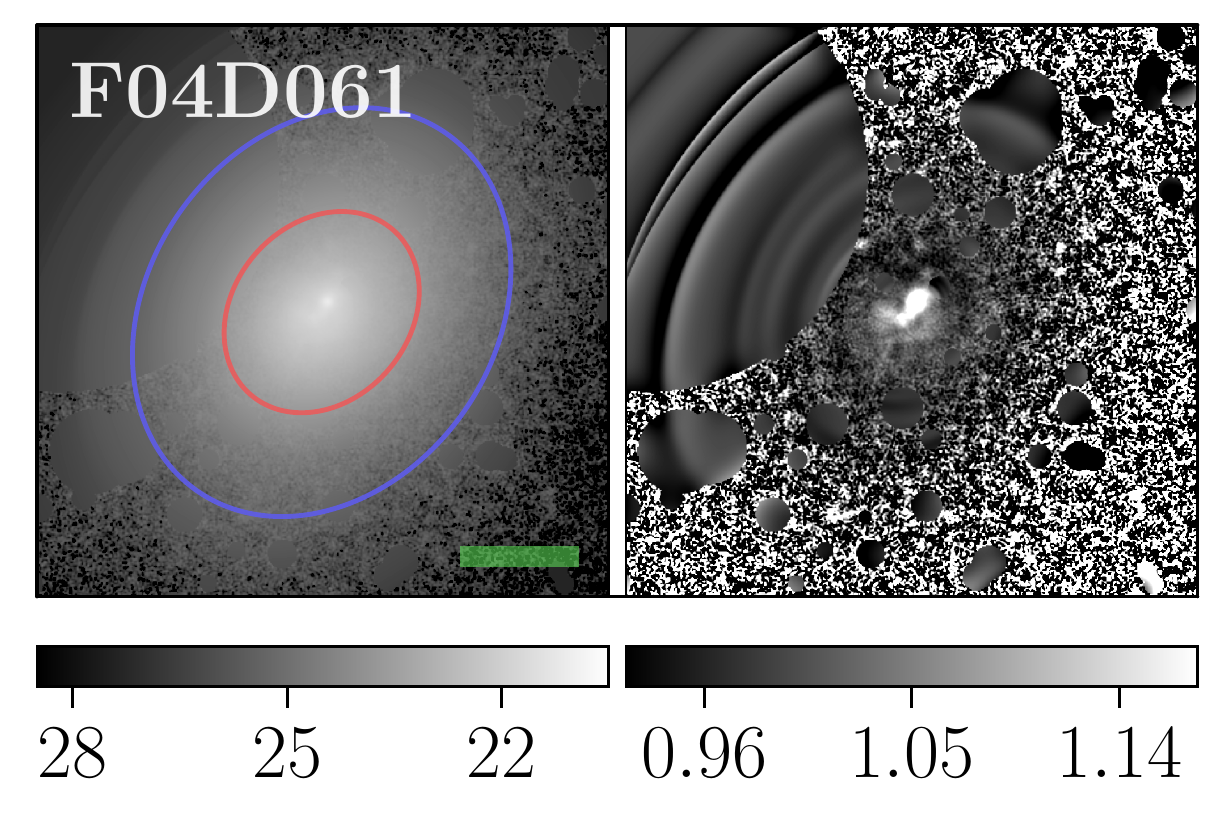}{0.29\textwidth}{}
          \hspace*{-40pt}
          \fig{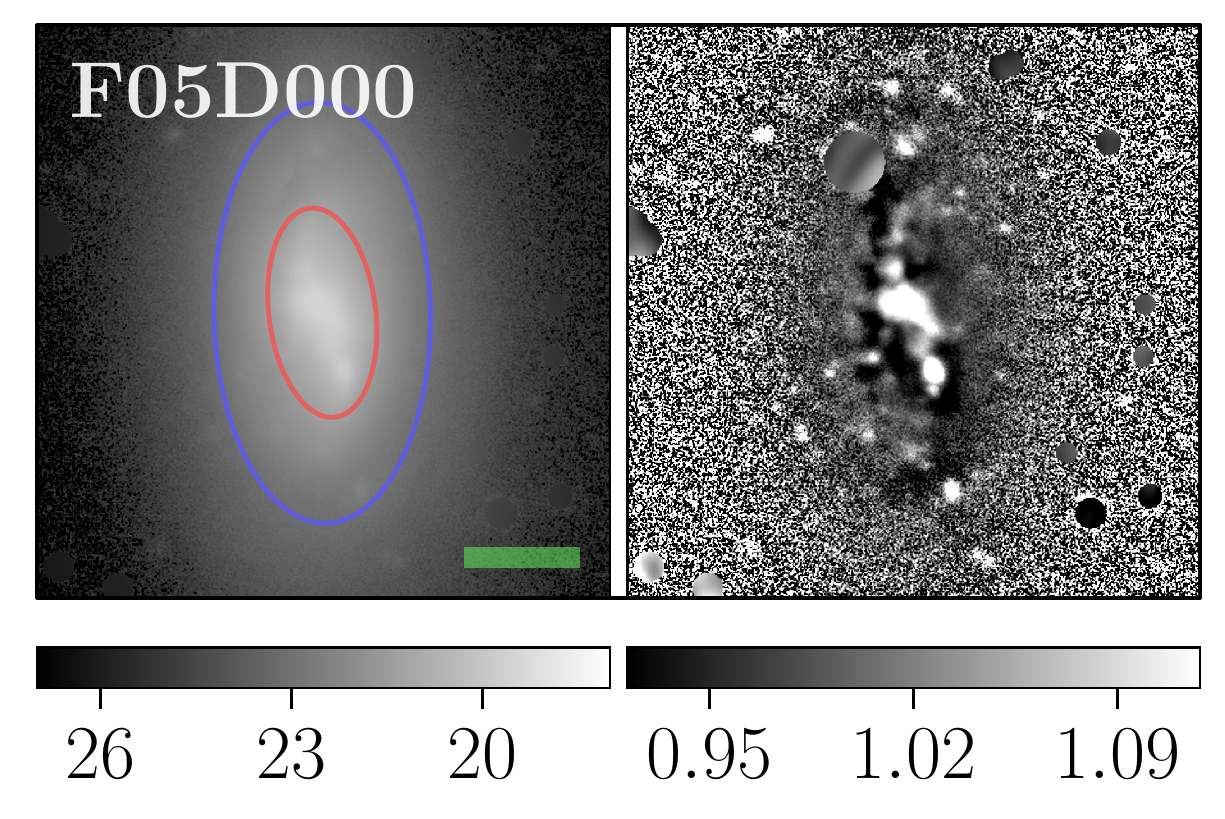}{0.29\textwidth}{}
          \hspace*{-40pt}
          \fig{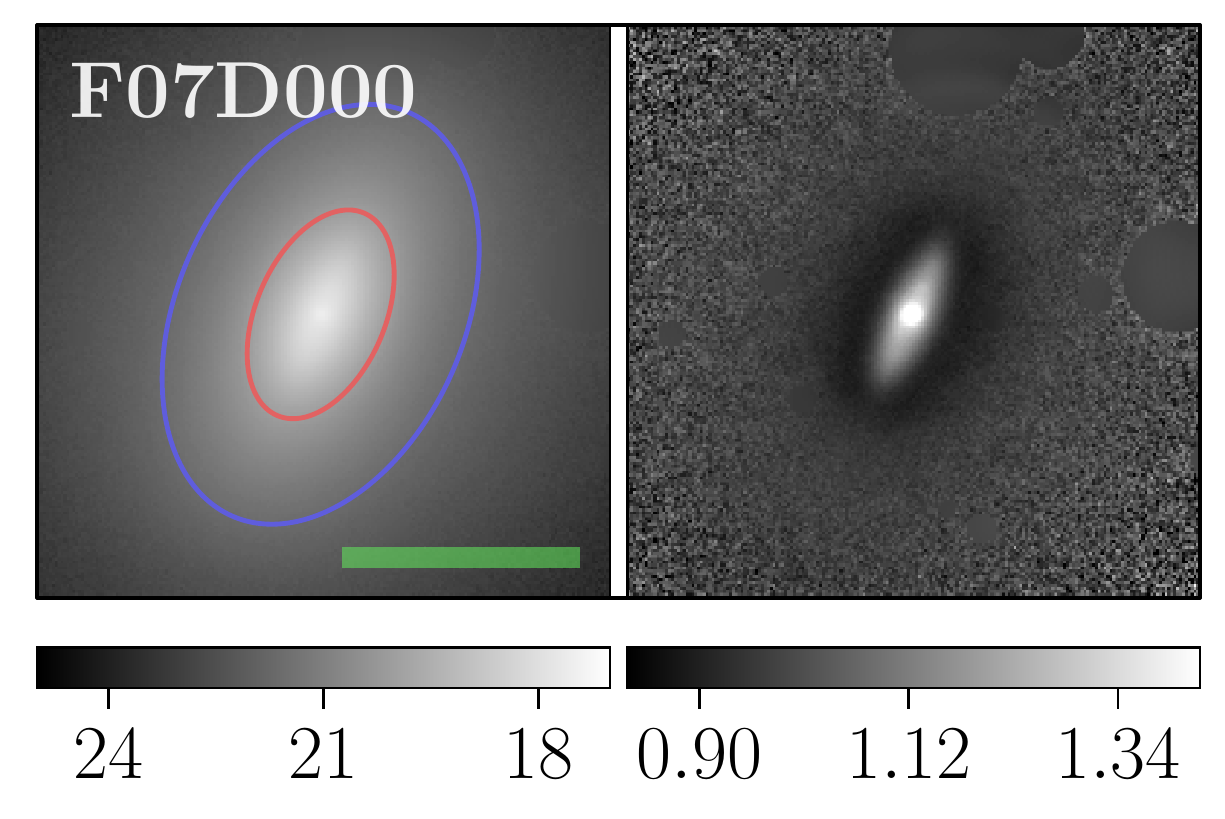}{0.29\textwidth}{}}
          \vspace*{-28.5pt}
\gridline{\fig{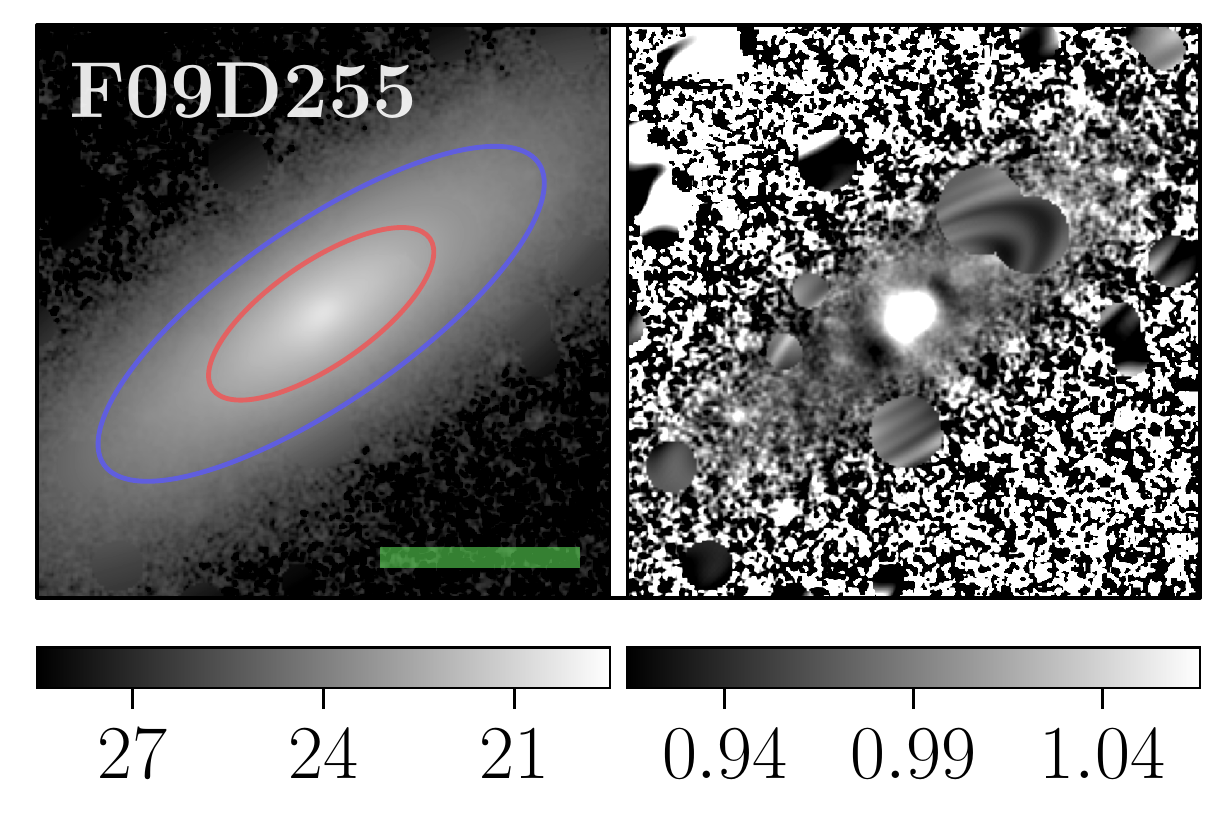}{0.29\textwidth}{}
          \hspace*{-40pt}
          \fig{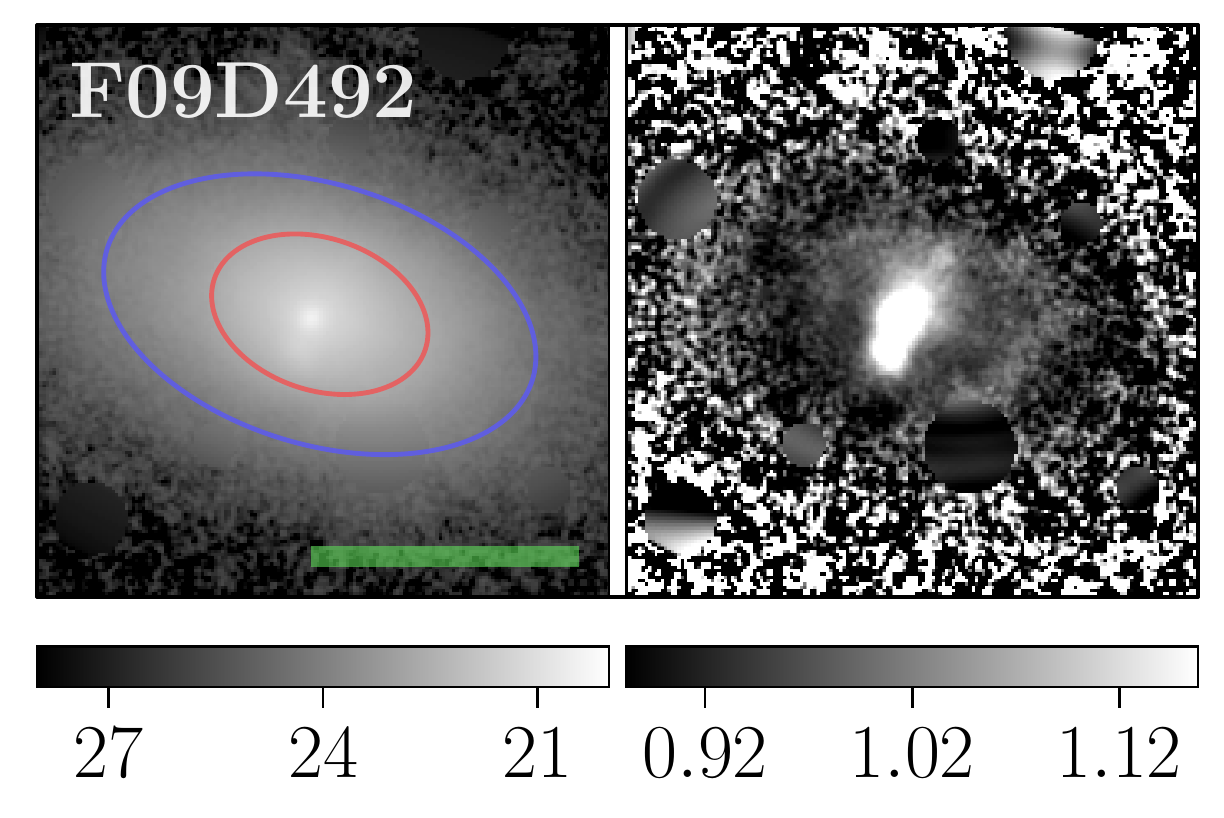}{0.29\textwidth}{}
          \hspace*{-40pt}
          \fig{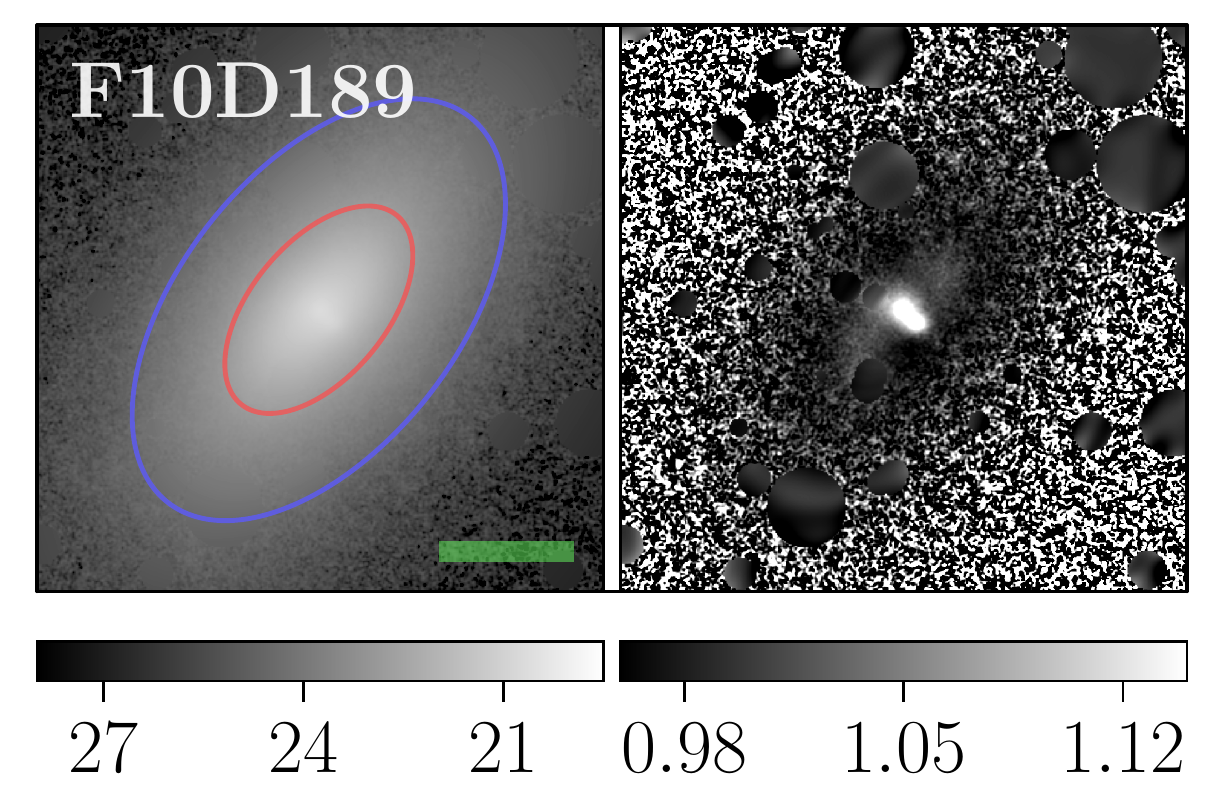}{0.29\textwidth}{}}
          \vspace*{-28.5pt}
\gridline{\fig{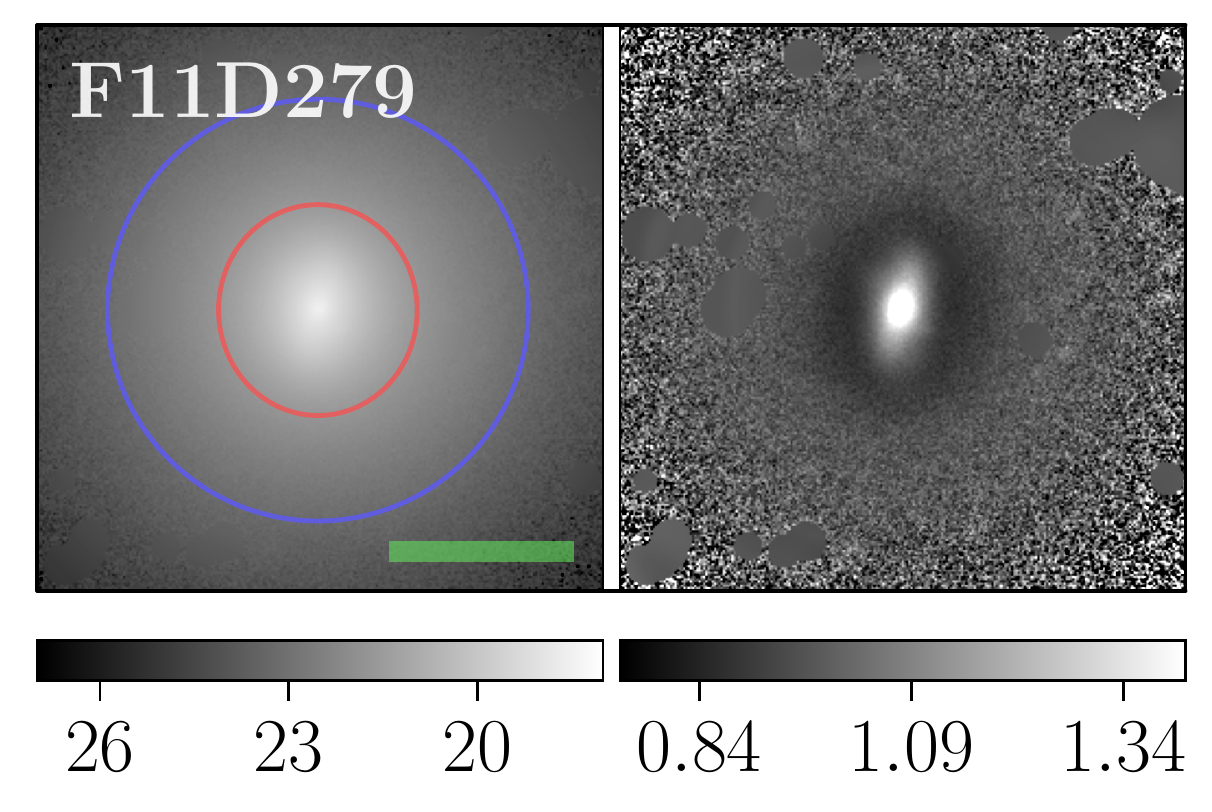}{0.29\textwidth}{}
          \hspace*{-40pt}
          \fig{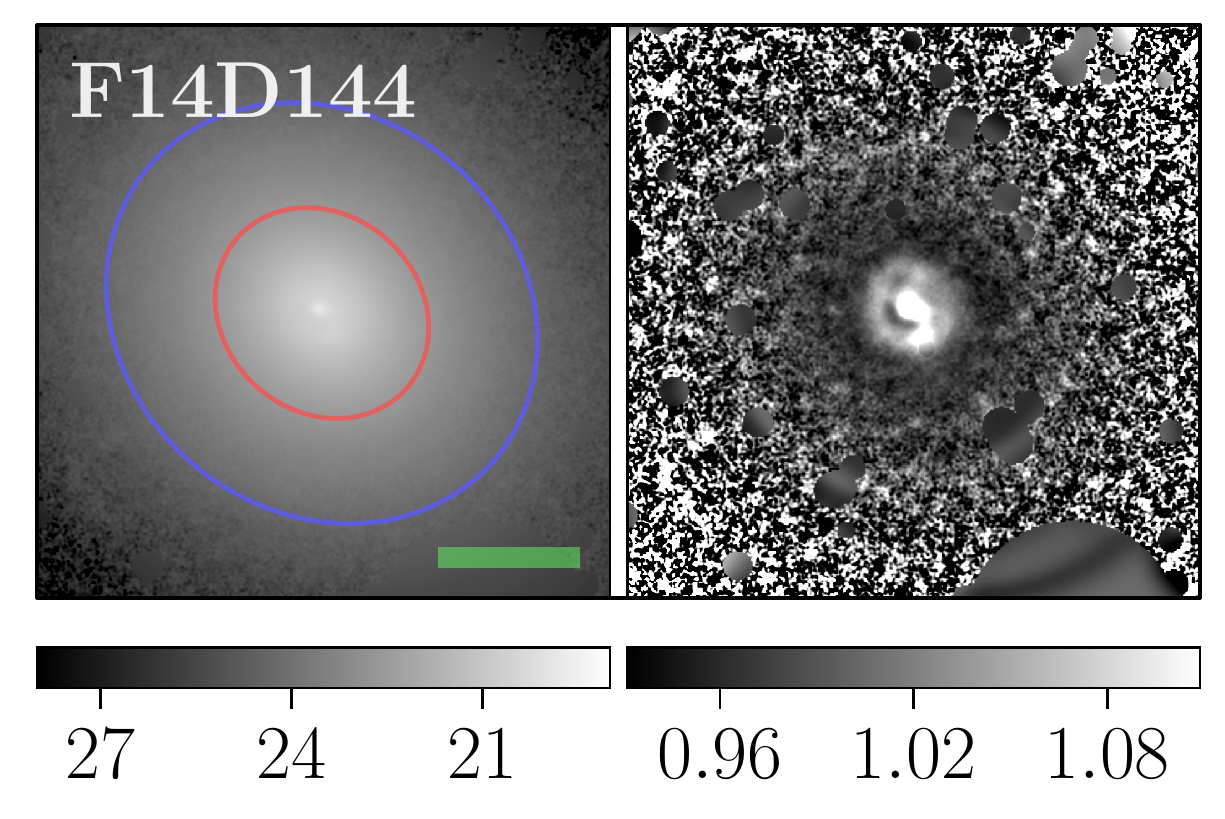}{0.29\textwidth}{}
          \hspace*{-40pt}
          \fig{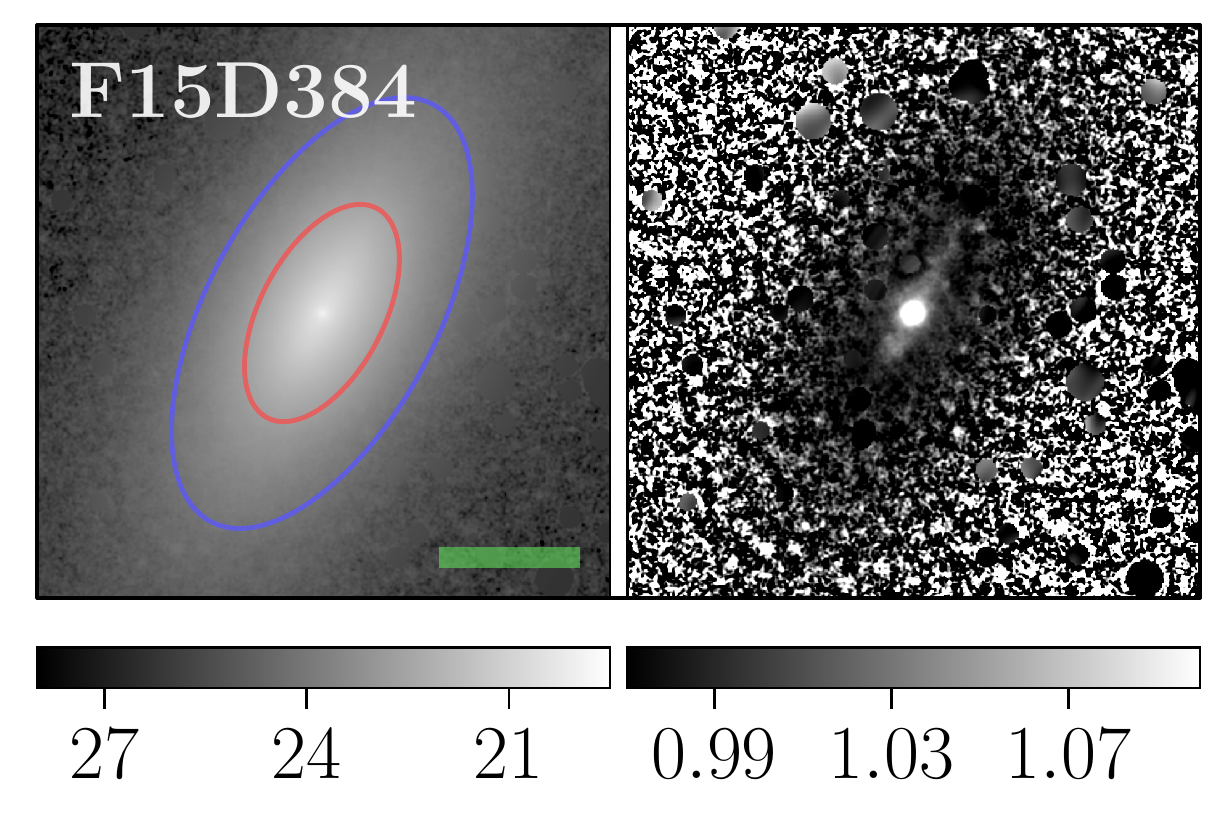}{0.29\textwidth}{}}
          \vspace*{-28.5pt}
\gridline{\fig{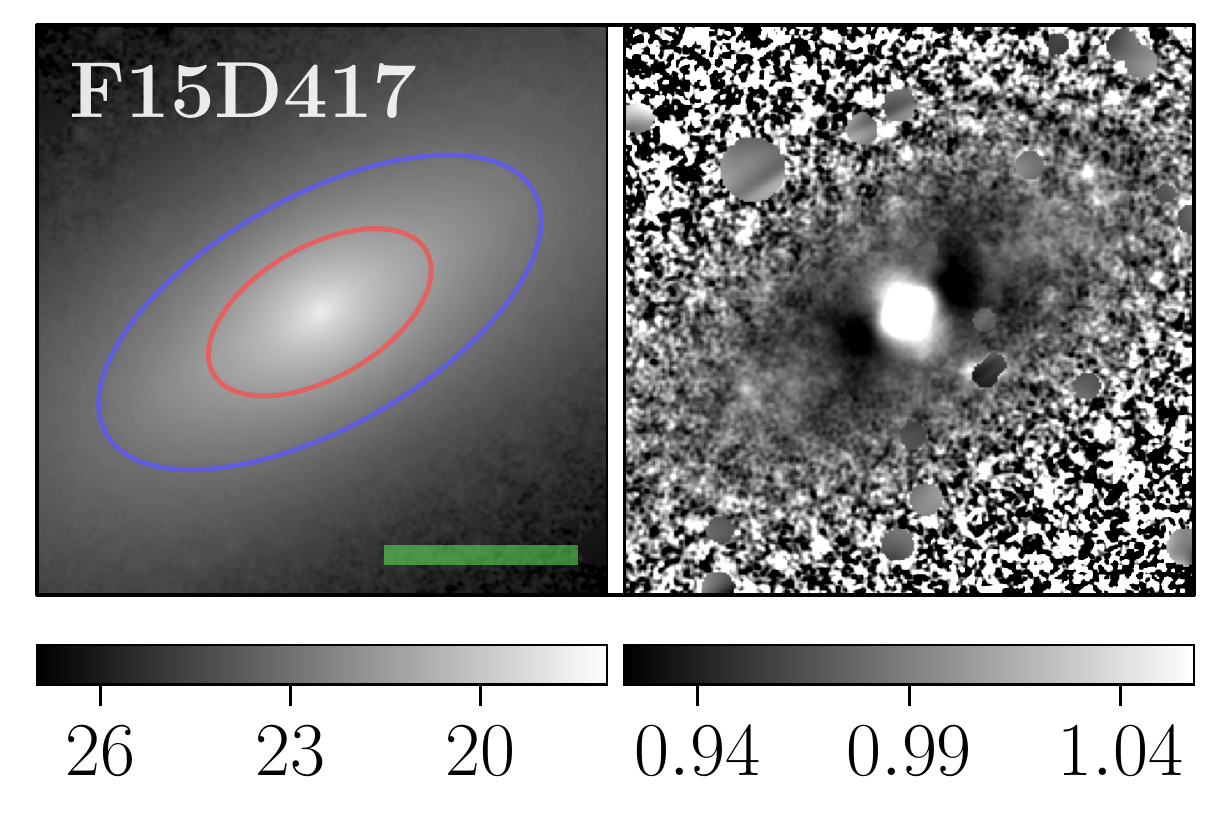}{0.29\textwidth}{}
          \hspace*{-40pt}
          \fig{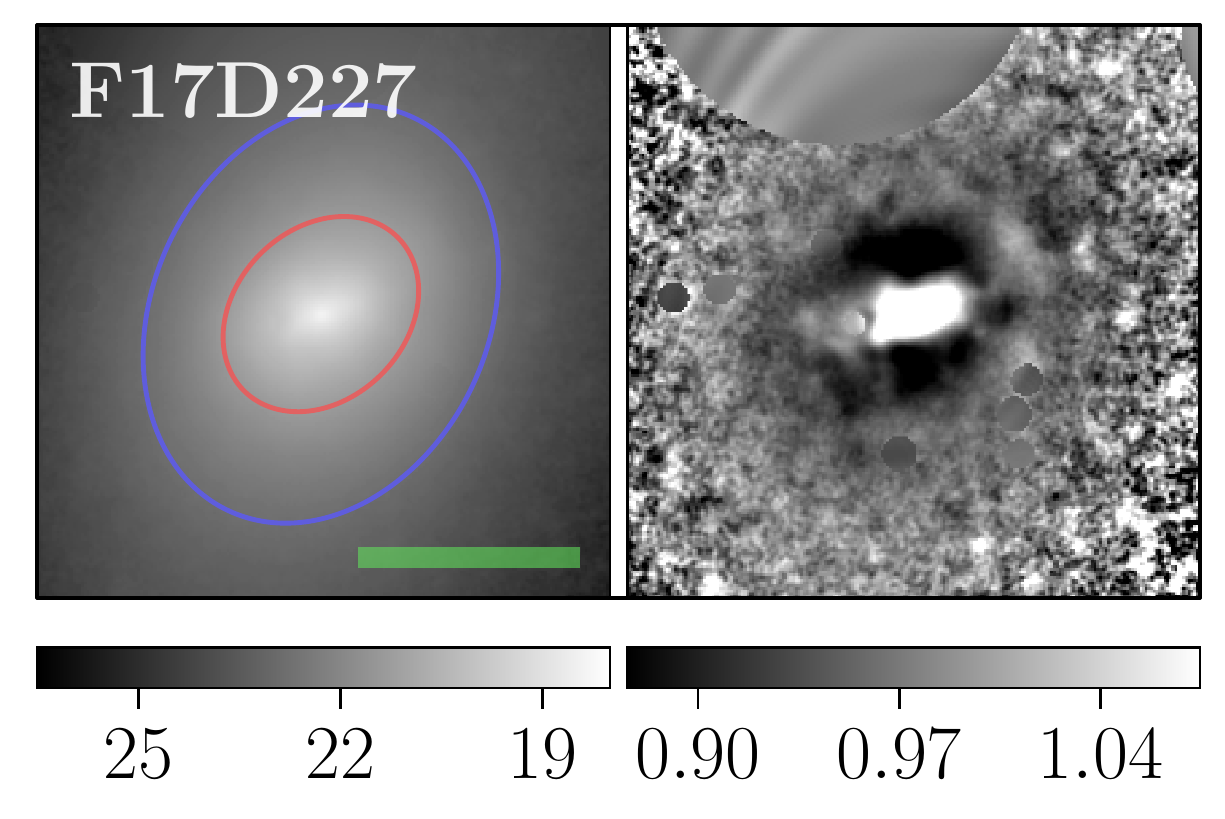}{0.29\textwidth}{}
          \hspace*{-40pt}
          \fig{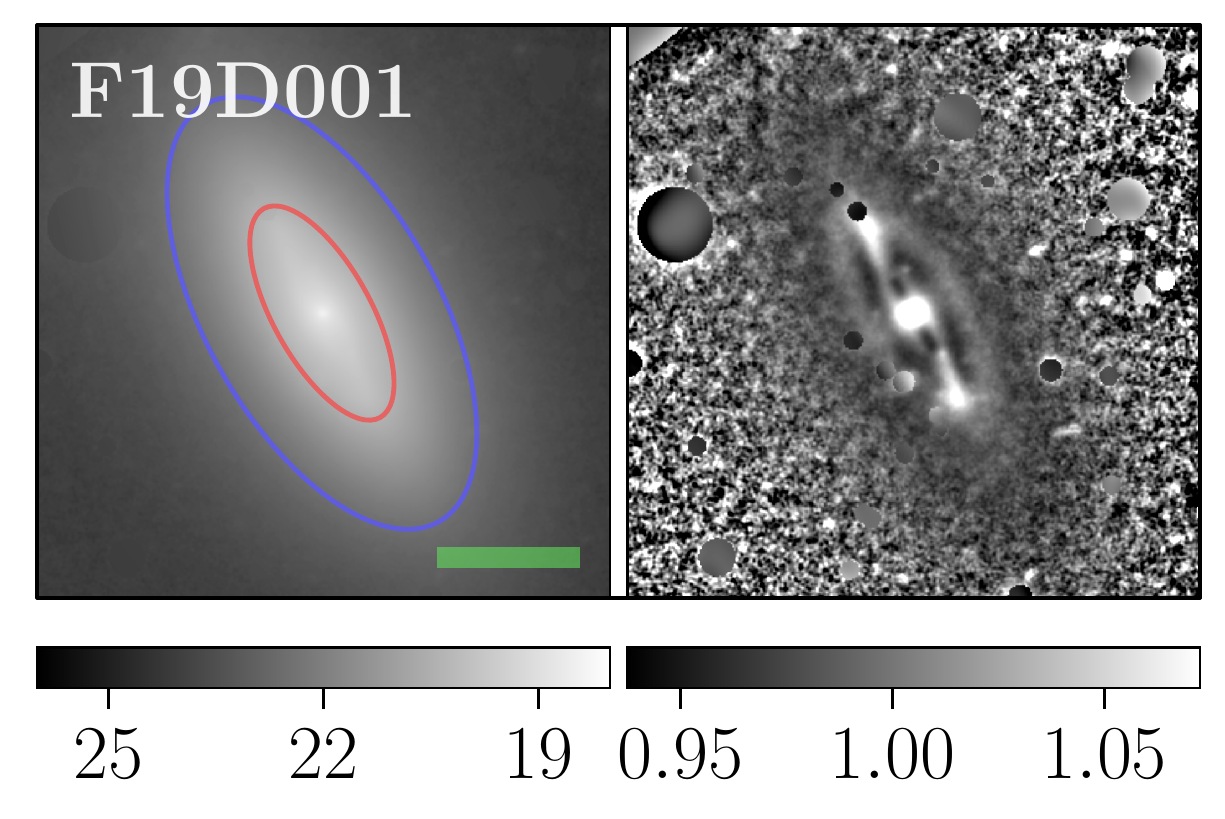}{0.29\textwidth}{}}
          \vspace*{-20pt}
\caption{Original galaxy images and unsharp mask images in the $r$ band of the Fornax dwarf ETG sample. Each image is labeled with the name of the corresponding dwarf galaxy, with its unsharp mask image displayed to its right. The galaxy isophotes at one and two effective radii are overlaid on the original image as red and blue ellipses, respectively. Interloping sources have been masked out. The original images are in units of mag arcsec$^{-2}$, while the unsharp masks are unitless due to being image ratios. The image scale is shown as a green line, and corresponds to $15$ arcsec ($1.5$ kpc). North is up, east is to the left.\label{fig:imgs_gals_uns}}
\end{center}
\end{figure*}

In order to visualize our Fornax dwarf ETG sample, we present the PSF-matched $r$-band galaxy images and their unsharp mask images in Figure \ref{fig:imgs_gals_uns}. A clean version of the galaxy images is shown, in which the interloping sources have been masked out. In some unsharp masks, wavy patterns appear on extended masked regions, but we note that these are only artifacts and not real features. The unsharp mask images are created by following the approach described in Section \ref{subsec:sample_uns}. For the size of the Gaussian smoothing kernel, we adopt a Gaussian standard deviation of size $4\times \sigma_{\text{PSF}}$ along its semi-major axis, derived from the FWHM of the $r$-band PSF. For its geometry, we adopt the ellipticity and position angle of the isophote at two effective radii the galaxy, which are given in Table \ref{tab:sample_struc}. We observe that the galaxy images have a predominantly smooth appearance, although in some cases it is already possible to discern some of their embedded substructure features. By removing the bright diffuse light of the galaxies, a rich variety of features is clearly revealed in their unsharp mask images. These include disk substructures, such as edge-on disks and rings; and clump substructures, such as irregular light overdensities and off-center nuclei. We note the possibility that the observed off-center nuclei could instead be nuclear star clusters, as those found in the central regions of bright dwarf galaxies \citep{Johnston:2020,Paudel:2020,Fahrion:2021}.

\addtocounter{figure}{-1}
\begin{figure*}[ht]
\begin{center}
\gridline{\fig{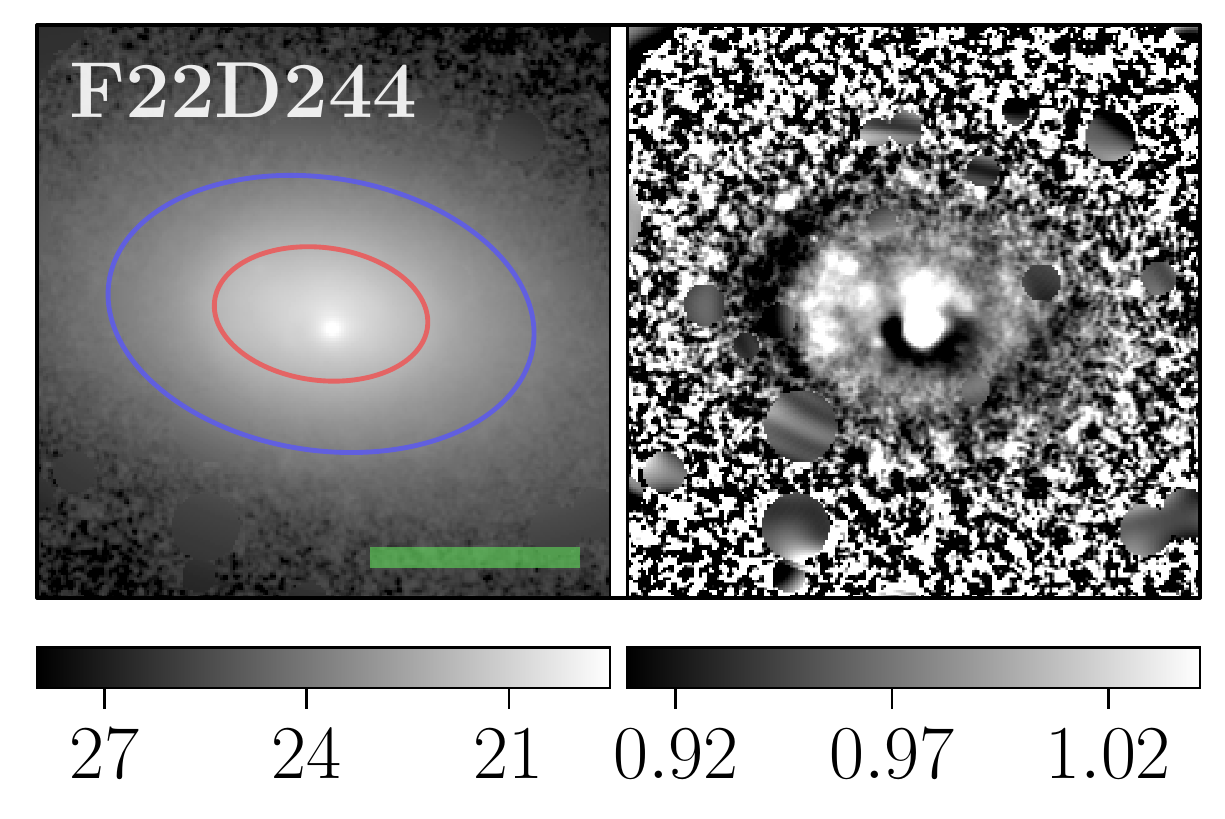}{0.29\textwidth}{}
          \hspace*{-40pt}
          \fig{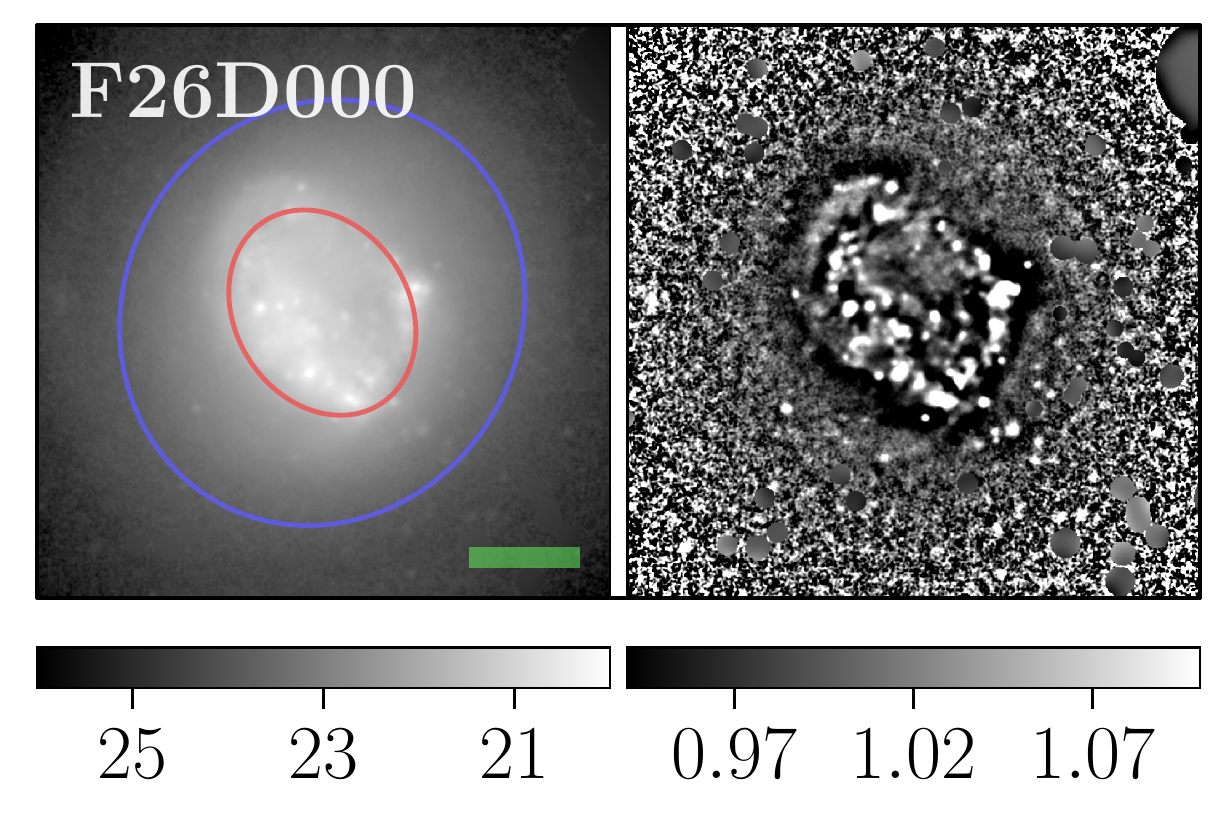}{0.29\textwidth}{}
          \hspace*{-40pt}
          \fig{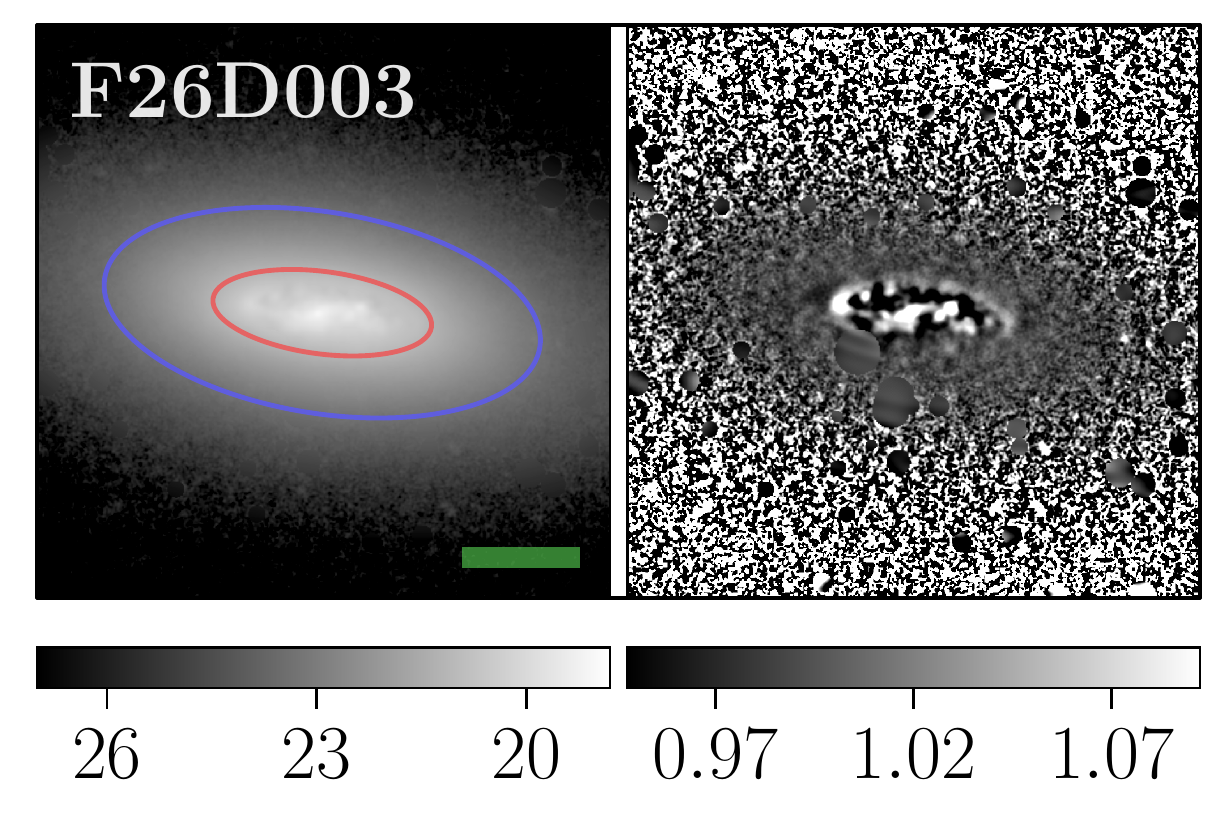}{0.29\textwidth}{}}
          \vspace*{-28.5pt}
\gridline{\fig{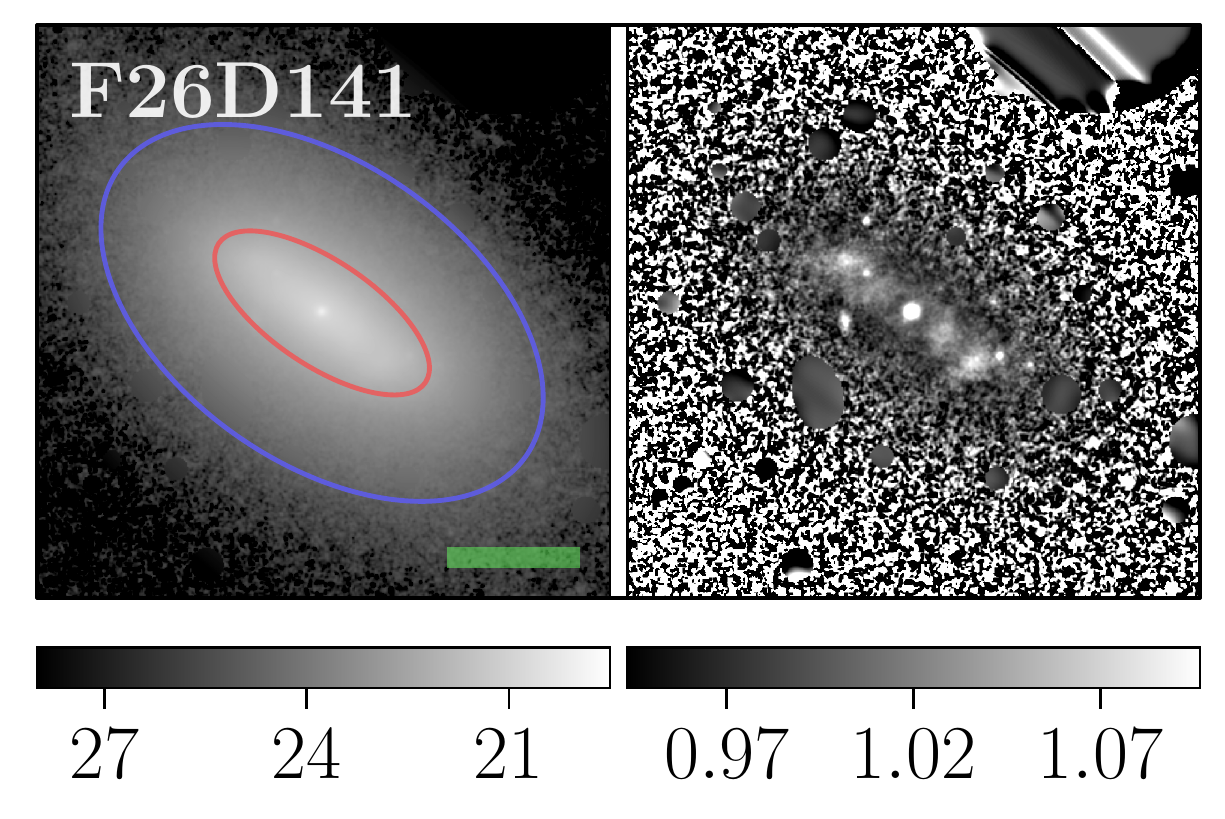}{0.29\textwidth}{}
          \hspace*{-40pt}
          \fig{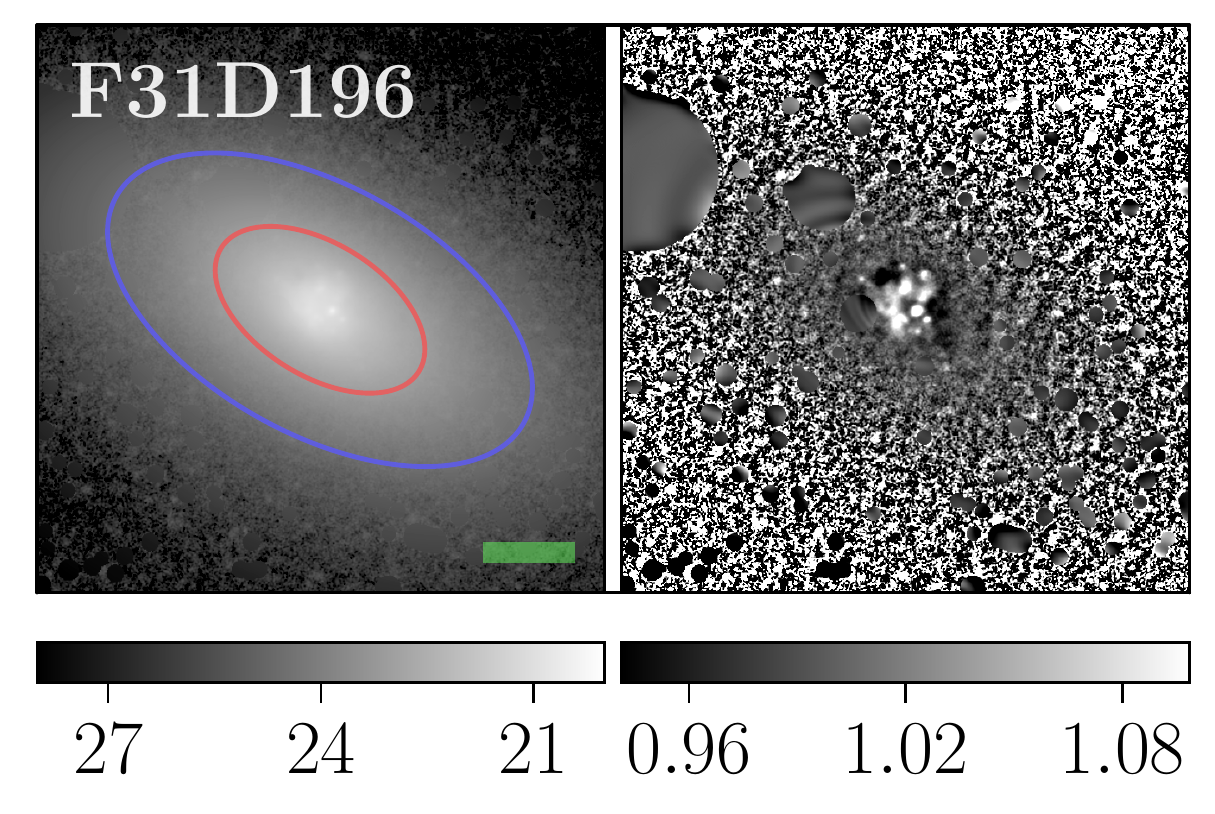}{0.29\textwidth}{}
          \hspace*{-40pt}
          \fig{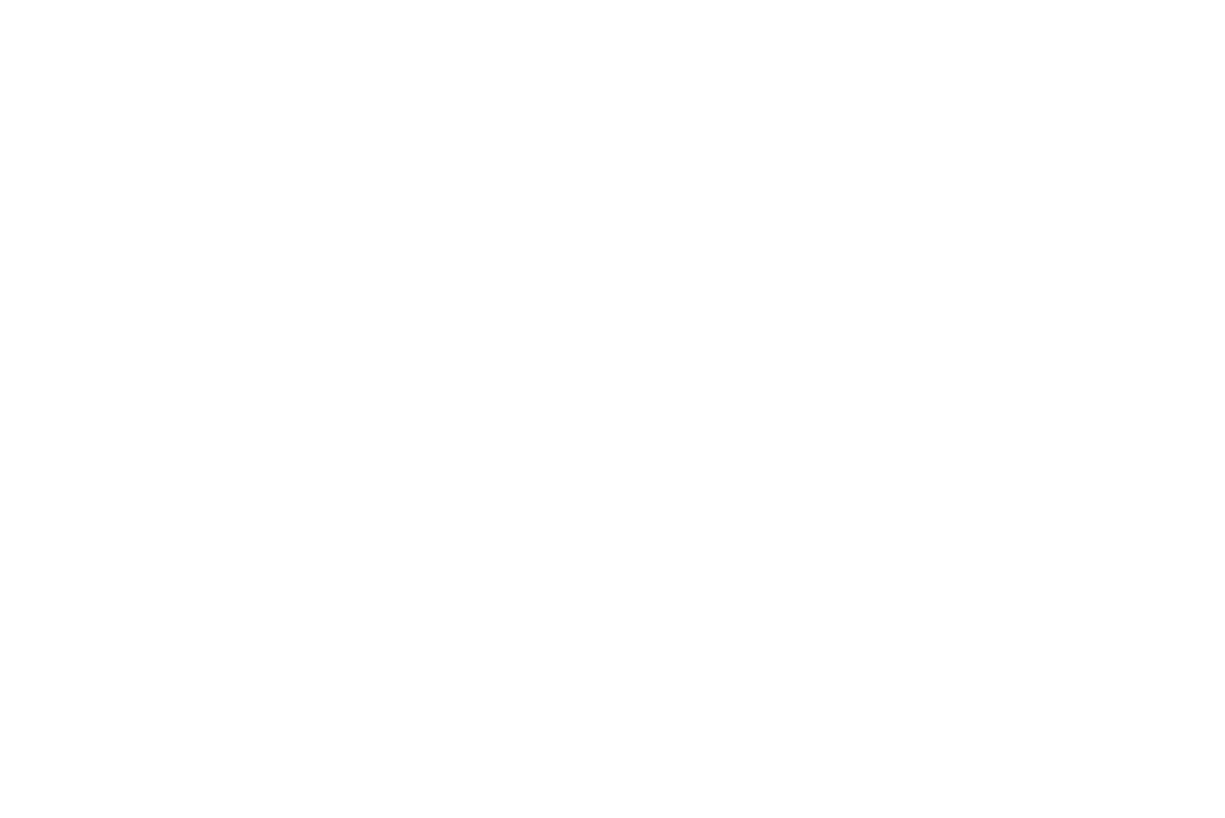}{0.29\textwidth}{}}
          \vspace*{-20pt}
\caption{Original galaxy images and unsharp mask images in the $r$ band of the Fornax dwarf ETG sample \textit{(cont.)}.}
\end{center}
\end{figure*}

\begin{figure*}[ht]
\begin{center}
\includegraphics[width=0.85\textwidth]{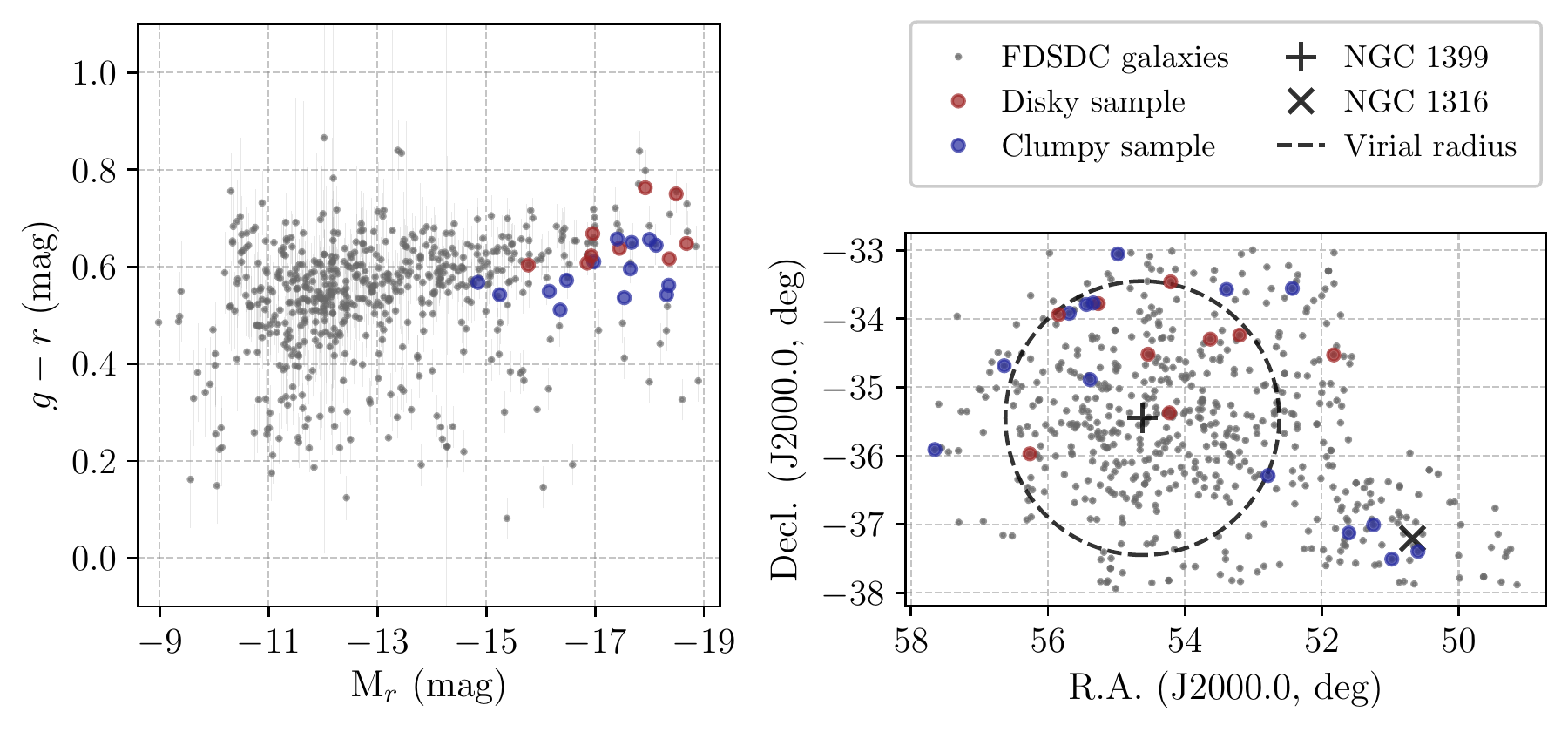}
\vspace*{-4pt}
\caption{Color-magnitude diagram and projected spatial distribution of the FDSDC dwarf galaxies and the dwarf ETGs in our sample. \textit{Left panel:} $g-r$ color measured within one effective radius vs. $r$-band total absolute magnitude of the galaxies. \textit{Right panel:} Projected spatial distribution of the galaxies in the ICRS. In both panels, all the FDSDC galaxies are plotted as gray points, while the galaxies that belong to our sample are highlighted as red and blue points, depending on their classification into the disky or clumpy categories, respectively. In the right panel, the central galaxy of the Fornax cluster (NGC 1399) is marked with a ``$+$'' sign, while the central galaxy of the Fornax A subgroup (NGC 1316) is marked with a ``$\times$'' sign. The dotted circle indicates the virial radius of the Fornax cluster \citep[R$_{\text{vir}} = 0.7$ Mpc;][]{Drinkwater:2001} centered on NGC 1399, which is equivalent to a radius of $2.0$ deg assuming a Fornax cluster distance of $20.0\pm1.4$ Mpc \citep{Blakeslee:2009}.\label{fig:plot_all_CMD_spatial}}
\end{center}
\end{figure*}

Finally, to understand the context of our selected dwarf ETGs, we provide a view into their color-magnitude and projected spatial distributions compared to the other FDSDC galaxies in Figure \ref{fig:plot_all_CMD_spatial}. Our sample is represented by red and blue points, based on the classification of their substructures into the disky and clumpy categories, respectively (according to criterion \textit{(2)} in Section \ref{subsec:sample_sel}). Our sample is consistent with being part of the red sequence defined by the dwarf galaxies of the Fornax cluster, with the disky subsample having in overall redder colors than the clumpy subsample. Additionally, the majority of the disky subsample lies inside the projected virial radius of the Fornax cluster, while the clumpy subsample tends to lie outside of it instead. Only galaxies belonging to the clumpy subsample are found in the projected vicinity of the Fornax A subgroup. When considered together, these differences could be an indication that dwarf ETGs with disk and clump substructures may constitute two different galaxy subpopulations. Or, alternatively, it may imply that the disk and clump subsamples are currently in different stages of their evolution. We will continue to address how the properties of the disk and clump subsamples differ from one another in the upcoming Sections \ref{sec:res_method} and \ref{sec:color}.

\section{Application of the Residual Method}\label{sec:res_method}

In our first paper of this series, Brought to Light I \citep{Michea:2021}, we presented the ``residual method'', a procedure we developed in order to separate a galaxy image into two distinct components. The primary (dominant, bright) component consists in the diffuse light of the galaxy, which is represented by a galaxy model image. The secondary (hidden, faint) component consists in the light of the galaxy contained in substructure features, which is represented by a galaxy residual image. The residual method achieves this separation gradually through a robust and iterative procedure, described in detail in \citet{Michea:2021}. In short, the galaxy model image is initially contaminated by some light that comes from the substructure features of the galaxy. The iterative aspect of the method allows it to progressively shift this extra light to the galaxy residual image instead. The iterations stop once the contamination of the model image is minimized, and the great majority of the substructure light is properly contained in the residual image. This way, the resulting residual image can be used to identify any substructure features embedded in the galaxy, and to quantify their relative contribution to the total galaxy light.

As an important caveat, throughout this work we define the substructure component as the excess, non-smooth light that is contained in the substructure features of a galaxy. Therefore, if disk features were to be present, the substructure component would not wholly represent the total amount of light that could potentially be contained in a physically thin embedded disk, as this thin disk could additionally have a smooth component of its own. This is a limitation of the residual method, as its goal is to separate smooth (axisymmetric) light from non-smooth (non-axisymmetric) light. Consequently, it does not allow us to distinguish nor identify different sources of smooth light ---such as the diffuse light originating from a thick disk from that of a thin disk.

In the next Subsections, we describe how we applied the residual method to the $g$ and $r$-band images of our sample.

\subsection{Residual Method Setup}\label{subsec:res_method_params}

In order to handle images in two bands, we first run the residual method on the PSF-matched $r$-band images of the sample closely following the steps described in \citet{Michea:2021}. This means that, for each galaxy, we obtain a series of $r$-band residual images, where each one corresponds to a particular configuration of the method's parameters. Next, to run the residual method on the $g$ band, we make use of the geometry information of the galaxy obtained during the run in the $r$ band. Specifically, for each parameter configuration set, we extract the radial shape and orientation of the diffuse component of the galaxy at the $r$-band stopping iteration. By using the \texttt{inellip} parameter of the IRAF \texttt{ellipse} task, this geometry is then imposed and kept fixed when running the method on the PSF-matched $g$-band images. The method is then allowed to iterate as usual, until a $g$-band stopping iteration is reached.

In summary, the residual method is first run freely on the $r$ band, then the geometry of the $r$-band diffuse component is extracted, and finally the method is run with this constrained geometry on the $g$ band. This way, the geometry of the diffuse component is kept fixed between the $g$ and $r$ bands, making any photometric measurements directly comparable. We note that a given parameter configuration set still has its own stopping iteration that is independent for each galaxy in each band. In the $g$ band, the iterative aspect of the method requires between one to three iterations, with a median of two iterations. In contrast, in the $r$ band, it requires between one to two iterations, with a median of one iteration.

With respect to the parameter configuration of the method, the same values are implemented in both the $g$ and $r$-band runs. The specific parameters that must be tuned to the properties of the data set correspond to the smoothing kernel size and the sampling step size.

On the one hand, the smoothing kernel size dictates the amount of smoothing the galaxy image is subjected to. It should be tuned to match the average half-width of the substructure features of the galaxies. By inspecting the unsharp mask images of our dwarf ETG sample (see Figure \ref{fig:imgs_gals_uns}), we identify the substructure features and measure that their average half-widths lie in the range of $1.2-4.7$ arcsec ($0.1-0.5$ kpc), which expressed in terms of the PSF FWHM of the data transforms into a Gaussian standard deviation corresponding to $3-8 \times \sigma_{\text{PSF}}$. Consequently, we adopt a Gaussian smoothing kernel with a standard deviation equal to $3,4,5,6,7,$ and $8 \times \sigma_{\text{PSF}}$.

On the other hand, the sampling step size dictates the frequency at which the radial light profile of the galaxy is sampled. It should be tuned based on the image resolution and provide the ability to sample the lower S/N regions at the galaxy outskirts. Taking into consideration that the resolution of the FDS data corresponds to $0.2$ arcsec pixel$^{-1}$, we carry out some test runs, and determine that the optimal separation between successive isophotes is given by a growth rate of the step size length in the range of $18-23\%$. Therefore, we adopt sampling step sizes with a growth rate equal to $18,19,20,21,22,$ and $23\%$.

Together, these parameter ranges create a $6\times 6$ configuration grid. Consequently, the residual method is independently run a total of $36$ times on the $g$ and $r$-band images of a galaxy, with each run adopting a particular smoothing-sampling parameter pair. As a result, small variations will be present throughout the multiple residual images being obtained for a single galaxy. When quantifying the contribution of the residual light to the total galaxy light, these variations then translate into the uncertainty range of the median measurement.

\subsection{Residual Method Results}\label{subsec:res_method_results}

\begin{figure*}[ht]
\centering
\includegraphics[width=0.84\textwidth]{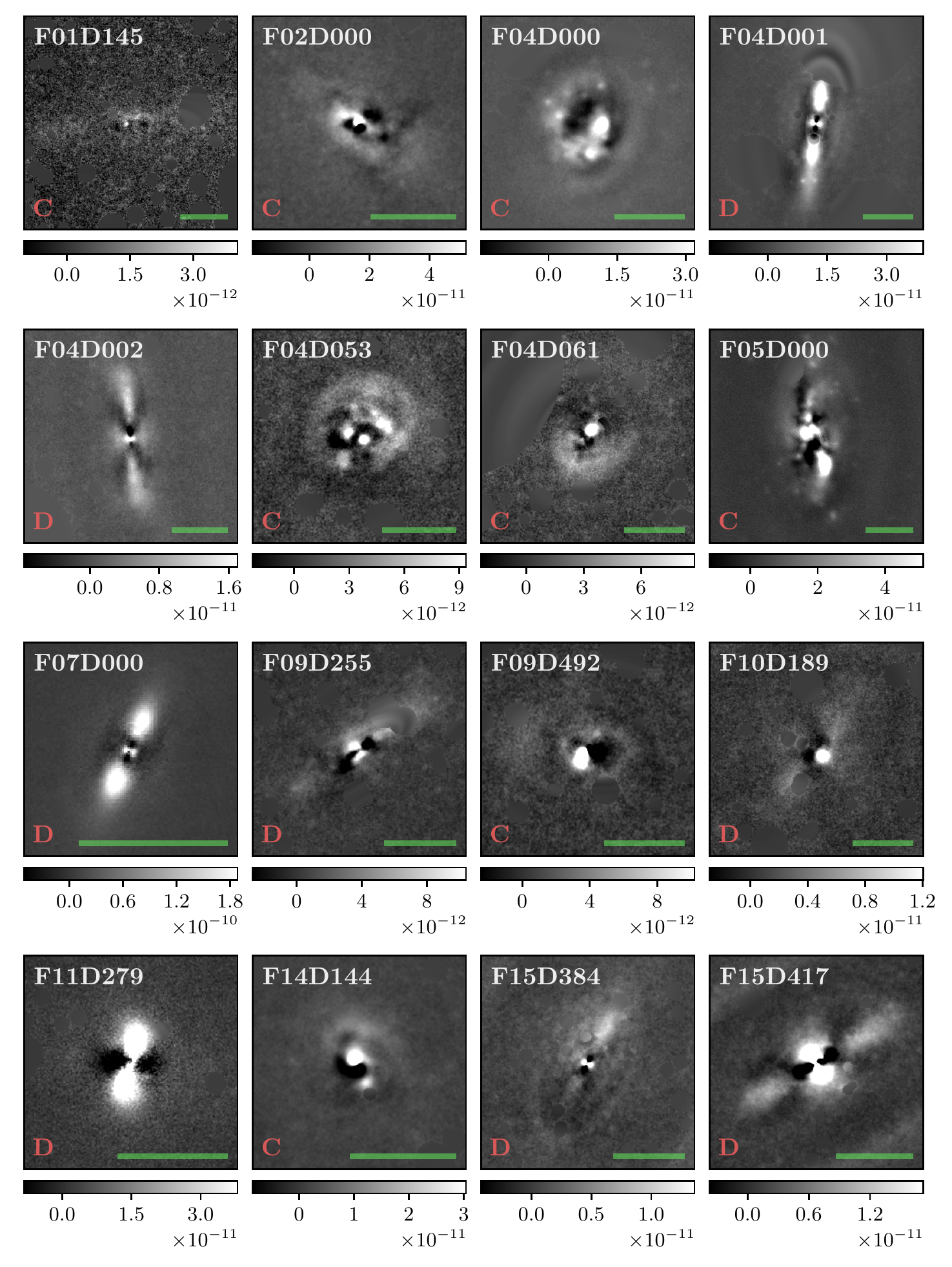}
\vspace*{-4pt}
\caption{Galaxy residual images in the $r$ band of our Fornax dwarf ETG sample. The parameter setup used to obtain these particular residual images corresponds to a smoothing kernel size equal to 5 times the Gaussian standard deviation of the PSF, and a sampling step size with a growth rate of 20\% between successive isophotes. Each image is labeled with the name of the corresponding dwarf galaxy, and has its own gray-scale bar in units of erg cm$^{-2}$ s$^{-1}$ Hz$^{-1}$. The red letter indicates the classification of the galaxy as either disky (``D'') or clumpy (``C''). The image scale is shown as a green line, and corresponds to 15 arcsec ($1.5$ kpc). North is up, east is to the left.\label{fig:imgs_residuals}}
\end{figure*}

By adopting the aforementioned parameter configuration, we proceed to apply the residual method to our sample of Fornax dwarf ETGs to measure the light in their substructure features. In Figure \ref{fig:imgs_residuals} we show $r$-band residual images of the sample. For illustration purposes, we do not show $g$-band residual images, as the appearance of the substructure features is extremely similar (if not completely the same) in both bands. We observe that the rich variety of substructures that were originally revealed in the unsharp mask images (see Figure \ref{fig:imgs_gals_uns}) have also been faithfully captured in the residual images. The appearance of the residuals supports our substructure classification into disk substructures and clump substructures described in criterion \textit{(2)} of Section \ref{subsec:sample_sel}. Out of the $23$ dwarf ETGs in our sample, we classify $9$ ($40\%$) as disky, and $14$ ($60\%$) as clumpy. The substructure classification of each galaxy is provided in column 2 of Table \ref{tab:rlfs}. We highlight that $10$ dwarf ETGs of our sample are also in the work of \citet{Hamraz:2019}, in which HST/ACS data is used to study Fornax galaxies. These images have higher spatial resolution compared to the FDS images. Notably, their unsharp mask images \citep[see Figure 8 of][]{Hamraz:2019} reveal very similar substructure features as the ones we observe in our unsharp mask and residual images that are based on the shallower FDS data, thus serving as further confirmation of our performed classification.

\addtocounter{figure}{-1}
\begin{figure*}[ht]
\centering
\includegraphics[width=0.84\textwidth]{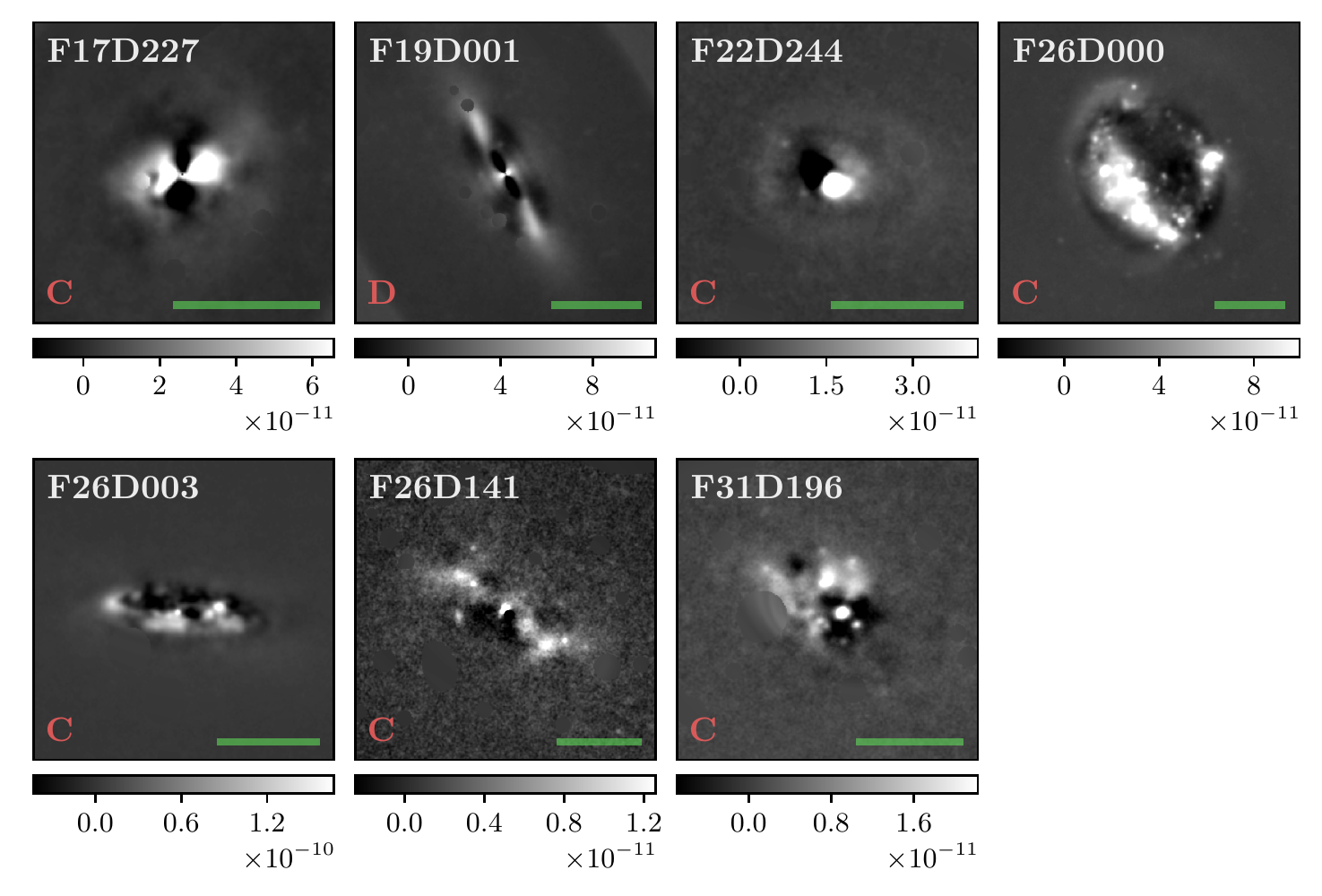}
\vspace*{-4pt}
\caption{Galaxy residual images in the $r$ band of our Fornax dwarf ETG sample \textit{(cont.)}.}
\end{figure*}

We note that some of our galaxies with embedded edge-on disks present artifacts in the central region of their residual images. These artifacts manifest as an hourglass-shaped light overdensity that is perpendicular to the major axis of the disk, and can be either very prominent (e.g., F04D002, F15D417) or quite small (e.g., F04D001, F07D000). These fake features arise as a direct consequence of smoothing something that is circular in shape with a highly elliptical kernel, such as a prominent spherical bulge or nucleus. In the residual method, the smoothing kernel adopts the shape and orientation of the galaxy isophote at two effective radii, as the objective is to match the overall geometry of the diffuse light and not the geometry of the most central region. Consequently, this makes the presence of such an artifact unavoidable in some cases. To address it, we carefully identify the affected regions in the residual images, and flag the pixels as bad in the corresponding BPM images. This way, all measurements and analyses we carry out consistently exclude these artificial features, whenever present.

To quantify the contribution of the substructure features to the total galaxy light, we compute the ``residual light fraction'' (RLF) within the isophotes at one and two effective radii of the galaxies. The residual light fraction is defined as the ratio of the residual light (contained in the galaxy residual image) to the total light (contained in the galaxy original image), and it is always quantified in its respective band. The results obtained are presented in Table \ref{tab:rlfs}. In the $g$ band, we find that the residual light fraction ranges between $2.7-22.3\%$ within one effective radius, with a median value of $6.7\%$, and between $4.7-20.0\%$ within two effective radii, with a median value of $8.7\%$. Similarly, in the $r$ band, it ranges between $2.9-20.8\%$ within one effective radius, with a median value of $4.6\%$, and between $4.5-16.9\%$ within two effective radii, with a median value of $5.3\%$. Overall, the RLF tends to be larger at two effective radii than at one effective radius, and larger in the $g$ band than in the $r$ band.

\begin{deluxetable*}{clccccc}[ht!]
\setlength{\tabcolsep}{6pt} 
\renewcommand{\arraystretch}{1} 
\tablewidth{0pt}
\tabletypesize{\small}
\tablecaption{Substructure classifications and residual light fractions of our Fornax dwarf ETG sample.\label{tab:rlfs}}
\tablehead{
\colhead{Dwarf Galaxy} & \colhead{Classification} & \multicolumn5c{Residual Light Fraction} \vspace*{1pt} \\
\cline{3-7} \\ [-2ex]
 & & \multicolumn2c{Within $1\,\text{R}_{e}$} & & \multicolumn2c{Within $2\,\text{R}_{e}$} \vspace*{1pt} \\
\cline{3-4} \cline{6-7} \\ [-3ex]
 & & \colhead{$g$ band} & \colhead{$r$ band} & & \colhead{$g$ band} & \colhead{$r$ band} \\
\colhead{(1)} & \colhead{(2)} & \colhead{(3)} & \colhead{(4)} & & \colhead{(5)} & \colhead{(6)} \vspace*{0.5pt}
}
\decimals
\startdata
F01D145 & Clumpy & $0.065^{+0.005}_{-0.006}$ & $0.038^{+0.006}_{-0.005}$ & & $0.110^{+0.007}_{-0.008}$ & $0.060^{+0.012}_{-0.006}$ \\ [0.8ex]
F02D000 & Clumpy & $0.067^{+0.006}_{-0.005}$ & $0.039^{+0.005}_{-0.004}$ & & $0.085^{+0.008}_{-0.007}$ & $0.046^{+0.006}_{-0.006}$ \\ [0.8ex]
F04D000 & Clumpy & $0.086^{+0.004}_{-0.019}$ & $0.050^{+0.003}_{-0.003}$ & & $0.087^{+0.005}_{-0.016}$ & $0.053^{+0.005}_{-0.005}$ \\ [0.8ex]
F04D001 & Disky & $0.051^{+0.009}_{-0.009}$ & $0.050^{+0.009}_{-0.009}$ & & $0.050^{+0.006}_{-0.007}$ & $0.051^{+0.006}_{-0.007}$ \\ [0.8ex]
F04D002 & Disky & $0.043^{+0.005}_{-0.006}$ & $0.032^{+0.004}_{-0.006}$ & & $0.063^{+0.011}_{-0.006}$ & $0.050^{+0.005}_{-0.006}$ \\ [0.8ex]
F04D053 & Clumpy & $0.101^{+0.007}_{-0.006}$ & $0.077^{+0.006}_{-0.006}$ & & $0.129^{+0.013}_{-0.009}$ & $0.095^{+0.009}_{-0.007}$ \\ [0.8ex]
F04D061 & Clumpy & $0.078^{+0.007}_{-0.007}$ & $0.065^{+0.005}_{-0.006}$ & & $0.109^{+0.008}_{-0.008}$ & $0.084^{+0.006}_{-0.007}$ \\ [0.8ex]
F05D000 & Clumpy & $0.107^{+0.009}_{-0.015}$ & $0.087^{+0.005}_{-0.009}$ & & $0.103^{+0.009}_{-0.015}$ & $0.092^{+0.008}_{-0.010}$ \\ [0.8ex]
F07D000 & Disky & $0.055^{+0.009}_{-0.009}$ & $0.057^{+0.009}_{-0.010}$ & & $0.052^{+0.007}_{-0.006}$ & $0.053^{+0.007}_{-0.007}$ \\ [0.8ex]
F09D255 & Disky & $0.050^{+0.010}_{-0.011}$ & $0.033^{+0.008}_{-0.007}$ & & $0.091^{+0.010}_{-0.010}$ & $0.063^{+0.010}_{-0.008}$ \\ [0.8ex]
F09D492 & Clumpy & $0.084^{+0.007}_{-0.005}$ & $0.046^{+0.004}_{-0.004}$ & & $0.099^{+0.008}_{-0.006}$ & $0.054^{+0.007}_{-0.004}$ \\ [0.8ex]
F10D189 & Disky & $0.062^{+0.004}_{-0.006}$ & $0.037^{+0.005}_{-0.004}$ & & $0.079^{+0.012}_{-0.004}$ & $0.045^{+0.011}_{-0.004}$ \\ [0.8ex]
F11D279 & Disky & $0.051^{+0.005}_{-0.006}$ & $0.051^{+0.007}_{-0.006}$ & & $0.053^{+0.006}_{-0.006}$ & $0.054^{+0.006}_{-0.006}$ \\ [0.8ex]
F14D144 & Clumpy & $0.051^{+0.007}_{-0.007}$ & $0.037^{+0.004}_{-0.004}$ & & $0.067^{+0.010}_{-0.010}$ & $0.045^{+0.006}_{-0.005}$ \\ [0.8ex]
F15D384 & Disky & $0.057^{+0.006}_{-0.005}$ & $0.037^{+0.005}_{-0.003}$ & & $0.083^{+0.007}_{-0.007}$ & $0.049^{+0.005}_{-0.004}$ \\ [0.8ex]
F15D417 & Disky & $0.027^{+0.010}_{-0.009}$ & $0.029^{+0.011}_{-0.010}$ & & $0.055^{+0.009}_{-0.007}$ & $0.052^{+0.008}_{-0.007}$ \\ [0.8ex]
F17D227 & Clumpy & $0.086^{+0.022}_{-0.018}$ & $0.064^{+0.008}_{-0.008}$ & & $0.085^{+0.018}_{-0.015}$ & $0.061^{+0.007}_{-0.007}$ \\ [0.8ex]
F19D001 & Disky & $0.032^{+0.004}_{-0.003}$ & $0.034^{+0.004}_{-0.003}$ & & $0.047^{+0.007}_{-0.005}$ & $0.048^{+0.006}_{-0.005}$ \\ [0.8ex]
F22D244 & Clumpy & $0.100^{+0.004}_{-0.023}$ & $0.068^{+0.021}_{-0.003}$ & & $0.101^{+0.006}_{-0.021}$ & $0.074^{+0.019}_{-0.004}$ \\ [0.8ex]
F26D000 & Clumpy & $0.223^{+0.009}_{-0.010}$ & $0.208^{+0.005}_{-0.059}$ & & $0.201^{+0.005}_{-0.008}$ & $0.169^{+0.008}_{-0.039}$ \\ [0.8ex]
F26D003 & Clumpy & $0.106^{+0.017}_{-0.013}$ & $0.068^{+0.005}_{-0.004}$ & & $0.102^{+0.015}_{-0.011}$ & $0.068^{+0.007}_{-0.006}$ \\ [0.8ex]
F26D141 & Clumpy & $0.070^{+0.006}_{-0.006}$ & $0.040^{+0.006}_{-0.005}$ & & $0.123^{+0.010}_{-0.008}$ & $0.053^{+0.008}_{-0.006}$ \\ [0.8ex]
F31D196 & Clumpy & $0.072^{+0.005}_{-0.005}$ & $0.045^{+0.003}_{-0.004}$ & & $0.091^{+0.006}_{-0.006}$ & $0.052^{+0.004}_{-0.006}$ \\ [0.3ex]
\enddata
\tablecomments{Col. (1): name of the dwarf galaxy. Col. (2): classification of the substructures into the disky or clumpy categories. Cols. (3) and (4): residual light fraction measurements within the one effective radius isophote in the $g$ and $r$ bands, respectively. Cols. (5) and (6): residual light fraction measurements within the two effective radii isophote in the $g$ and $r$ bands, respectively.}
\end{deluxetable*}

\begin{figure*}[ht]
\centering
\includegraphics[width=0.84\textwidth]{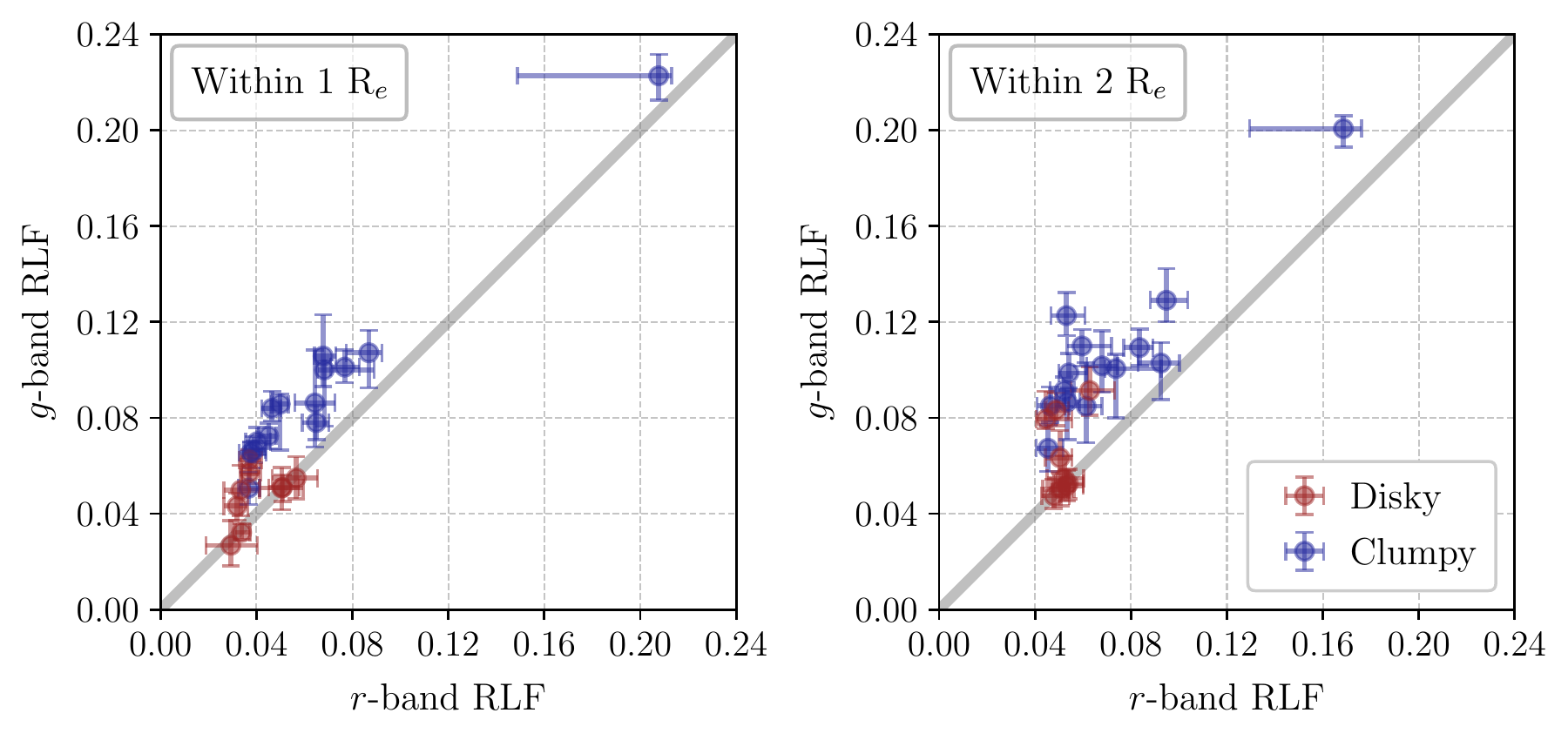}
\vspace*{-4pt}
\caption{Residual light fractions in the $g$ band vs. the $r$ band of our Fornax dwarf ETG sample. \textit{Left panel:} residual light fractions measured within one effective radius. \textit{Right panel:} residual light fractions measured within two effective radii. Galaxies classified as disky are shown as red points, while galaxies classified as clumpy are shown as blue points. The error bars are given by the 16th and 84th percentiles of the distributions. The one-to-one relation is shown as a gray line.\label{fig:plot_RLF_g_vs_r}}
\end{figure*}

To visualize these results, and to establish if there are any differences between the disky and clumpy subsamples, we plot the $g$-band versus the $r$-band residual light fractions in Figure \ref{fig:plot_RLF_g_vs_r}. It is evident that there is a strong correlation between the measurements: if the substructures have a high relative brightness in the $g$ band (i.e., a large RLF), then they will also appear bright in the $r$ band, and vice versa. To quantify this correlation, we compute the Pearson correlation coefficient $r_{P}$ of the distributions and its associated $p$-value. We find that the distribution within one effective radius (left plot) has a Pearson correlation coefficient $r_{P}=0.94$ and a $p$-value $=1 \times 10^{-11}$. Similarly, the distribution within two effective radii (right plot) has an $r=0.84$ and a $p$-value $=5 \times 10^{-7}$. Thus, the $g$ and $r$-band residual light fractions are strongly correlated within both one and two effective radii of the galaxies. However, we also note that for a given $r$-band RLF, the $g$-band RLF tends to be comparatively larger (i.e., above the one-to-one relation, the gray line in Figure \ref{fig:plot_RLF_g_vs_r}). This difference is accentuated when dividing the dwarf ETGs into the disky and clumpy subsamples. Overall, the disky subsample presents smaller RLFs in both bands, and also smaller differences between its $g$ and $r$-band measurements. In contrast, the clumpy subsample presents larger RLFs, with its $g$-band measurements being larger than the $r$-band measurements. This is a possible indication that the substructure features are comparatively bluer in the clumpy subsample than in the disky subsample.

We would like to note that dwarf galaxy F26D000, located in the upper right regions of Figure \ref{fig:plot_RLF_g_vs_r}, appears to be an outlier, as it presents a residual light fraction that is significantly larger compared to the rest of the sample. However, our choice of including this galaxy in our sample is well justified: it fulfills the three selection criteria of Section \ref{subsec:sample_sel}, and it is not even the brightest nor the bluest among the sample galaxies. Based on Figures \ref{fig:imgs_gals_uns} and \ref{fig:imgs_residuals}, F26D000 appears to have quite bright, numerous clumpy substructures, which logically translate into bright residuals, and thus into a large RLF.

\begin{figure*}[ht]
\centering
\includegraphics[width=0.84\textwidth]{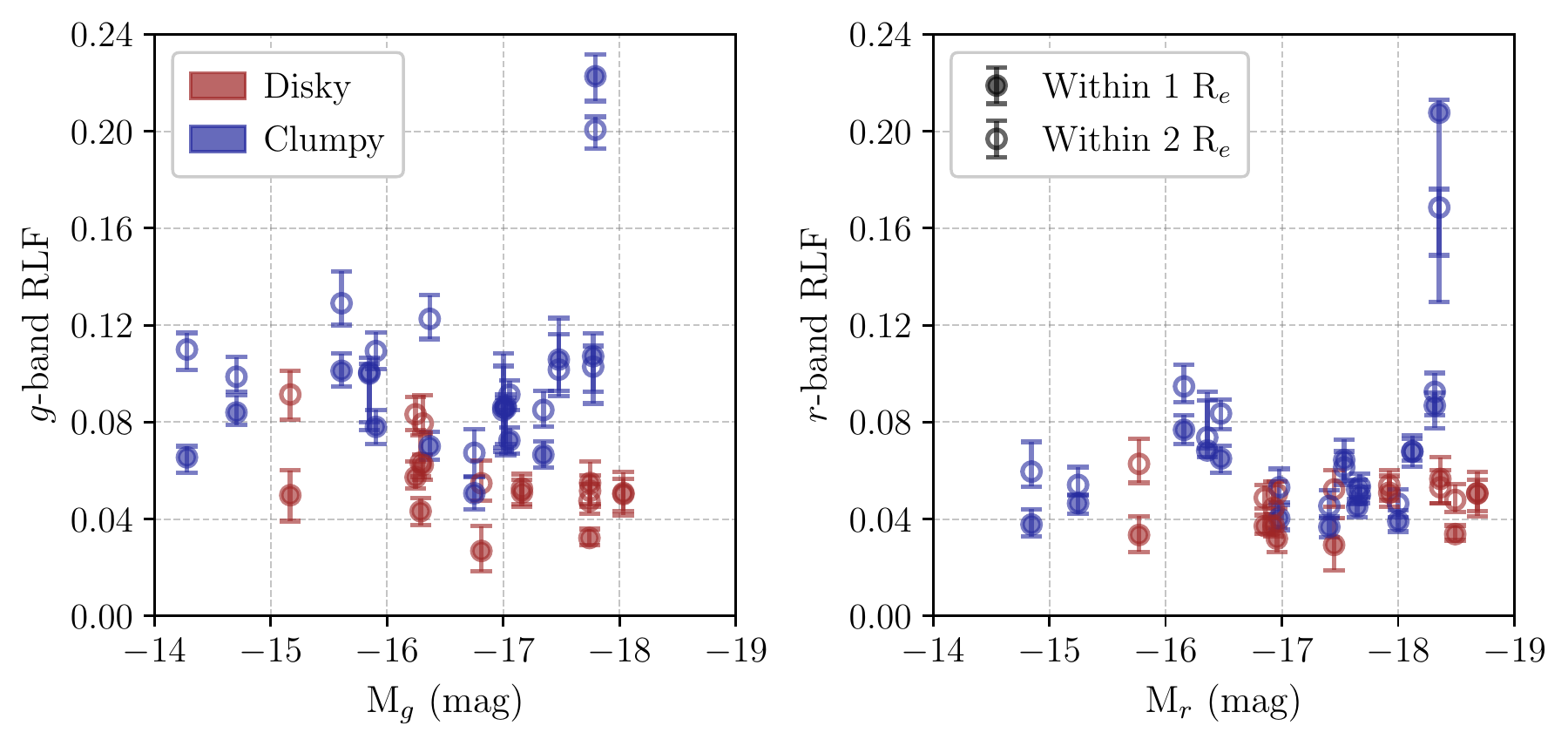}
\vspace*{-4pt}
\caption{Residual light fractions of our Fornax dwarf ETG sample as a function of their total absolute magnitude. \textit{Left panel:} $g$-band residual light fraction vs. total $g$-band absolute magnitude. \textit{Right panel:} $r$-band residual light fraction vs. total $r$-band absolute magnitude. Galaxies classified as disky are shown as red points, while galaxies classified as clumpy are shown as blue points. The measurements are performed within one and two effective radii of the galaxies, shown as filled and empty circles, respectively. The error bars are given by the 16th and 84th percentiles of the distributions.\label{fig:plot_RLF_vs_mag}}
\end{figure*}

In Figure \ref{fig:plot_RLF_vs_mag}, we plot the $g$ and $r$-band residual light fractions versus the $g$ and $r$-band total absolute magnitude of the galaxies, respectively. Two main things are noticeable. First, there appear to be no significant trends, which we corroborate by computing the Pearson correlation coefficient $r_{P}$ of the distributions and its associated $p$-value. However, if we exclude the outlier galaxy F26D000, we find that there is one weak correlation: the $g$-band RLF measured within two effective radii slightly correlates with the total $g$-band absolute magnitude, with fainter galaxies having larger RLFs. This distribution has a Pearson correlation coefficient $r_{P}=0.55$ with a $p$-value $=8 \times 10^{-3}$. We note that, similarly, there was no significant trend between the residual light fraction and the galaxy luminosity in the Virgo dwarf ETGs analyzed in Brought to Light I \citep{Michea:2021}. The second noticeable thing is that the distributions have a larger scatter in the $g$ band than in the $r$ band. Overall, we recognize the same behavior we observed in Figure \ref{fig:plot_RLF_g_vs_r}: the clumpy subsample tends to have larger residual light fractions than the disky subsample in both the $g$ and $r$ bands, albeit the difference is more substantial in the $g$ band.

\begin{figure*}[ht]
\centering
\includegraphics[width=0.84\textwidth]{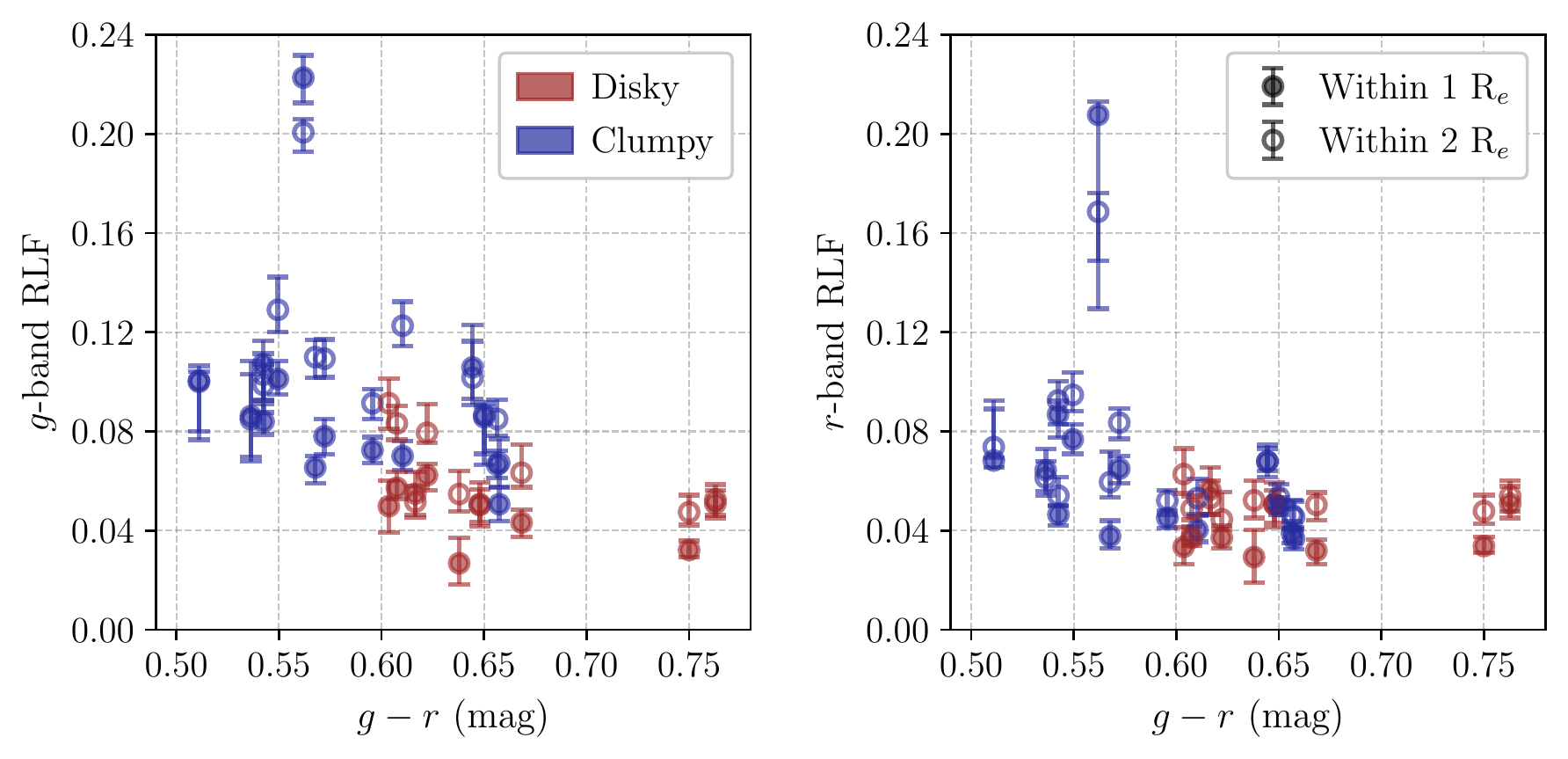}
\vspace*{-4pt}
\caption{Residual light fractions of our Fornax dwarf ETG sample as a function of their total $g-r$ color. \textit{Left panel:} $g$-band residual light fraction vs. total $g-r$ color. \textit{Right panel:} $r$-band residual light fraction vs. total $g-r$ color. Galaxies classified as disky are shown as red points, while galaxies classified as clumpy are shown as blue points. The measurements are performed within one and two effective radii of the galaxies, shown as filled and empty circles, respectively. The error bars are given by the 16th and 84th percentiles of the distributions.\label{fig:plot_RLF_vs_color}}
\end{figure*}

In Figure \ref{fig:plot_RLF_vs_color}, we plot the $g$ and $r$-band residual light fractions versus the total $g-r$ color of the galaxies. We can clearly appreciate how the residual light fraction in both bands increases with bluer galaxy colors, although the correlation is stronger in the $g$ band than in the $r$ band. In the $g$ band, the distributions have a Pearson correlation coefficient $r_{P}=-0.51$ ($r=-0.60$) and a $p$-value $=1 \times 10^{-2}$ ($p$-value $=2 \times 10^{-3}$) for the measurements at one and two effective radii, respectively. The correlations are less significant in the $r$ band, with a Pearson correlation coefficient $r_{P}=-0.37$ ($r=-0.45$) and a $p$-value $=8 \times 10^{-2}$ ($p$-value $=3 \times 10^{-2}$) for the measurements at one and two effective radii, respectively. The main result, however, is how the disky and clumpy subsamples become separated by the $g-r$ color of the galaxies. The disky subsample is comparatively redder ($g-r \geq 0.6$ mag), being relegated to the right half of the plots; while the clumpy subsample is comparatively bluer ($g-r \leq 0.65$ mag), and is thus relegated to the left half of the plots. In the upcoming Section \ref{sec:color}, we provide a more in-depth view into these color differences, as we compute and analyze the $g-r$ colors of the diffuse and substructure components of the galaxies separately.

\begin{figure}[ht]
\centering
\includegraphics[width=1.01\columnwidth]{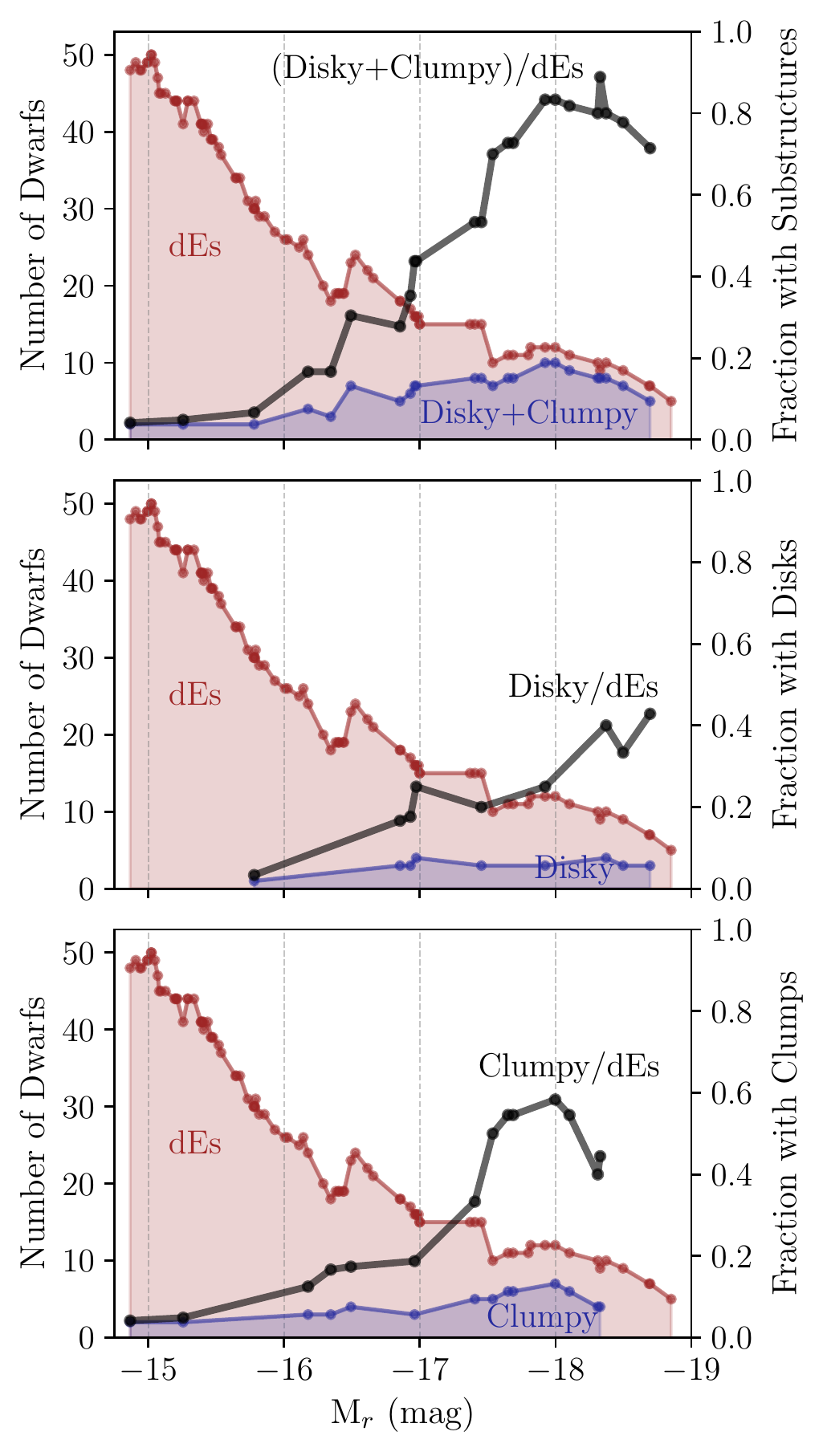}
\vspace*{-2em}
\caption{Luminosity function and fraction with substructures, disks, and clumps of our Fornax dwarf ETG sample. In red, the histogram of all Fornax dwarf ETGs (``dEs'') in the FDSDC as a function of their total $r$-band absolute magnitude (left $y$-axis). Similarly, in blue, the histogram of our galaxy sample that have either disk or clump substructures (\textit{top panel}; ``Disky+Clumpy''), disk substructures (\textit{middle panel}; ``Disky''), and clump substructures (\textit{bottom panel}; ``Clumpy''). The black lines correspond to the ratio of both histograms, and thus represent the relative fraction of each sub-population (right $y$-axis). The histogram bins are calculated at the position of each galaxy, where each bin includes all the galaxies that lie within a range of $\pm0.5$ magnitudes. The histograms were created following the same technique and binning used in Figure $12$ of \citet{Lisker:2006a}, such that the distribution of our Fornax sample can be directly compared to their Virgo sample (see upcoming Section \ref{subsec:discussion_morph}). The luminosity functions are complete, and are plotted out to the faintest galaxy in our sample.}\label{fig:plot_hist_subfrac}
\end{figure}

As a way to compare our sample of dwarf ETGs with substructures with the whole dwarf ETG population of the Fornax cluster, in Figure \ref{fig:plot_hist_subfrac} we plot the brightness distribution of both populations. As a reminder, we have defined the Fornax dwarf ETG population as all of the galaxies classified as early types in the FDSDC plus all of the galaxies in our sample (see Section \ref{subsec:sample_sel}). As the main result, we find that the brighter a dwarf ETG is, the higher the probability of it having embedded substructure features, either disky or clumpy. Thus, as shown in the top panel, the overall fraction of dwarf ETGs with substructures can be as high as $80\%$ on the bright end (M$_{r} \approx -19$ mag), dropping to $5\%$ on the faint end (M$_{r} \approx -15$ mag). If we only consider disk substructures, as shown in the middle panel, the fraction of dwarf ETGs with disk-like features can reach up to $40\%$ on the bright end, dropping to $5\%$ already by M$_{r} \approx -16$ mag. We provide a comparison with the disk fraction of dwarf ETGs in the Virgo cluster in the discussion in Section \ref{sec:discussion}. If now we only consider clump substructures, as shown in the bottom panel of Figure \ref{fig:plot_hist_subfrac}, the fraction of dwarf ETGs with clump-like features apparently follow a similar behavior. However, we caution against a physical interpretation of this result, as it is likely that our selection criteria (which require galaxies to have an early-type morphology) has removed a lot of the lower luminosity dwarf galaxies where clumpy structures are likely even more common (e.g., normal dwarf irregulars), but where the clumps became too bright for the galaxies to be considered early types. Nonetheless, we highlight the fact that substructures, in either the form of disk features or clump features, are highly probable to be present in a dwarf ETG if the galaxy is bright enough (M$_{r} \leq -18$ mag).

\section{Color Analysis}\label{sec:color}

As a means of showcasing the potential of the residual method, we analyze the integrated $g-r$ colors as well as the $g-r$ color profiles of the diffuse and substructure components of the galaxies in our sample.

\subsection{Integrated Colors of the Diffuse and Substructure Components}\label{subsec:color_subs}

As explained in Section \ref{subsec:res_method_params}, the residual method provides us with the $g$ and $r$-band images of the diffuse and substructure components for each galaxy in our sample. When running the method, we imposed the $r$-band galaxy geometry on the $g$ band, thus making any photometric measurements directly comparable in both bands on a pixel-by-pixel (and isophote-by-isophote) basis. Consequently, for each galaxy, we are able to robustly compute the $g-r$ color of its diffuse component out to three galactocentric radii by comparing the $g$ and $r$-band model images, and likewise the color of its substructure component by comparing the $g$ and $r$-band residual images.

We now proceed with a strong note of caution. While our method succeeds in isolating the excess light coming from substructure features in a residual image, the light contained in each pixel still consists of a mix of substructure and diffuse light. Therefore, there is no way to completely separate the photons from substructures and photons from diffuse component in order to measure their true, uncontaminated color, without some kind of additional information that is not available in photometry alone, or without making artificial assumptions about the light profile shape where the substructure and diffuse light are combined. Indeed, in some of the radial light profiles (see Appendix Figure \ref{fig:profiles_app}), the substructure color becomes slightly more blue at radii where the residual light becomes strongest. We cannot be certain if this is due to contamination from substructure features, or because of the complex color profiles that dwarf ETGs are known to have, even in the absence of substructures \citep{Urich:2017}. Nevertheless, some color contamination could arise, especially if the substructures have a much bluer color than the diffuse component. In such a case, we predict that the true color of the substructure may be even bluer than the values we measure, and the diffuse component's color may be slightly redder. However, as the residual fraction is generally only a few percent, the diffuse component's color is unlikely to be significantly altered by the presence of the substructure. Also, in cases where the substructure and residual are observed to have a similar color, obviously we do not have to worry about color contamination.

\begin{deluxetable*}{ccccrr}[ht!]
\setlength{\tabcolsep}{6pt} 
\renewcommand{\arraystretch}{1} 
\tablewidth{0pt}
\tabletypesize{\small}
\tablecaption{$g-r$ color of the diffuse and substructure components of our Fornax dwarf ETG sample.\label{tab:comp_colors}}
\tablehead{
\colhead{Dwarf Galaxy} & \multicolumn5c{$g-r$ Color}  \vspace*{1pt} \\
\cline{2-6}\\ [-2ex]
 & \multicolumn2c{Diffuse Component} & & \multicolumn2c{Substructure Component} \vspace*{1pt} \\
\cline{2-3} \cline{5-6}\\ [-3ex]
 & \colhead{Within $1\,\text{R}_{e}$} & \colhead{Within $2\,\text{R}_{e}$} & & \colhead{Within $1\,\text{R}_{e}$} & \colhead{Within $2\,\text{R}_{e}$} \\
  & \colhead{(mag)} & \colhead{(mag)} & & \colhead{(mag)} & \colhead{(mag)} \\
\colhead{(1)} & \colhead{(2)} & \colhead{(3)} & & \colhead{(4)} & \colhead{(5)} \vspace*{0.5pt}
}
\decimals
\startdata
F01D145 & $0.639^{+0.002}_{-0.020}$ & $0.670^{+0.003}_{-0.031}$ & & $-0.105^{+0.362}_{-0.076}$ & $-0.163^{+0.370}_{-0.038}$ \\ [0.8ex]
F02D000 & $0.732^{+0.002}_{-0.002}$ & $0.734^{+0.003}_{-0.002}$ & & $0.115^{+0.038}_{-0.029}$ & $0.032^{+0.041}_{-0.044}$ \\ [0.8ex]
F04D000 & $0.701^{+0.001}_{-0.030}$ & $0.701^{+0.002}_{-0.031}$ & & $0.029^{+0.405}_{-0.012}$ & $0.094^{+0.405}_{-0.030}$ \\ [0.8ex]
F04D001 & $0.685^{+0.002}_{-0.002}$ & $0.675^{+0.001}_{-0.001}$ & & $0.695^{+0.031}_{-0.040}$ & $0.688^{+0.018}_{-0.030}$ \\ [0.8ex]
F04D002 & $0.682^{+0.018}_{-0.001}$ & $0.689^{+0.027}_{-0.001}$ & & $0.491^{+0.025}_{-0.456}$ & $0.539^{+0.020}_{-0.428}$ \\ [0.8ex]
F04D053 & $0.500^{+0.022}_{-0.002}$ & $0.552^{+0.034}_{-0.002}$ & & $0.250^{+0.016}_{-0.236}$ & $0.263^{+0.013}_{-0.263}$ \\ [0.8ex]
F04D061 & $0.595^{+0.002}_{-0.002}$ & $0.599^{+0.004}_{-0.001}$ & & $0.400^{+0.020}_{-0.017}$ & $0.294^{+0.008}_{-0.013}$ \\ [0.8ex]
F05D000 & $0.543^{+0.006}_{-0.028}$ & $0.560^{+0.004}_{-0.032}$ & & $0.226^{+0.275}_{-0.032}$ & $0.373^{+0.314}_{-0.016}$ \\ [0.8ex]
F07D000 & $0.734^{+0.001}_{-0.001}$ & $0.714^{+0.001}_{-0.001}$ & & $0.760^{+0.013}_{-0.015}$ & $0.744^{+0.008}_{-0.012}$ \\ [0.8ex]
F09D255 & $0.661^{+0.005}_{-0.022}$ & $0.672^{+0.005}_{-0.030}$ & & $0.090^{+0.441}_{-0.205}$ & $0.090^{+0.382}_{-0.056}$ \\ [0.8ex]
F09D492 & $0.559^{+0.007}_{-0.006}$ & $0.598^{+0.005}_{-0.004}$ & & $-0.164^{+0.079}_{-0.046}$ & $-0.123^{+0.055}_{-0.046}$ \\ [0.8ex]
F10D189 & $0.721^{+0.001}_{-0.002}$ & $0.720^{+0.002}_{-0.002}$ & & $0.125^{+0.052}_{-0.029}$ & $0.031^{+0.084}_{-0.055}$ \\ [0.8ex]
F11D279 & $0.815^{+0.001}_{-0.001}$ & $0.787^{+0.001}_{-0.001}$ & & $0.832^{+0.017}_{-0.022}$ & $0.828^{+0.012}_{-0.019}$ \\ [0.8ex]
F14D144 & $0.724^{+0.011}_{-0.010}$ & $0.728^{+0.015}_{-0.015}$ & & $0.358^{+0.229}_{-0.222}$ & $0.288^{+0.252}_{-0.242}$ \\ [0.8ex]
F15D384 & $0.666^{+0.003}_{-0.003}$ & $0.677^{+0.003}_{-0.003}$ & & $0.142^{+0.017}_{-0.035}$ & $0.021^{+0.029}_{-0.017}$ \\ [0.8ex]
F15D417 & $0.660^{+0.002}_{-0.001}$ & $0.646^{+0.001}_{-0.001}$ & & $0.816^{+0.082}_{-0.052}$ & $0.642^{+0.025}_{-0.020}$ \\ [0.8ex]
F17D227 & $0.473^{+0.012}_{-0.026}$ & $0.540^{+0.009}_{-0.029}$ & & $-0.028^{+0.361}_{-0.013}$ & $-0.021^{+0.392}_{-0.015}$ \\ [0.8ex]
F19D001 & $0.784^{+0.001}_{-0.001}$ & $0.796^{+0.001}_{-0.001}$ & & $0.845^{+0.025}_{-0.033}$ & $0.806^{+0.013}_{-0.013}$ \\ [0.8ex]
F22D244 & $0.486^{+0.029}_{-0.005}$ & $0.525^{+0.031}_{-0.005}$ & & $0.311^{+0.016}_{-0.338}$ & $0.374^{+0.024}_{-0.345}$ \\ [0.8ex]
F26D000 & $0.546^{+0.067}_{-0.007}$ & $0.588^{+0.054}_{-0.004}$ & & $0.427^{+0.027}_{-0.327}$ & $0.369^{+0.029}_{-0.306}$ \\ [0.8ex]
F26D003 & $0.709^{+0.021}_{-0.018}$ & $0.705^{+0.021}_{-0.019}$ & & $0.179^{+0.191}_{-0.185}$ & $0.217^{+0.212}_{-0.210}$ \\ [0.8ex]
F26D141 & $0.661^{+0.002}_{-0.003}$ & $0.712^{+0.003}_{-0.004}$ & & $0.032^{+0.070}_{-0.067}$ & $-0.273^{+0.059}_{-0.063}$ \\ [0.8ex]
F31D196 & $0.667^{+0.002}_{-0.002}$ & $0.675^{+0.002}_{-0.002}$ & & $0.118^{+0.014}_{-0.019}$ & $0.025^{+0.020}_{-0.045}$ \\ [0.3ex]
\enddata
\tablecomments{Col. (1): name of the dwarf galaxy. Cols. (2) and (3): $g-r$ color of the diffuse component measured within the isophotes at one and two effective radii, respectively. Cols. (4) and (5): $g-r$ color of the substructure component measured within the isophotes at one and two effective radii, respectively.}
\end{deluxetable*}

In Table \ref{tab:comp_colors} we provide the integrated $g-r$ colors of the diffuse and substructure components of our dwarf ETG sample. These color measurements are integrated within the isophotes at one and two effective radii of the galaxies, for which we derive a median color value and its associated uncertainties from the 16th and 84th percentiles of the distributions. We find that the color of the diffuse component lies in the range between $0.47-0.82$ mag within one effective radius, with a median $g-r = 0.67$ mag; and in the range between $0.53-0.80$ mag within two effective radii, with a median $g-r = 0.68$ mag. In contrast, the substructure component covers a wider color range and is comparatively much bluer on average: its color lies in the range between $-0.16$ and $0.84$ mag within one effective radius, with a median $g-r = 0.23$ mag; and in the range between $-0.27$ and $0.83$ mag within two effective radii, with a median $g-r = 0.26$ mag. It would appear that, when considering the color of an individual component, the measurements within one and two effective radii are quite similar and thus comparable.

\begin{figure*}[ht]
\centering
\includegraphics[width=0.84\textwidth]{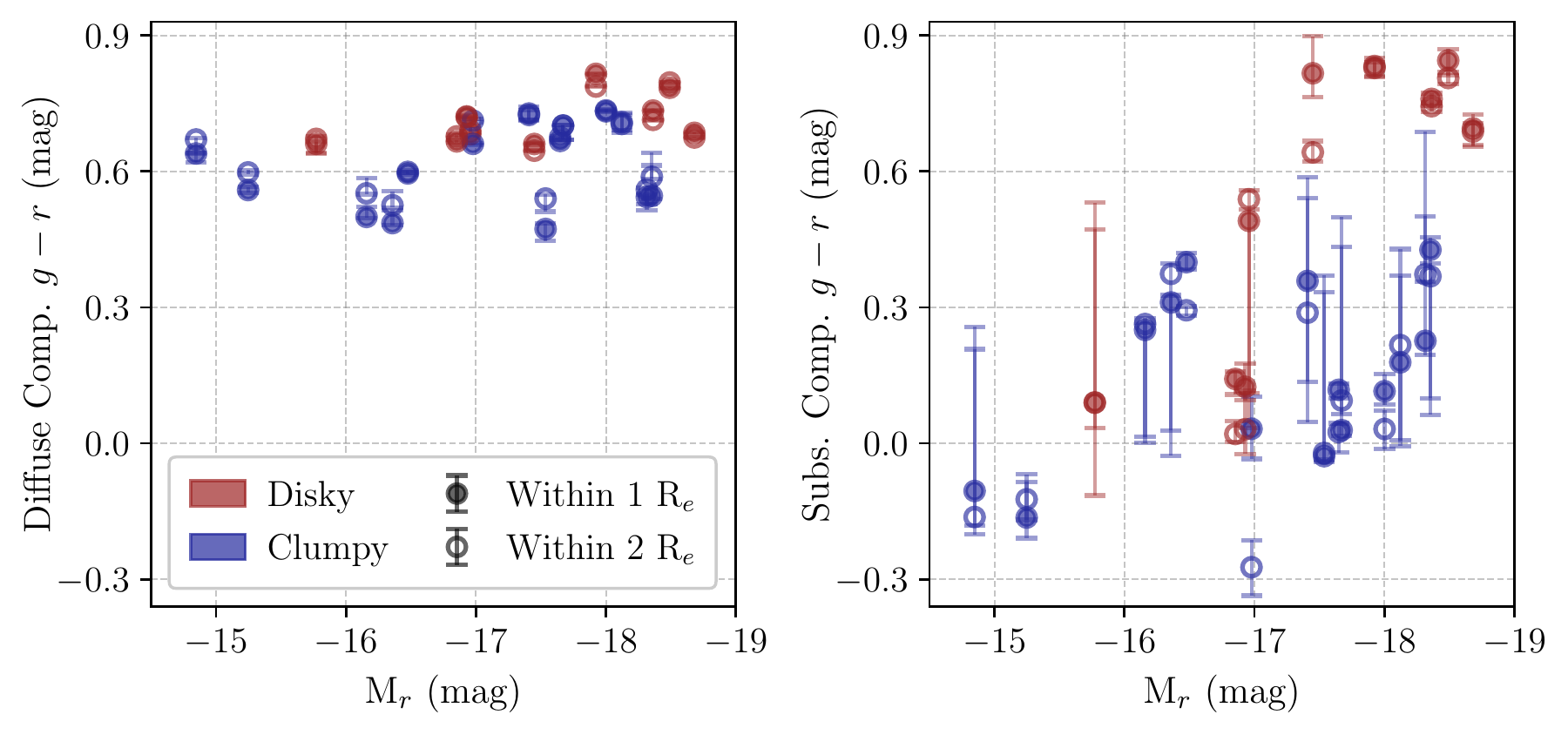}
\vspace*{-4pt}
\caption{$g-r$ color of the diffuse and substructure components of our Fornax dwarf ETG sample as a function of the total $r$-band absolute magnitude of the galaxies. \textit{Left panel:} $g-r$ color of the diffuse component vs. total $r$-band absolute magnitude. \textit{Right panel:} $g-r$ color of the substructure component vs. total $r$-band absolute magnitude. Galaxies classified as disky are shown as red points, while galaxies classified as clumpy are shown as blue points. The measurements are performed within one and two effective radii of the galaxies, shown as filled and empty circles, respectively. The error bars are given by the 16th and 84th percentiles of the distributions.\label{fig:plot_comps_color_vs_mag}}
\end{figure*}

As a way to visualize these results, in Figure \ref{fig:plot_comps_color_vs_mag} we plot the integrated $g-r$ color of the diffuse and substructure components versus the total $r$-band absolute magnitude of the galaxies. We can clearly appreciate that while the diffuse component tends to have more consistently red colors, the substructure component displays a richer diversity, with the majority adopting bluer colors than their diffuse counterparts. Overall, there are weak correlations between the component color and the galaxy brightness: brighter galaxies tend to have redder components. For the diffuse component, we compute a Pearson correlation coefficient of $r_{P}=-0.37$ ($r=-0.32$) and an associated $p$-value $=8 \times 10^{-2}$ ($p$-value $=1 \times 10^{-1}$) for the measurements at one and two effective radii, respectively. In comparison, the correlations are slightly stronger for the substructure component, having an $r=-0.57$ and a $p$-value $=5 \times 10^{-3}$ for both measurements at one and two effective radii. Additionally, we observe that both components in the clumpy subsample tend to have comparatively bluer colors than in the disky subsample. This color difference, however, is significantly more pronounced in the substructure component than in the diffuse component. In other words, clump substructures are on average bluer than disk substructures. If the color of the diffuse component were heavily altered by contamination from the substructures, we might expect that those galaxies with high residual fractions and blue substructures would deviate further downwards from the sequence seen in the left-hand panel. We tested for this possibility but saw little evidence for a dependency on residual fraction and/or color difference, which supports the notion that the color of the diffuse component is not strongly altered by the presence of substructures.

\begin{figure*}[ht]
\centering
\includegraphics[width=0.68\textwidth]{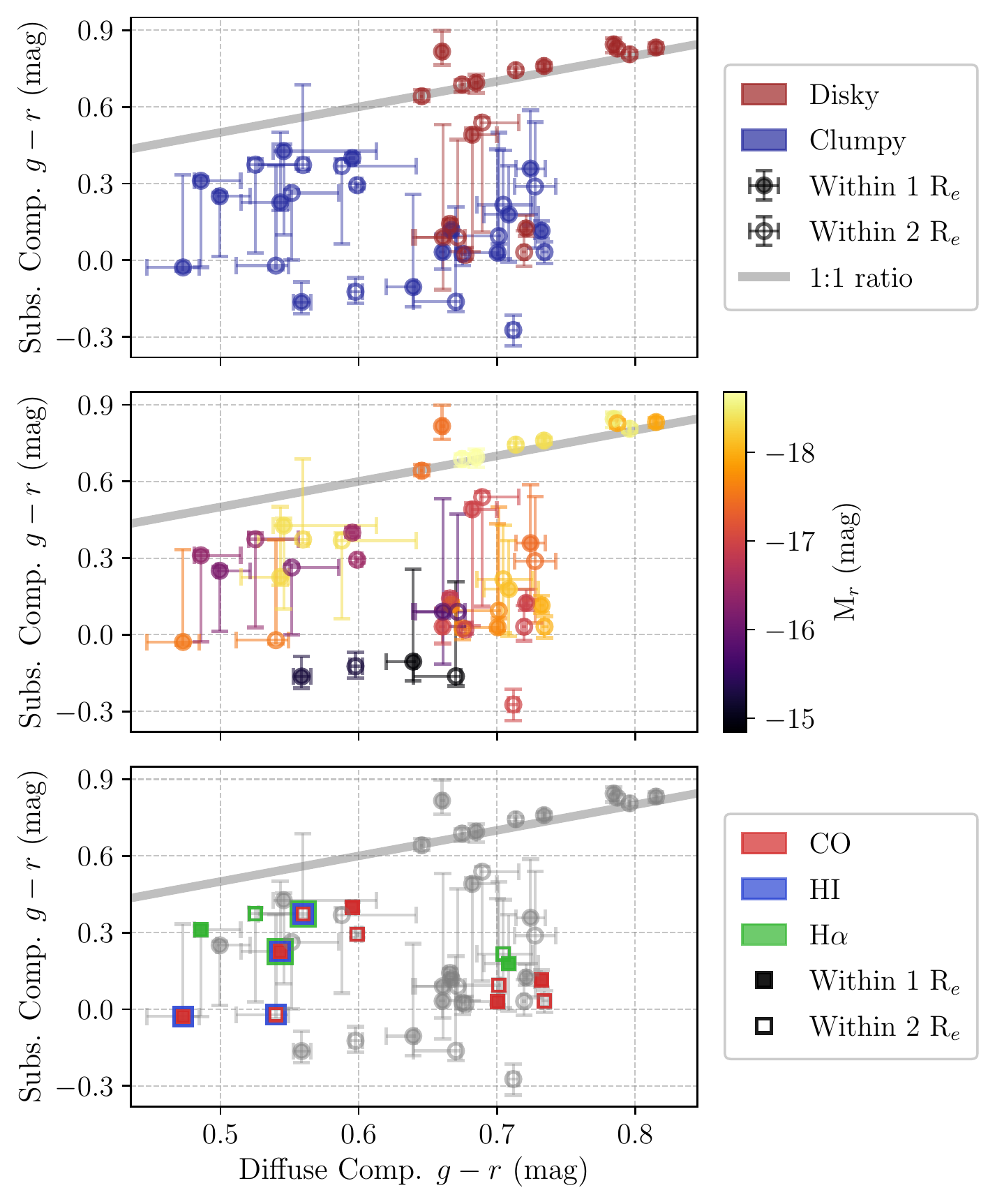}
\vspace*{-4pt}
\caption{Substructure component vs. diffuse component $g-r$ colors of our Fornax dwarf ETG sample. \textit{Top panel:} the galaxies classified as disky are shown as red points, while galaxies classified as clumpy are shown as blue points. \textit{Middle panel:} the galaxies are color coded based on their total $r$-band absolute magnitude, with lighter colors corresponding to brighter galaxies. \textit{Bottom panel:} the galaxies with CO, HI, and H$\alpha$ detections are highlighted by red, blue, and green squares, respectively. In all panels, the measurements are performed within one and two effective radii of the galaxies, shown as filled and empty markers, respectively. The error bars are given by the 16th and 84th percentiles of the distributions. The one-to-one relation is shown as a gray line.\label{fig:plot_comps_color_mod_vs_res}}
\end{figure*}

To provide a different perspective, in Figure \ref{fig:plot_comps_color_mod_vs_res} we plot the integrated $g-r$ colors of the substructure component versus the diffuse component. The colors of the components appear to be weakly correlated. We compute a Pearson correlation coefficient  $r_{P}=0.42$ with an associated $p$-value $=4 \times 10^{-2}$ for the measurements within one effective radius, while no correlation of any significance is obtained for the measurements within two effective radii. In the upper panel, we can appreciate the striking difference between our two subsamples. On the one hand, the clumpy subsample has a substructure component that is always bluer than the diffuse component (i.e., below the one-to-one relation shown in gray). The colors of both components also display a large scatter. On the other hand, the disky subsample is divided into two: five galaxies present a disk substructure as red as the diffuse component, while for four galaxies the disk substructure is notably bluer. We tested but saw no evidence for a dependence between substructure color and clustercentric distance. When considering our dwarf ETG sample as a whole, it is clear that the substructure features are never redder than the diffuse light of a galaxy, a point that will be discussed further in Section \ref{sec:discussion}. Next, in the middle panel, we color-code the measurements based on the total $r$-band absolute magnitude of the galaxies. Corroborating our findings of Figure \ref{fig:plot_comps_color_vs_mag}, we can indeed observe that brighter galaxies tend to present redder colors in both of their components, albeit with a large scatter. Finally, the results shown in the lower panel are addressed in the upcoming Section \ref{subsec:color_ssp}.

\subsection{Color Profile of the Diffuse and Substructure Components}\label{subsec:color_prof}

\begin{figure*}[ht]
\centering
\includegraphics[width=0.85\textwidth]{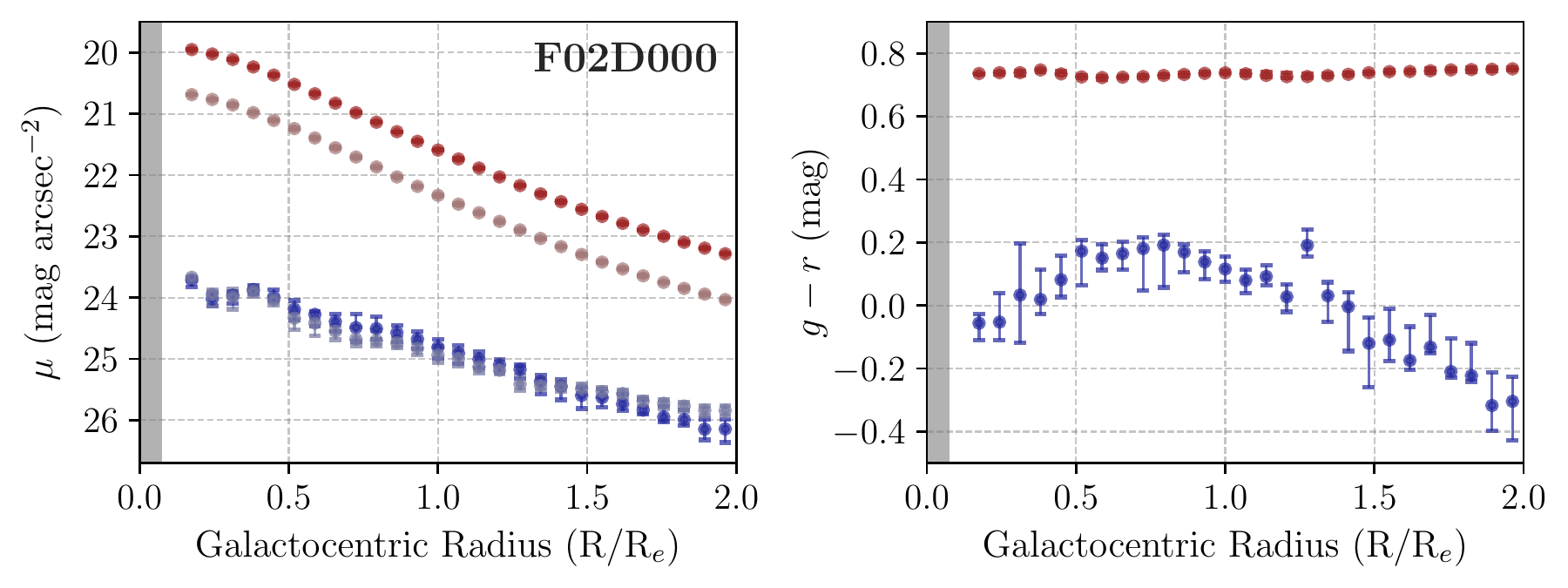}
\vspace*{-4pt}
\caption{Surface brightness and $g-r$ color profiles of the diffuse component (red points) and substructure component (blue points) of the Fornax dwarf galaxy F02D000. \textit{Left panel:} surface brightness profiles in the $g$ and $r$ bands. The bright red and bright blue points correspond to the $r$-band profile, while the faint red and faint blue points correspond to the $g$-band profile. \textit{Right panel:} $g-r$ color profiles. The measurements are performed in elliptical annuli of constant width that match the geometry of the isophotes of the diffuse component of the galaxy out to two effective radii. The central region is excluded due to the effects of the PSF, and disregard an amount equal to 1.5 times the PSF FWHM. The gray-shaded area indicates the extension of the PSF FWHM. For Figures of the complete sample, see Appendix \ref{sec:appendix}.\label{fig:plot_comp_profiles}}
\end{figure*}

To further characterize the diffuse and substructure components, we construct their $g$ and $r$-band surface brightness profiles and $g-r$ color profiles. For the radial profiles to be comparable on both bands, we sample the model and residual images of a galaxy using a fixed set of isophotes. These isophotes follow the geometry of the diffuse component, in order to minimize any effect that the substructure component may have in driving the galaxy geometry. Specifically, we construct concentric elliptical annuli of a constant $5$ pixels ($1$ arcsec) in width that follow the ellipticity and position angle of the isophotes of the $r$-band galaxy model image. These annuli are then used to sample the galaxy model and galaxy residual images in both the $g$ and $r$ bands, extract their light profiles, and construct their surface brightness and color profiles. As an example, we present these radial profiles for the case of the dwarf galaxy F02D000 in Figure \ref{fig:plot_comp_profiles}, while corresponding Figures for the rest of our dwarf ETG sample are provided in Appendix \ref{sec:appendix}.

In regard to the surface brightness profiles of our sample, we observe that the diffuse component is brighter than the substructure component at all galactocentric radii. From the galaxy center out to two effective radii, the diffuse component in the $g$ band is on average between $2.0-3.3$ mag brighter than the substructure component, with a median of $2.6$ mag. Similarly, in the $r$ band, it is on average between $2.2-3.5$ mag brighter, with a median of $3.0$ mag. The diffuse component tends to have a smooth profile that becomes steadily fainter with increasing radius, while in contrast the substructure component presents a more irregular profile with fluctuations, which declines less steeply with increasing radius.

To obtain a more quantitative assessment of these observations, we attempt to model the surface brightness profile of the diffuse component, but not of the substructure component. Due to the fluctuating nature of the latter, it would be very difficult to model it properly. Consequently, we choose to only run the 2D fitting algorithm GALFIT \citep{Peng:2002} on the galaxy model images in the $r$ band, using the $r$-band PSFs constructed in Section \ref{subsec:sample_psfmatch}. As a result, we find that the $r$-band diffuse components of our dwarf ETG sample appear to be described well by a single S\'ersic profile \citep{Sersic:1968} of  index $n$ ranging between $0.60-2.28$, with a median value of $n=1.33$. Clearly, the diffuse components of our sample are more consistent with being described by an exponential disk profile ($n=1$) than by a de Vaucouleurs profile \citep[$n=4$;][]{DeVaucouleurs:1948}. Furthermore, when separated into subsamples, we find that the S\'ersic indices of the disky subsample range between $1.04-2.20$, with a median of $n=1.61$; while the S\'ersic indices of the clumpy subsample range between $0.60-2.28$, with a median of $n=1.16$. Therefore, the diffuse component of the disky subsample presents on average larger S\'ersic indices than the clumpy subsample. In other words, the disky subsample has a comparatively steeper inner profile with more extended wings on the outer profile. These results are consistent with the work of \citet{Su:2021}, who perform multi-component decompositions of the FDS galaxies with GALFIT. Interestingly, for our dwarf ETG sample, the S\'ersic indices they obtain by modeling the $r$-band galaxy images are very similar to the ones we obtain by modeling our $r$-band model images. This is likely because the residual fractions are low enough not to significantly change the overall light distribution.

Finally, similarly to the surface brightness profile, the color profile of the diffuse component tends to be smooth, while the color profile of the substructure component usually presents strong fluctuations. For the majority of the galaxies in our sample, their substructure component is bluer than their diffuse component at all galactocentric radii, although there are also cases in which both components have approximately the same radial color. In conclusion, for our sample, their substructure component is always either bluer than or the same color as their diffuse component at all radii within the two effective radii isophote.

\subsection{Stellar Populations}\label{subsec:color_ssp}

To go one step further, we use the $g-r$ color information we have extracted to shed some light on the properties of the stellar populations of the diffuse and substructure components of our dwarf ETG sample. However, we proceed with caution, as a single color is usually not enough to break the age-metallicity degeneracy \citep{Worthey:1994}.

While taking this degeneracy into account, we make use of the stellar population synthesis models based on the \href{http://research.iac.es/proyecto/miles/}{Medium-Resolution Isaac Newton Telescope Library of Empirical Spectra} \citep[MILES;][]{Vazdekis:2010}, an empirical stellar spectra library. In particular, we use the Extended MILES models \citep[E-MILES;][]{Vazdekis:2012,Vazdekis:2016}, which cover stellar Spectral Energy Distributions (SEDs) in a wider spectral range ($168-5000$ nm). From the E-MILES SEDs, photometric predictions are derived for a variety of photometric filters, encompassing a large range of ages and metallicities. As the only caveat, these photometric predictions assume simple stellar populations (SSPs), meaning that all of the stars are assumed to have formed instantaneously in an initial burst of star formation. This clearly is a simplified approach to a possibly more complex star formation history of a real galaxy but, nonetheless, we use SSPs as a first-order approximation to the problem.

Photometric predictions in the form of AB magnitudes are available in the SDSS $g$ and $r$ bands, allowing us to construct $g-r$ color models that we can directly compare with our observational data. Following \citet{Venhola:2019}, we select SSPs with the following properties. For the initial mass function (IMF), we choose a Kroupa Universal \citep[KU;][]{Kroupa:2001} IMF, which is a multi-part power-law function with a logarithmic slope of $1.30$. For the construction of the evolutionary tracks of the SSPs, we use the Padova 2000 isochrones \citep{Girardi:2000}. Then, we consider all available ages within the approximate age of the Universe, from $0.063$ Gyr to $14$ Gyr, increasing by $\sim 12\%$ between successive steps. Finally, low-mass galaxies ($\text{M}/\text{M}_{\odot} \lesssim 10^{9}$) are expected to have sub-solar metallicities \citep[$\log_{10}(\text{Z}/\text{Z}_{\odot})\approx-0.6$;][]{Gallazzi:2005,Toloba:2015,Sybilska:2017}. However, to provide a wider range as a reference, we consider sub-solar to solar metallicities: $\log_{10}(\text{Z}/\text{Z}_{\odot}) = -1.3$, $-0.7$, $-0.4$, and $0.0$. We note that the observed colors are corrected for Galactic foreground extinction but not for dust extinction intrinsic to the galaxies. Hence, the age estimates derived here are to be considered upper limits. This is particularly true for the age estimates of the substructures whose colors are possibly still contaminated by the dominant, redder diffuse components.

\begin{figure*}[ht]
\centering
\includegraphics[width=1\textwidth]{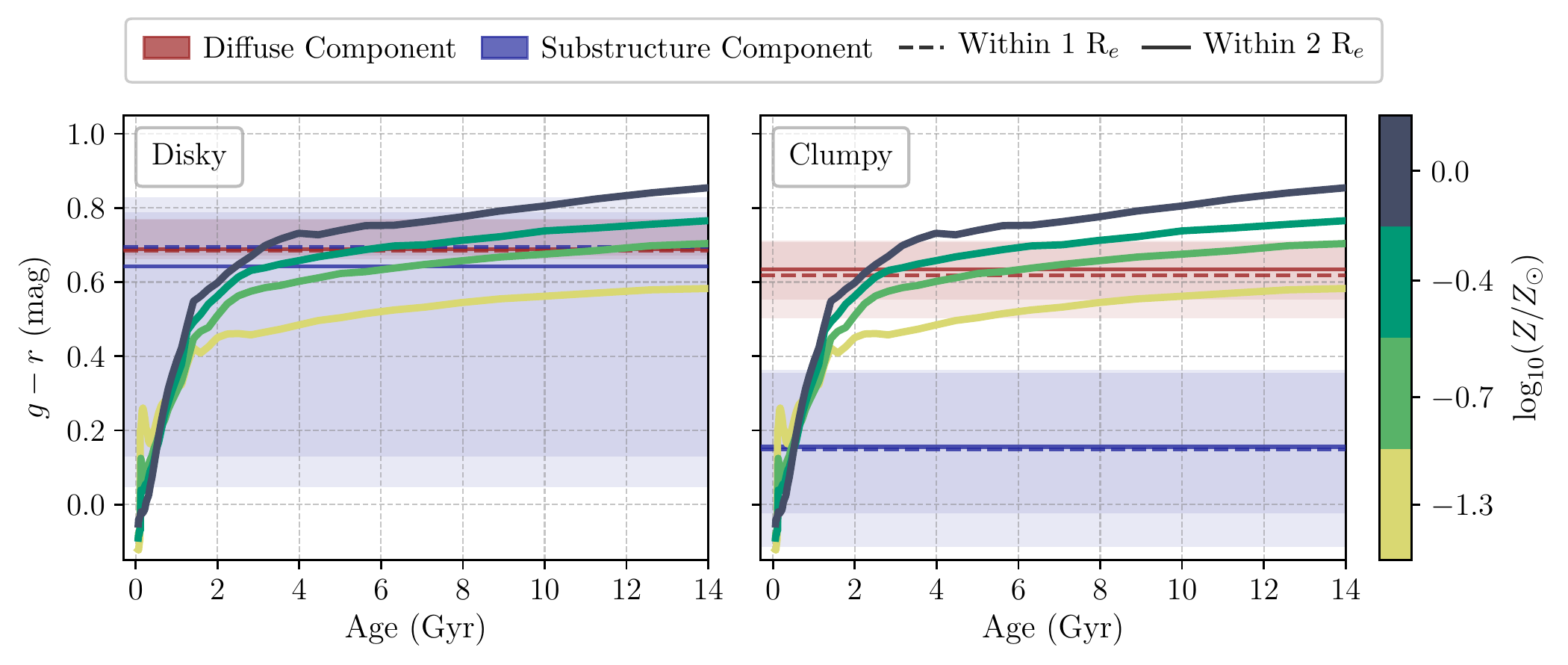}
\vspace*{-15pt}
\caption{Evolution of the $g-r$ color with time, based on the photometric predictions of E-MILES SEDs. The SSP models assume a Kroupa-like IMF and make use of the Padova 2000 isochrones. Four evolutionary tracks with metallicities $\log_{10}(\text{Z}/\text{Z}_{\odot}) = -1.3$, $-0.7$, $-0.4$, and $0.0$ in the age range of $0$ to $14$ Gyr, with ages starting at $0.063$ Gyr and increasing by $\sim 12\%$ between successive age steps, are shown as thick color-coded lines. We also superimpose the average $g-r$ colors of the diffuse and substructure components of our Fornax dwarf ETG sample in red and blue colors, respectively. \textit{Left panel}: average $g-r$ colors of the diffuse and substructure components of the subsample of disky dwarf ETGs. \textit{Right panel}: average $g-r$ colors of the diffuse and substructure components of the subsample of clumpy dwarf ETGs. The median value of the color measurements within one effective radius is shown by a dashed line, while the median value within two effective radii is shown by a continuous line. The shaded regions correspond to the 16th and 84th percentiles of the distributions.\label{fig:plot_emiles}}
\end{figure*}

The above setup allows us to construct four $g-r$ color models, each one with a specific metallicity, evolving from $0$ to $14$ Gyr. In Figure \ref{fig:plot_emiles}, we plot these evolutionary tracks and superimpose the median integrated $g-r$ colors of the diffuse and substructure components (dashed and solid lines in red and blue, respectively) of the disky (left-hand panel) and clumpy (right-hand panel) galaxies in our sample, as well as the 16th and 84th percentiles of their color distributions (shown as shaded red and blue regions, respectively). If we compare the behavior of the evolutionary tracks, we observe that there are no significant differences in color in the first $1$ Gyr. After that, stellar populations that are metal-richer transition into redder colors at a faster rate, and are able to reach redder colors in overall. The age-metallicity degeneracy is evident: a young, metal-rich stellar population can have the same $g-r$ color as a comparatively older, metal-poorer stellar population. Despite this effect, we find that the $g-r$ color of the galaxy components can still give us some insight on their stellar populations. 

On the one hand, the diffuse components of the disky and clumpy galaxies display a small color scatter, and share a similar median $g-r$ color ($0.69$ mag and $0.63$ mag, respectively) within 1$\sigma$. Such a color is consistent with either intermediate-age ($2-4$ Gyr) stellar populations at solar metallicity, or older ($3-14$ Gyr) stellar populations at sub-solar metallicity. On the other hand, the disk substructure components exhibit a large color scatter, since $40\%$ are as blue as the clump substructures (refer to Figure \ref{fig:plot_comps_color_mod_vs_res}). According to the evolutionary tracks in Figure \ref{fig:plot_emiles}, the red ($g-r \geq 0.6$ mag) disks have similar stellar populations as the diffuse components, whereas the blue disks (median $g-r \approx 0.2$ mag) are consistent with stellar populations younger than $\sim 1-2$ Gyr. The latter also applies to the clump substructure components, whose median color is $g-r = 0.16$ mag.

Another way to acquire some knowledge on the stellar populations of galaxies is by considering their gas content. The presence of gas serves as an indicator of possible recent or ongoing star formation, and thus of the existence of a young stellar population. Therefore, we match our sample with the ALMA Fornax Cluster Survey \citep[AlFoSC;][]{Zabel:2019} targeting the CO$(1 - 0)$ emission line, the ATCA HI Survey \citep{Loni:2021}, and the SAMI-Fornax Dwarfs Survey \citep[SAMI-FDS;][]{Scott:2020} for the detection of H$\alpha$ emission. We note that while H$\alpha$ is a strong indicator of recent star formation, HI and CO may just sit there passively without leading to star formation. Conversely, in galaxies with only HI or CO, low-level star formation may be going on leading to the formation of a low number of massive stars, which may not be enough to significantly ionize the interstellar medium (ISM).

\begin{deluxetable*}{ccccccccc}[ht!]
\rotate
\setlength{\tabcolsep}{6pt} 
\renewcommand{\arraystretch}{1} 
\tablewidth{0pt}
\tabletypesize{\small}
\tablecaption{CO, HI, and H$\alpha$ detections in our Fornax dwarf ETG sample.\label{tab:gas_detections}}
\tablehead{
\colhead{Dwarf Galaxy} & \multicolumn2c{CO} & & \multicolumn2c{HI} & & \multicolumn2c{H$\alpha$} \vspace*{1pt}\\
\cline{2-3}\cline{5-6}\cline{8-9}\\ [-3ex]
 & \colhead{Targeted?} & \colhead{Detection} & & \colhead{Targeted?} & \colhead{Detection} & & \colhead{Targeted?} & \colhead{Detection} \vspace*{1pt}\\
\colhead{(1)} & \colhead{(2)} & \colhead{(3)} & & \colhead{(4)} & \colhead{(5)} & & \colhead{(6)} & \colhead{(7)} \vspace*{0.5pt}
}
\decimals
\startdata
F02D000 & Yes & Detected, disturbed & & Yes & Undetected & & No & $-$ \\ [0.8ex]
F04D000 & Yes & Detected, disturbed & & Yes & Undetected & & No & $-$ \\ [0.8ex]
F04D061 & Yes & Detected, disturbed & & Yes & Undetected & & No & $-$ \\ [0.8ex]
F05D000 & Yes & Detected, disturbed & & Yes & Detected, deficient & & Yes & Detected \\ [0.8ex]
F07D000 & No & $-$ & & No & $-$ & & Yes & Undetected \\ [0.8ex]
F10D189 & No & $-$ & & No & $-$ & & Yes & Undetected \\ [0.8ex]
F11D279 & No & $-$ & & No & $-$ & & Yes & Undetected \\ [0.8ex]
F15D384 & No & $-$ & & No & $-$ & & Yes & Undetected \\ [0.8ex]
F15D417 & No & $-$ & & No & $-$ & & Yes & Undetected \\ [0.8ex]
F17D227 & Yes & Detected, disturbed & & Yes & Detected, deficient, disturbed & & No & $-$ \\ [0.8ex]
F22D244 & No & $-$ & & No & $-$ & & Yes & Detected \\ [0.8ex]
F26D003 & No & $-$ & & No & $-$ & & Yes & Detected \\ [0.3ex]
\enddata
\tablecomments{Col. (1): name of the dwarf galaxy. Cols. (2) and (3): detection of the CO$(1-0)$ line by the ALMA Fornax Cluster Survey \citep[AlFoCS;][]{Zabel:2019}. Targeted galaxies by the AlFoCS were previously detected in either HI by the HI Parkes All Sky Survey \citep[HIPASS;][]{Barnes:2001,Waugh:2002} or in the far-infrared by the Herschel Fornax Cluster Survey \citep[HeFoCS;][]{Fuller:2014}. The AlFoCS contains five galaxies of our sample. The morphology and kinematics of the molecular gas of our five galaxies is identified as disturbed, as opposed to regular. Cols. (4) and (5): detection of the HI line by the ATCA HI Survey \citep{Loni:2021}, which covers an area that goes slightly beyond the virial radius of the Fornax cluster \citep[R$_{\text{vir}} = 0.7$ Mpc;][]{Drinkwater:2001}. Two out of our five galaxies also contained in AlFoCS have detections in HI, while the remaining three were not detected. The two detections are deficient in HI, with one of them also having a disturbed HI morphology. Cols. (6) and (7): detection of the H$\alpha$ line by the SAMI-Fornax Dwarfs Survey \citep[SAMI-FDS;][]{Scott:2020}, to be published in Peletier et al. (in prep.). The SAMI-FDS targeted eight galaxies of our sample, of which only three have detections in H$\alpha$ (R. F. Peletier, private communications).}
\end{deluxetable*}

As summarized in Table \ref{tab:gas_detections}, we find that some kind of ISM has been detected in $7$ of our dwarf ETGs: there are $5$ CO, $2$ HI, and $3$ H$\alpha$ detections. Concerning the $5$ CO detections, the morphology and kinematics of the molecular gas of the galaxies is classified as disturbed, as opposed to regular. In regard to the $2$ HI detections, only one galaxy shows a disturbed HI morphology. Interestingly, all of the galaxies in our sample with either CO, HI, or H$\alpha$ detections have been classified in this work as having clump substructures. This means that half of the the galaxies in the clumpy subsample ($7/14$ galaxies) have been confirmed to contain gas, while in contrast in none of the galaxies in the disky subsample ($0/9$ galaxies) any gas was detected.

These results can be visualized in the lower panel of Figure \ref{fig:plot_comps_color_mod_vs_res}. We observe that the galaxies with gas detections have a substructure component that is always bluer than their diffuse component. Therefore, we can hypothesize that their diffuse component is mostly quenched; or, alternatively, their star formation rate is too low to make a difference in the observed $g-r$ color. Given the comparatively bluer color of the substructure component, we would instead expect that the galaxies are forming stars in their clumpy substructure regions, although the resolution of the surveys is too low to confirm a possible spatial correlation. This would imply that the bluer colors of their substructures are caused by the presence of young stellar populations.

\section{Discussion}\label{sec:discussion}

In the first paper of this series \citep[Brought to Light I,][]{Michea:2021} we presented the residual method, a new approach to identify, extract and measure substructure features embedded in dwarf ETGs. We applied it to nine bright dwarf ETGs in the Virgo cluster, which were imaged with a white filter, and revealed the presence of bars and spiral arms contributing between $2.2-6.4\%$ of the total galaxy light as measured within the galaxy two effective radii and in the $r$-band equivalent.

Here we have taken that kind of analysis a step further, and applied our residual method to the $g$ and $r$-band images of $23$ dwarf ETGs in the Fornax cluster, with the aim of deriving not only the morphology and the contributed light, but also the $g - r$ color of their substructure features and diffuse components. Our targeted galaxies were extracted from the Fornax Deep Survey Dwarf Catalog of \citet{Venhola:2018}, according to their overall early-type morphology, red integrated $g - r$ color, and the presence of disk/clump substructure features in their unsharp mask images.

\subsection{Implications of Substructure Morphology}\label{subsec:discussion_morph}

As part of the analysis of our Fornax dwarf ETG sample (see Section \ref{subsec:res_method_results}), we showed how the fraction of galaxies that display disk-like features is dependent on the galaxy brightness, steeply decreasing as the dwarf ETGs become fainter. As illustrated in the middle panel of Figure \ref{fig:plot_hist_subfrac}, we find that the disk fraction of Fornax dwarf ETGs can be as high as $40\%$ on the bright galaxy end, dropping to a low $5\%$ on the faint galaxy end, where we define the location of the faint end as three magnitudes below from the bright end. In comparison, while \citet{Lisker:2006a} observe that the same overall behavior is followed by the disky dwarf ETGs of the Virgo cluster, they find a comparatively higher disk fraction. As shown in the top panel of Figure 12 of \citet{Lisker:2006a}, the disk fraction of the Virgo dwarf ETGs corresponds to about $60\%$ on the bright end, dropping to $5\%$ once reaching the faint end (again defined to be three magnitudes below). This indicates that dwarf ETGs with disk substructures are in relative terms more numerous in Virgo compared to Fornax.

When comparing the results of \citet{Michea:2021} for nine dwarf ETGs in Virgo with those obtained here for $23$ dwarf ETGs in Fornax, we notice an interesting difference. In the Virgo sample, we detected numerous examples of spectacular spiral-like features, such as tightly wrapped grand-design spiral arms. Meanwhile, in the Fornax sample we generally do not detect these dramatic spiral patterns. We do detect some edge-on disks that form the basis of our disky sample, and there are two or three cases that appear to contain bars. However, the lack of tightly wrapped spiral arm structures is unexpected, and might indicate some fundamental difference, either in the dwarf internal properties or in their external environment, between the Fornax and Virgo samples.

Another less physical reason for this discrepancy could be a selection bias, as the Virgo galaxies analyzed by \citet{Michea:2021} were hand-picked based on the appearance of spiral features in their SDSS images. Nevertheless, several more galaxies like these ones have been reported to exist in the Virgo cluster \citep{Lisker:2006a}. Another explanation could be an observational bias. Although not detailed here, we ran a series of tests on the spatial resolution and depth of the data, and found that the better resolution and shallower depth of the Fornax (FDS) images compared to the Virgo (SDSS) ones is not fundamentally hindering our ability to detect substructure features of any kind, including spiral-like features. We should thus be capable to detect substructures of similar brightness and appearance in Fornax as the ones observed in our Virgo sample. We also believe it rather unlikely that all of the disky Fornax dwarf ETGs with spiral arms are systematically projected edge-on with respect to our line of sight.

The differences between the disky dwarf ETG populations in Virgo and Fornax could therefore be physical, and possibly related to the different properties of these two clusters. As a matter of fact, Virgo is more massive ($4 \times 10^{14}$ M$_{\odot}$) than Fornax \citep[$6 \times 10^{13}$ M$_{\odot}$,][]{McLaughlin:1999,Drinkwater:2001,Weinmann:2011}, has a denser and hotter intracluster medium \citep{Jones:1997,Schindler:1999,Paolillo:2002,Janz:2021}, and is dynamically younger with a larger fraction of star-forming galaxies \citep{Ferguson:1989b}. This would make environmental effects such as galaxy-cluster harassment \citep{Moore:1996}, ram-pressure stripping \citep{Gunn:1972}, and thermal evaporation \citep{Cowie:1977a,Cowie:1977b} comparatively more efficient in triggering disk substructure features in Virgo dwarf ETGs.

In the second paper of this series \citep[Brought to Light II,][]{Smith:2021} we simulated dwarf ETGs along different orbits in a Virgo-like cluster potential. We found that tidal triggering can reproduce the disk substructures observed in Virgo dwarf ETGs only when a galaxy contains a cold and thin disk embedded in a more massive and hot disk, and experiences very plunging pericenter passages ($r_{\text{peri}} < 0.25 \,r_{\rm \text{vir}}$). Therefore, perhaps the dwarf ETGs in Virgo contain more cold stellar disk substructures, and/or their orbits could be more plunging than in Fornax. In a forthcoming paper of this series (Smith et al. in prep.), we will carry out a more in-depth analysis of substructures in Virgo dwarf ETGs using deep multi-wavelength survey data \citep[rather than the small sample of pointed observations that were used in][]{Michea:2021} for comparison with the Fornax dwarf ETG population.

\subsection{Implications of Substructure Color}\label{subsec:discussion_color}

While the disky dwarf ETG populations in Virgo and Fornax seem to differ in the morphology of their substructure features, they appear instead to be comparable in their residual light fraction in the $r$-band at fixed M$_r$ range \citep[compare our Figure \ref{fig:plot_RLF_vs_mag} with Figure 5 of][]{Michea:2021}. This would exclude a dependence of the stellar light and mass of the disk substructures on the properties of the host environment. 

For what concerns our Fornax sample alone, we note that the disky dwarf ETGs tend to comprise a brighter magnitude range, and have redder integrated $g-r$ colors. In comparison, the clumpy dwarf ETG tend to comprise a wider magnitude range that can reach fainter magnitudes, and have bluer (``less red'') global $g-r$ colors. Moreover, disk substructures tend to contribute a smaller fraction of the total galaxy light (i.e., have smaller RLFs in both $g$ and $r$ bands) independently of the integrated $g-r$ color of their host galaxy. In comparison, clump substructures tend to contribute a larger fraction of the total galaxy light (i.e., have larger RLFs) which increases with bluer $g-r$ colors of their host galaxy. We also note that some of these clumpy dwarf ETGs have been detected either in H$\alpha$, HI or CO, hence they are relatively more gas-rich than their disky counterparts.

In all cases, we measure that substructures are either similar or bluer in color than their diffuse components. If we assume their color to indicate the ages of the stellar populations (considering all caveats discussed throughout Section \ref{sec:color}), we might conclude that the substructures were either formed at the same time as the main disk (in the case of matching color), or were subsequently added to an existing galaxy at some later stage (in the case of bluer substructures). The color of the clump substructures is consistent with them being regions of on-going or recent star formation in host galaxies that are likely on the way to quench their star formation activity. This picture is also supported by the detected presence of gas, either cold or ionized. We note that, as addressed in Section \ref{subsec:color_subs}, all color measurements that we provide are not free from contamination, as the residual method does not allow us to fully disentangle the true underlying color of each component based on imaging alone. As such, we believe that our analysis would deeply benefit if supplemented with spectroscopic data, which would provide additional insight on the stellar populations (and also the dynamics) of the galaxies. Several studies have already shown how Integral Field Unit (IFU) spectroscopy can be used to thoroughly dissect and extract the properties of galaxies of the Fornax cluster \citep{Iodice:2019,Johnston:2020,Lara-Lopez:2022}, including dwarf galaxies. This is something we would like to incorporate in our future work.

Interestingly, the clumpy dwarf ETGs in our sample are spatially located, in projection, at the Fornax virial radius or outside it (see Figure \ref{fig:plot_all_CMD_spatial}). Their spatial distribution could indicate that the clumpy dwarf ETGs were originally star-forming galaxies, such as dwarf irregulars. One possibility is that these dwarf irregulars are having their star formation suppressed, which might explain why clumps make up only a small fraction of their total light compared to the more dominant star forming regions of normal dwarf irregulars. If so, the quenching process appears to have occurred beyond the cluster virial radius, perhaps because their low masses make them more susceptible to environmental effects. These may be examples of backsplash galaxies \citep{Gill:2004}, objects that previously entered the cluster but are now located beyond the virial radius. Alternatively, their quenching process may have begun in substructures beyond the cluster \citep[so-called ``pre-processing'';][]{Mihos:2004}. Indeed, four of them appear to be associated with the infalling Fornax A subgroup.

Conversely, all of the disky dwarf ETGs except one are found at or within the cluster virial radius, and none of them are at the position of the Fornax A subgroup. As they are deeper inside the Fornax cluster, they could have been subjected to stronger environmental effects such as strangulation and ram-pressure stripping, which could explain why a large fraction of them show red colors, as if their star formation has been quenched. However, we saw no evidence that the blue disky sample is spatially distributed in a different way from those dwarf ETGs with red colors. The progenitors of these galaxies could have been simple dwarf ETGs that later gathered a new gas disk, perhaps via a merger of cosmic web gas accretion. Alternatively, their progenitors may have been late-type dwarf spiral galaxies that already contained a thick and thin disk component since formation.

In summary, we hypothesize that dwarf ETGs with clumpy substructures are likely a transition population, caught in the transformation from normal star-forming dwarf irregulars to non-star-forming dwarf ETGs. The origin of the disky substructures is less clear, and they may have simply formed that way or later formed the substructures via gas accretion or a merger. Follow-up spectroscopy is needed in order to use their stellar properties, such as age and metallicity, and their star formation histories to better understand their evolutionary path(s). Nevertheless, although we do see the presence of thin disk substructures, and in rare cases bars, we notice an absence of the tightly wrapped spiral features that were seen in our previous study of Virgo dwarf ETGs \citep{Michea:2021}. We will investigate this apparent difference in an upcoming study.

\section{Summary}\label{sec:summary}

In this work, we analyzed the $g$ and $r$-band images of $23$ dwarf ETGs with disk and clump substructure features that are part of the Fornax galaxy cluster.

As described in Section \ref{sec:data}, the imaging data consist of a deep, multi-band ($u$, $g$, $r$, and $i$) survey centered on the Fornax cluster and out to the virial radius, which additionally includes the Fornax A infalling subgroup. We also had access to the Fornax dwarf galaxy catalog of \citet{Venhola:2018}, comprised by $564$ dwarf galaxies that were identified as cluster members.

In Section \ref{sec:sample}, we described how we processed the $g$ and $r$-band images of all the dwarf galaxies in the catalog. First, we constructed PSF models of each field in the survey, and matched the PSFs of the $g$ and $r$-band galaxy cutouts. Then, we coadded the PSF-matched $g$ and $r$-band galaxy cutouts, and created a series of unsharp mask images for each one. Finally, we inspected the $g+r$ galaxy images and their unsharp mask images in order to identify and select the dwarf ETGs with substructure features, and constructed our dwarf ETG sample. We divided our sample into dwarf ETGs with disk-like features and dwarf ETGs with clump-like features.

We applied the residual method to our dwarf ETGs sample in Section \ref{sec:res_method}. Here, we explained how we adapted the method for the application in two different bands (the $g$ and $r$ bands), and specified the adopted parameter configuration. As a result, we obtained that the $g$-band residual light fraction has a median value of $6.7\%$ and $8.7\%$ within one and two effective radii, respectively. In contrast, the $r$-band residual light fraction is comparatively smaller, with a median value of $4.6\%$ and $5.3\%$ within one and two effective radii, respectively. We found that the $g$ and $r$-band residual light fractions of a galaxy strongly correlate with each other. We also observed a different behavior between the disky and clumpy sub-classes: disk substructures tend to be fainter, and are found in galaxies with redder $g-r$ global colors, while clump substructures tend to be brighter, and are found in galaxies with bluer $g-r$ global colors. We also found that the fraction of Fornax dwarf ETGs that have substructures can be as high as $80\%$ on the bright galaxy end, and these can be broadly split in half into clumpy and disk substructures. At the faint end, located only three magnitudes away, this fraction steeply drops to $5\%$. Meanwhile, the fraction of an individual galaxy's light that is in the form of substructures is not a strong function of galaxy luminosity.

In Section \ref{sec:color}, we performed a color analysis on the dwarf ETG sample, in which we considered the diffuse and substructure components of the galaxies separately. We computed both integrated $g-r$ colors and radial $g-r$ color profiles. We found that the diffuse component tends to have significantly redder $g-r$ colors than the substructure component. In particular, disky dwarf ETGs have overall redder diffuse components, and their two components tend to have similar colors. Clumpy dwarf ETGs have overall bluer diffuse components, and their substructure component tends to be bluer than their diffuse component. Finally, we carried out a basic stellar population analysis of the two components. We found that the median $g-r$ colors of the diffuse component suggest the dominance of middle-aged to old stellar populations ($>3$ Gyr). In contrast, for the substructure component they suggest significantly younger stellar populations ($\leq 1-2$ Gyr) for both clumpy substructures and blue disky substructures, while red disky substructures share a similar age as their diffuse component. However, we note the caveat that we cannot rule out a change in the substructure color due to flux contamination from the diffuse component, thus our age estimates of the substructures are likely upper limits (even more so, as we did not correct the observed colors for the galaxy intrinsic extinction).

Finally, in Section \ref{sec:discussion} we compared the morphology of substructures in dwarf ETGs of the Virgo and Fornax clusters, and noted that disky dwarf ETGs appear to be more numerous in Virgo and to more often show spiral arms than dwarf ETGs in Fornax. These differences may be due to the different properties of the two clusters, and clearly call for a more in-depth investigation, which we will carry out in a forthcoming paper. We then made use of the spatial and color distribution of our sample to discuss the possible evolutionary paths of disky and clumpy dwarf ETGs in Fornax. On the one hand, clumpy dwarf ETGs could be normal dwarf irregulars that are caught in the transition phase towards passive objects. They are spatially located in the cluster outskirts and in Fornax A subgroup. If their quenching is environmentally driven, then it seems that intermediate density environmental mechanisms are sufficient to drive the transition, unless these galaxies are backsplash galaxies that used to be much deeper inside the cluster in the past. On the other hand, for the disky dwarf ETGs, the origins of their substructures are less clear. The presence of thin disk substructures that are bluer than their diffuse component could suggest that a thin disk was later added to an already existing thick disk, for example by the addition of new gas from a merger or the cosmic web. In those cases where the thin disk has a similar color to the diffuse component, its origin is less certain. Perhaps the thin disk was added so long enough ago that it has since stopped forming stars, and now shares a similar optical color as its diffuse component. Or, perhaps, the two components were formed simultaneously. Differentiating these possibilities requires a more detailed analysis of their separate stellar populations than is not possible with optical imaging alone, but a spectroscopic analysis would be most valuable for a better understanding of both clumpy and disky dwarf ETGs.

\acknowledgments

We thank the anonymous referee for their thoughtful comments that helped to improve the manuscript. J.M. acknowledges the support of the Deutscher Akademischer Austauschdienst (DAAD) through a doctoral scholarship, and the funding provided by Universit\"at Heidelberg and the Max-Planck-Institut f\"ur Astronomie. We also acknowledge support from the European Union’s Horizon 2020 research and innovation program under the Marie Sklodowska-Curie grant agreement no. 721463 to the SUNDIAL ITN network.

\software{Astropy \citep{Astropy:2013, Astropy:2018}, IRAF \citep{IRAF:1986, IRAF:1993}, NumPy \citep{Numpy:2006, Numpy:2011}, Matplotlib \citep{Matplotlib:2007, Matplotlib:2020}, pandas \citep{pandas:2010, pandas:2020}, SAOImage DS9 \citep{ds9:2003, ds9:2019}, scikit-image \citep{skimage:2014}, SciPy \citep{Scipy:2020}, STSDAS \citep{stsdas:1994}.}

\vspace*{2.85em}
\bibliographystyle{aasjournal}{}
\bibliography{references}

\appendix

\vspace{-5pt}
\section{Radial Profiles of the Diffuse and Substructure Components}\label{sec:appendix}

\begin{figure}[ht]
\centering
\includegraphics[width=0.85\textwidth]{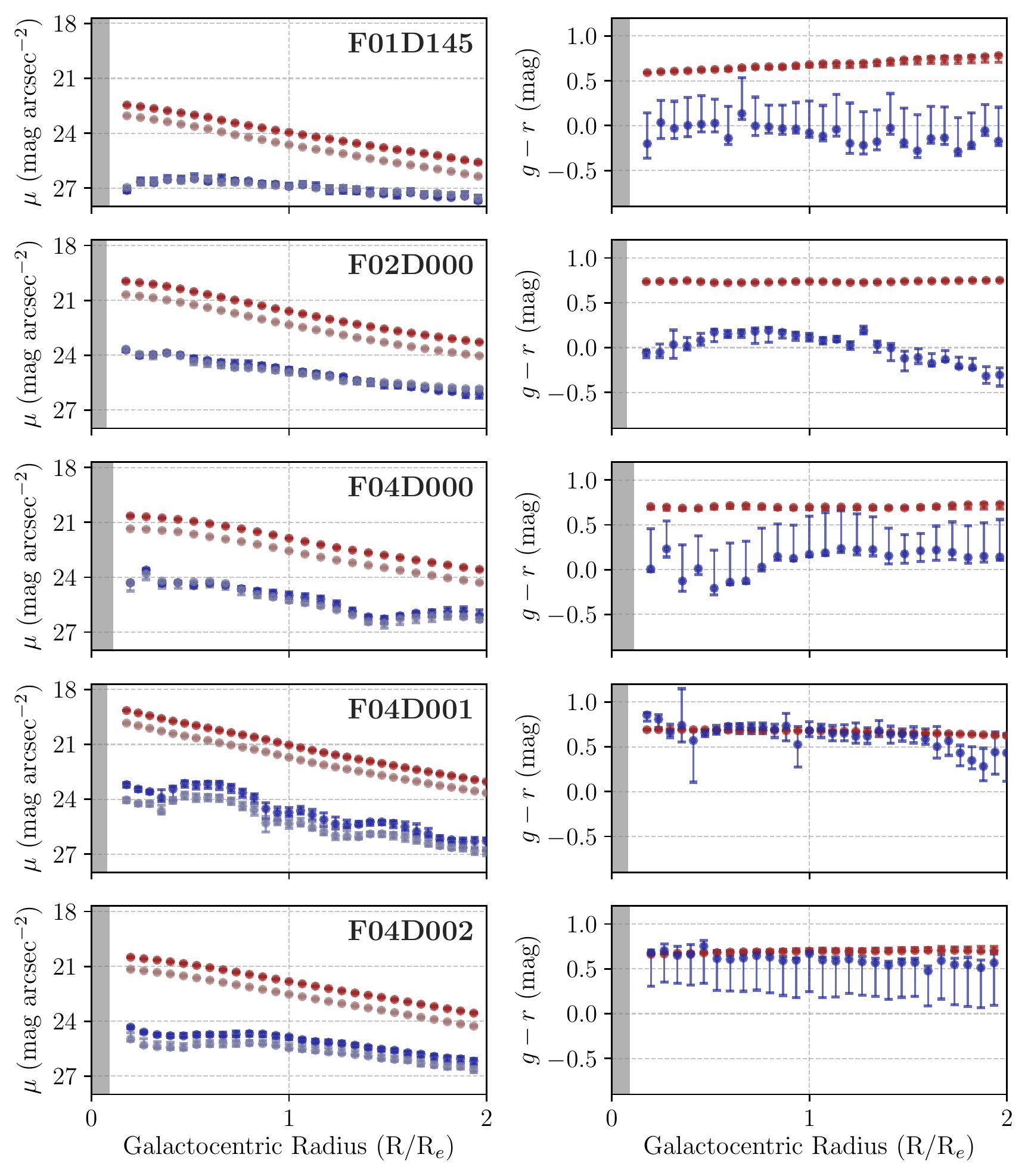}
\vspace*{-4pt}
\caption{Surface brightness and $g-r$ color profiles of the diffuse component (red points) and substructure component (blue points) of our Fornax dwarf ETG sample. \textit{Left panels:} surface brightness profiles in the $g$ and $r$ bands. The bright red and bright blue points correspond to the $r$-band profile, while the faint red and faint blue points correspond to the $g$-band profile. \textit{Right panels:} $g-r$ color profiles. The measurements are performed in elliptical annuli of constant width that match the geometry of the isophotes of the diffuse component of the galaxies out to two effective radii. The central region is excluded due to the effects of the PSF, and disregard an amount equal to 1.5 times the PSF FWHM. The gray-shaded area indicates the extension of the PSF FWHM.\label{fig:profiles_app}}
\end{figure}

\addtocounter{figure}{-1}
\begin{figure}[ht]
\centering
\includegraphics[width=0.85\textwidth]{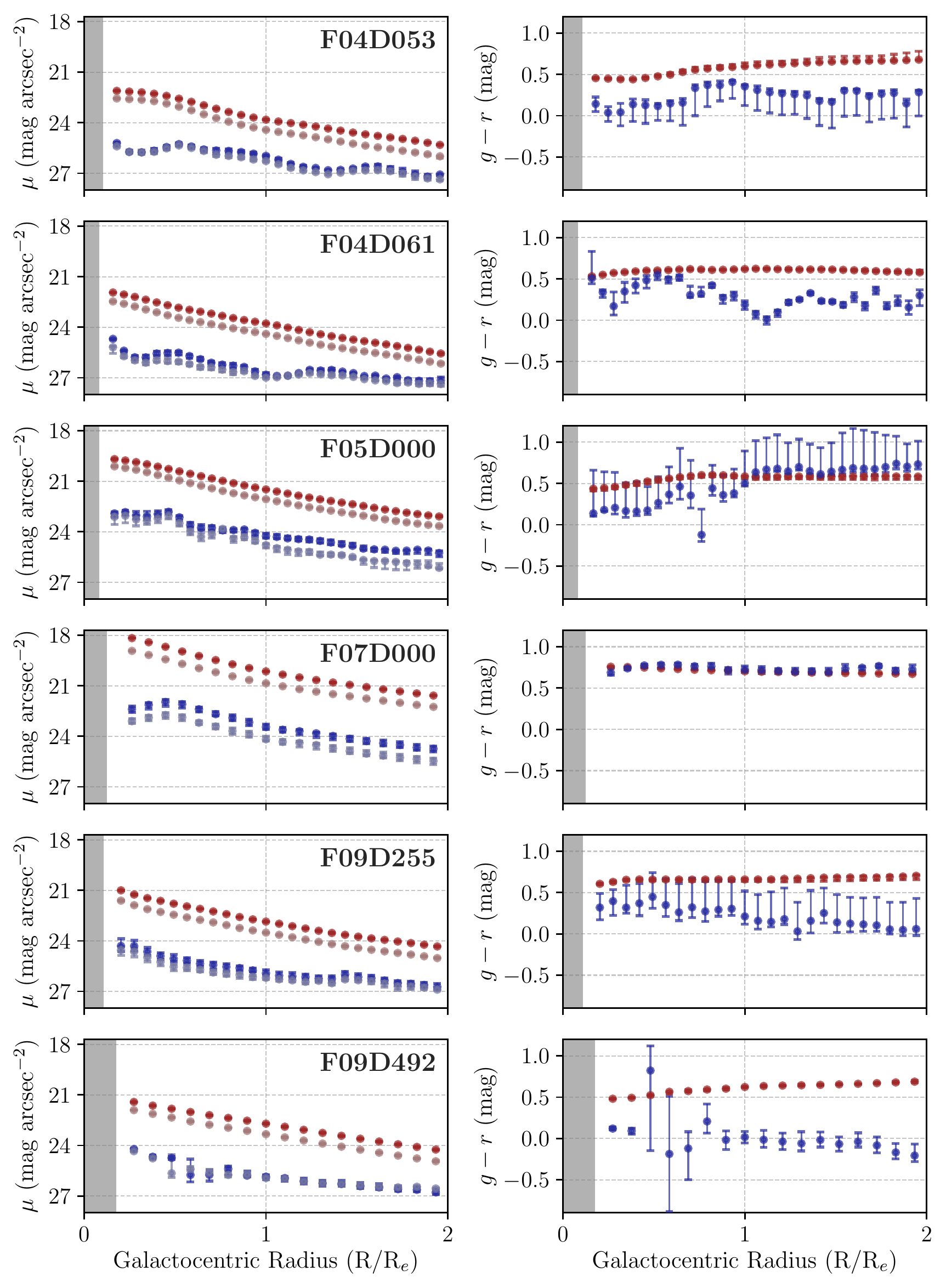}
\vspace*{-4pt}
\caption{Surface brightness and $g-r$ color profiles of the diffuse and substructure components of our Fornax dwarf ETG sample \textit{(cont.)}.}
\end{figure}

\addtocounter{figure}{-1}
\begin{figure}[p]
\centering
\includegraphics[width=0.85\textwidth]{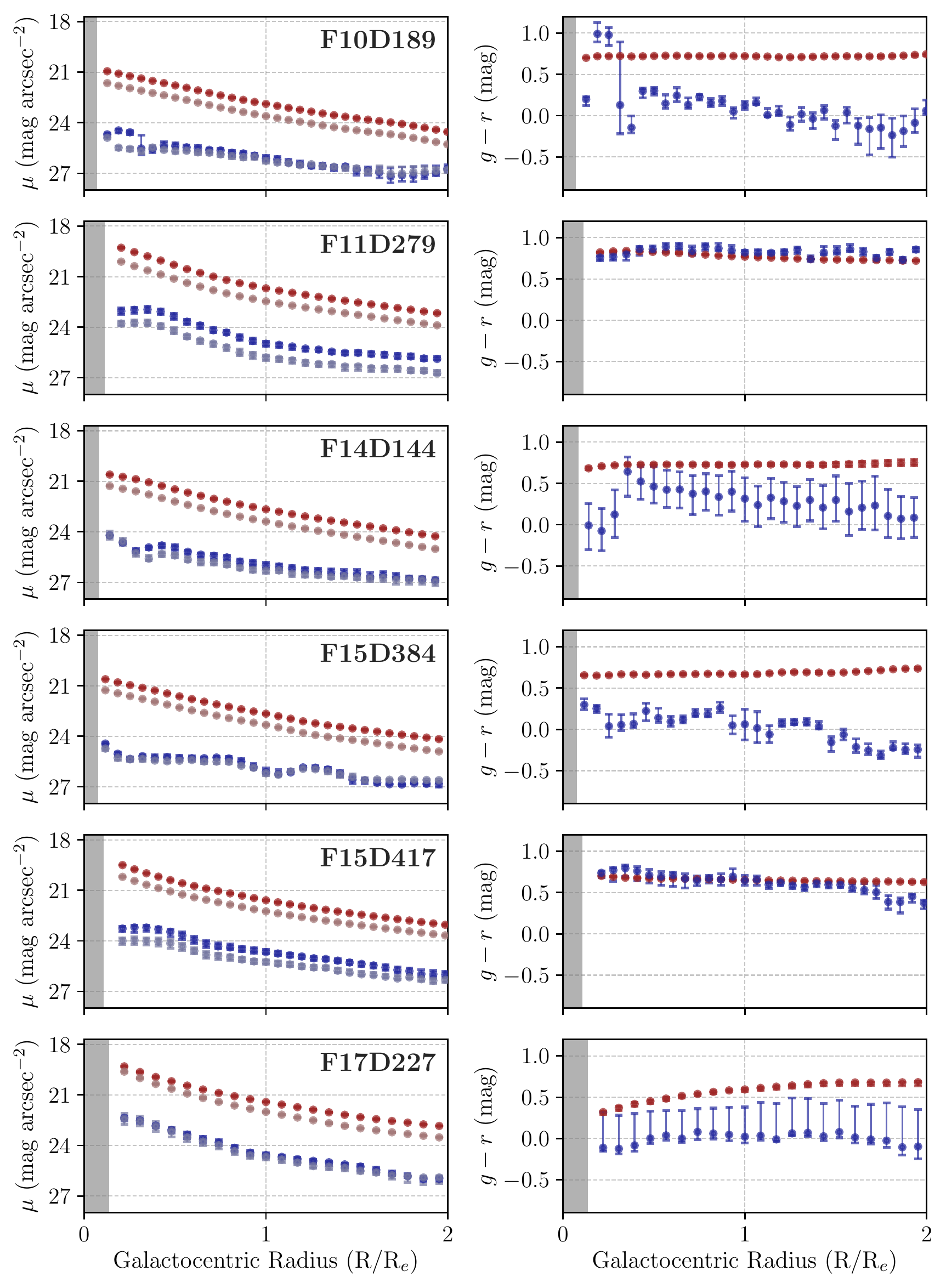}
\vspace*{-4pt}
\caption{Surface brightness and $g-r$ color profiles of the diffuse and substructure components of our Fornax dwarf ETG sample \textit{(cont.)}.}
\end{figure}

\addtocounter{figure}{-1}
\begin{figure}[p]
\centering
\includegraphics[width=0.85\textwidth]{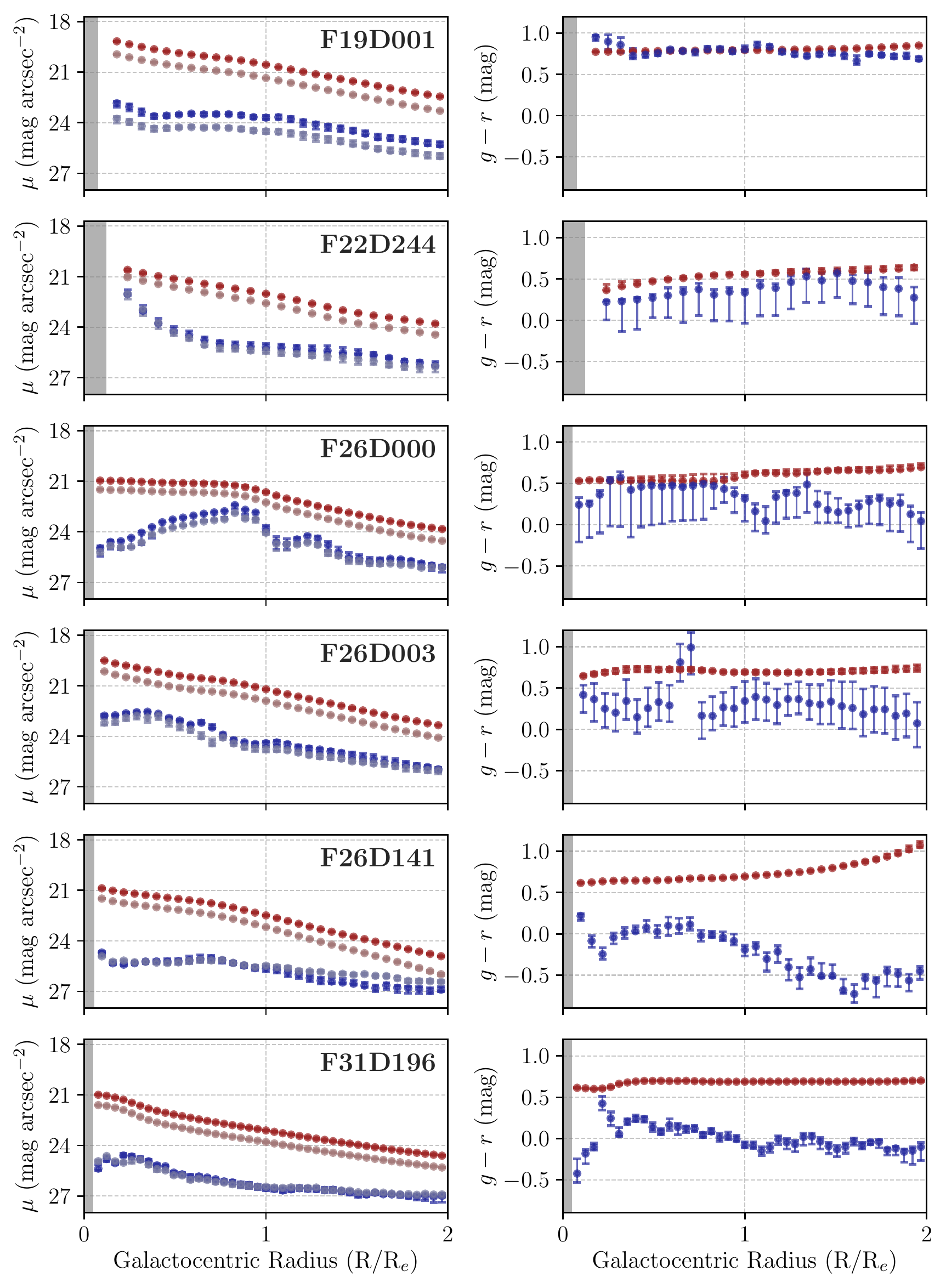}
\vspace*{-4pt}
\caption{Surface brightness and $g-r$ color profiles of the diffuse and substructure components of our Fornax dwarf ETG sample \textit{(cont.)}.}
\end{figure}

\end{document}